\documentclass[12pt]{article}
\usepackage{latexsym}
\usepackage{amssymb}
\usepackage{amsmath}
\usepackage{graphicx}
\usepackage{enumitem}
\usepackage{slashed}
\usepackage{hyperref}
\usepackage{xcolor}

\def\be{\begin{equation}}
\def\ee{\end{equation}}
\newcommand\bra[1]{{\langle {#1}|}}
\newcommand\ket[1]{{|{#1}\rangle}}

\def\dd{\mbox{d}}

\def\bra{\langle}
\def\ket{\rangle}
\def\a{\alpha}
\def\b{\beta}
\def\g{\gamma}
\def\d{\delta}

\def\g{\gamma}
\def\G{\Gamma}
\def\e{\epsilon}

\def\f{\phi}

\def\k{\kappa}
\def\l{\lambda}
\def\L{\Lambda}
\def\m{\mu}
\def\n{\nu}
\def\s{\sigma}
\def\Si{\Sigma}

\def\r{\rho}
\def\t{\tau}

\def\pa{\partial}
\newcommand{\ti}[1]{\tilde{#1}}

\newcommand{\sm}[1]{\mbox{\scriptsize #1}}
\newcommand{\tn}[1]{\mbox{\tiny #1}}
\renewcommand{\@}[1]{\sqrt{#1}}
\renewcommand{\le}[1]{\label{#1}\end{eqnarray}}
\newcommand{\bea}{\begin{eqnarray}}
\newcommand{\eea}{\end{eqnarray}}
\newcommand{\eq}[1]{(\ref{#1})}
\def\nn{\nonumber\\}
\def\nm{\nonumber}
\def\na{\nabla}
\def\half{{1\over2}\,}

\begin{document}

\pagestyle{plain}

\centerline{{\Large \bf Noether's Theorems and Energy in General Relativity}}
\vskip.5cm
\centerline{{\Large\bf }}
\vskip.7cm

\begin{center}
{\large Sebastian De Haro}\\
\vskip .7truecm
{\it Institute for Logic, Language and Computation, University of Amsterdam}\\
{\it Institute of Physics, University of Amsterdam}\\
{\tt s.deharo@uva.nl}

\end{center}

\vskip .7truecm

\begin{center}
\today
\end{center}

\vskip 2truecm

\begin{center}
\textbf{\large \bf Abstract}
\end{center}

This paper has three main aims: first, to give a pedagogical introduction to Noether's two theorems and their implications for energy conservation in general relativity, which was a central point of discussion between Hilbert, Klein, Noether and Einstein. Second, it introduces and compares two proposals for gravitational energy and momentum, one of which is very influential in physics: and, so far as I know, neither of the two has been discussed in the philosophical literature. Third, it assesses these proposals in connection with recent philosophical discussions of energy and momentum in general relativity.

After briefly reviewing the debates about energy conservation between Hilbert, Klein, Noether and Einstein, I give Noether's two theorems. I show that Einstein's gravitational energy-momentum pseudo-tensor, including its superpotential, is fixed, through Noether's theorem, by the boundary terms in the action. That is, the freedom to add an arbitrary superpotential to the gravitational pseudo-tensor corresponds to the freedom to add boundary terms to the action without changing the equations of motion. This freedom is fixed in the same way for both problems. I also review two proposals for energy and momentum in GR, of which one is a quasi-local alternative to the local expressions, and the other builds on Einstein's local pseudo-tensor approach. I discuss the recent philosophical literature on the conservation of energy and momentum in general relativity, and I assess and compare the two proposals in the light of this literature: especially, in light of questions about diffeomorphism invariance and background-independence.

\vskip 1.5cm
\noindent Forthcoming in: J.~Read, N.~Teh and B.~Roberts (Eds.), {\it The Philosophy and Physics of Noether's Theorems,} Cambridge University Press, 2021.

\newpage

\tableofcontents

\newpage

\section{Introduction}\label{intro}

This paper has three main aims: first, to give a pedagogical introduction to Noether's two theorems: and in particular to the role that her second theorem played in important discussions between Hilbert, Klein, Noether and Einstein about the conservation of energy and momentum in general relativity (GR). These discussions have been the subject of recent scholarship, and are still valuable today.\footnote{See for example Kastrup (1987:~pp.~116-125), Brading (2005:~pp.~125-127), Brading and Brown (2003:~pp.~91, 06), Kosmann-Schwarzbach (2011:~pp.~37-46), Rowe (1999:~pp.~209-221; 2019:~pp.~7-11, 29-34, 2019a). 
} 

Second, I introduce for a philosophical audience two proposals for the gravitational energy-momentum tensor: namely, Einstein's pseudo-tensor (and the quasi-local expressions derived from it\footnote{Especially in recent work by Chang et al.~(1999) and Chen et al.~(2017, 2018).}) and the quasi-local definition by Brown and York, especially as modified in the recent literature on the holographic principle. I argue that arguments from gauge-gravity duality support the quasi-local quantities as correctly expressing the energy and momentum of general relativity.

Third, I discuss and critically assess the various proposals in light of a recent philosophical discussion of energy and momentum in general relativity, by Hoefer (2000), Pitts (2010), Lam (2011), Duerr (2019a), and Read (2020).

General relativity suffers from a number of problems regarding its local conservation laws for energy and momentum. This was the subject of a crucial discussion between Hilbert, Klein, Noether and Einstein between 1915 and 1918. What is perhaps less well-known is that the problem of the definition and conservation of energy\footnote{For brevity, I will sometimes follow the literature in using `energy' or `energy-momentum' for the longer `energy-momentum and angular momentum', especially in expressions like `conservation of energy-momentum and angular momentum'.} played a crucial role in both: the genesis of general relativity and Noether's reasons for proving her two theorems. 

To define gravitational energy and momentum, Einstein introduced a gravitational {\it pseudo-tensor,} i.e.~a quantity that is non-covariant, and is in some ways analogous to the Christoffel symbols. However, unlike the Christoffel symbols, pseudo-tensors are not geometric objects. The question of energy and momentum, and the pseudo-tensor proposal, are discussed in virtually all of Einstein's papers on the subject.\footnote{See for example Janssen and Renn (2007:~p.~844).} Part of the reason it played such a prominent role, including in his November (1915, 1915a, 1915b) papers, is because the pseudo-tensor is intimately linked to Einstein's field equations and to the contracted Bianchi identities. 

Noether's own work was directly motivated by the question of energy-momentum conservation, and prior to publishing her two theorems she had worked on comparing Hilbert's and Einstein's approach to it. According to Hilbert, Klein, and Noether, the conservation law in general relativity is {\it improper,} because it is given in terms of a superpotential whose conservation law is an identity. On Klein's interpretation, the conservation law does not make direct use of the equations of motion, and so it has no physical content.

Einstein's pseudo-tensor, and thus the energy, are known to have a number of other problems: the pseudo-tensor is non-covariant, even non-geometric, and there appears to be no unique expression for it. For many different expressions have appeared in the literature. This non-uniqueness stems from the ability to add to the pseudo-tensor the derivative of an arbitrary `superpotential'. The superpotential is such that its divergence is identically zero, and so an arbitrary superpotential can be added into the conservation law without modifying it. Thus, the very definition of the pseudo-tensor appears to be ambiguous.

The philosophical interest in energy and momentum in general relativity has so far been moderate.\footnote{For a detailed analysis of Hilbert's (1916, 1917) two papers on the `Foundations of Physics', see Renn and Stachel (2007:~pp.~878-902, 914-957). Hilbert's first paper, including the changes that he introduced into the proofs, is discussed, and compared with Einstein's four November 1915 papers, in Sauer (1999:~pp.~539-557), Renn and Stachel (2007:~pp.~872-933), and Corry et al.~(1997:~pp.~1271-1273). A brief account of the interactions between Einstein and Hilbert in 1915 is in Corry (1999:~pp.~177-179).} For example, Hoefer (2000), Lam (2011), and Duerr (2019, 2019a)
have criticised, and Read (2020) has defended, local definitions of energy and momentum, in particular those involving the energy pseudo-tensor. I will give considerable attention to this recent philosophical discussion in Sections \ref{phildis} and \ref{scep}.

There are two aspects that these recent discussions have not focussed on, and which I will address in this paper: namely, the alleged non-uniqueness of the superpotential, and the quasi-local (as opposed to local) definitions of energy and momentum.

I will show in Section \ref{NoeT} that, once the Lagrangian and its variation are fixed, Noether's second theorem fixes the superpotential. This holds not only for general relativity, but for any theory with a local symmetry. In general relativity, the gravitational pseudo-tensor, which is the Noether current corresponding to the local translation symmetry, is thus fixed, through Noether's theorem, by the choice of the action. In particular, the freedom to add an arbitrary superpotential corresponds to the freedom to add boundary terms to the action without changing the equations of motion.\footnote{In a different context, Chang et al.~(1999:~pp.~3-4) and Chen et al.~(2018:~p.~4) have also made a related connection between the superpotential and boundary conditions in the Hamiltonian, although they do not mention Noether's theorem in that connection.} Thus the superpotential is not arbitrary, but is fixed, through Noether's theorem, by the choice of boundary terms in the action. Although this is known to the experts, I have not seen these consequences of Noether's theorem discussed in the literature: in fact, the ``problem of non-uniqueness'' is still frequently cited without a discussion of the implications of the choice of boundary conditions and boundary terms. For example, even Brading and Brown (2003:~pp~101-102), whose formulation of Noether's second theorem I largely follow, have the superpotential implicit in their formulas, but do not seem to recognise it as such, and they do not comment on the definition and alleged non-uniqueness of the pseudo-tensor in connection with Noether's theorems.\footnote{For a discussion of the literature on this point, see Section \ref{2con}.}

The detailed consideration of Noether's second theorem also allows one to address Klein's worry about the alleged `lack of physical content' of the local energy conservation law of general relativity. Indeed, I will argue that the distinction between `proper' and `improper' conservation laws, in the context of comparing general relativity with classical mechanics, is a red herring: and that the claim that local energy conservation laws lack physical content is misleading.

I will also introduce two different quasi-local expressions: one, due to Chang et al.~(1999), Nester (2004), and Chen et al.~(2017, 2018, 2018a), directly obtained from the total energy-momentum (i.e.~the energy-momentum tensor plus the gravitational pseudo-tensor) suitably integrated over a volume. These authors show that, under assumptions that I will discuss in Sections \ref{Revisited} and \ref{scep}, this expression can be obtained from the covariant Hamiltonian formalism of general relativity. Indeed, part of their aim appears to be to justify the use of the pseudo-tensor from the Hamiltonian formalism.

The second expression is the Brown-York (1993) quasi-local stress-energy tensor, which is directly defined on a two-dimensional surface. This stress-energy tensor has been studied extensively in the physics literature, especially in the presence of a cosmological constant (positive or negative), and has been shown to have many of the desired properties of the quasi-local stress-energy, as I will also review. 

I also compare, in Section \ref{scep}, the two quasi-local proposals. While quasi-local expressions have sometimes been easily dismissed (or avoided altogether) in the philosophical literature, I will argue that neither dismissal nor avoidance are justified. Also, we will see that the analysis of the Brown-York stress-energy tensor in the gauge-gravity literature gives crucial insights into the correct definition and properties of this tensor.

One obstruction to the construction of normal conservation laws for energy and momentum in curved spacetimes is the presence of a covariant derivative in the conservation law of the energy-momentum tensor.\footnote{Notice that this fact is not specific to general relativity: for any translation-invariant classical field theory on a curved spacetime has a covariantly-conserved, two-index, energy-momentum tensor. But GR offers a specific interpretation of the source or sink terms in terms of a dynamical gravitational field.} One can integrate such a law to get an integral form of the conservation law that is properly covariant only if the metric admits a timelike Killing vector field, i.e.~if the metric has a timelike isometry. 

An important property of the quasi-local energy-momentum tensors is that they can lead to covariant integral conservation laws under weaker conditions than the energy-momentum tensor does. For, while the latter requires a timelike Killing vector field in the whole spacetime, the quasi-local expression from Section \ref{BYQL} only requires a (conformal) Killing vector field at the quasi-local surface, i.e.~usually at asymptotic spatial or null infinity. I show this in Section \ref{hst}.

In physics, there are well-known global expressions (called the Bondi and ADM masses) for the total energy at infinity, depending on whether they are integrated over spheres at spatial infinity (in the case of asymptotically flat and AdS cases) or null infinity (in the case of the Bondi mass). The two kinds of quasi-local quantities discussed in this paper reproduce these expressions. But these expressions, including the quasi-local quantities, do not necessarily have well-defined local densities.\footnote{This is often further explicated by an appeal to the principle of equivalence, that a local observer cannot detect a gravitational field at a single point. This often stated, infinitesimal or strong, form of the principle of equivalence is in Pauli (1958:~p.~145); see also Brown (2005:~pp.~169-170). For its role in Einstein's own thinking about general relativity, and its relation with the Einstein equivalence principle, see Lehmkuhl (2019:~pp.~2-7, 19-22) and Janssen (2014:~pp.~174-182). The strong equivalence principle is criticised by Norton (1985:~pp.~207-208, 239-242, 245) and Curiel (2019:~p.~91).} 
Chen et al.~(1999:~p.~2) discuss a quasi-local formula that {\it can} be localised: its localisation gives rise to the gravitational pseudo-tensor, which depends on the choice of coordinates. They view the interplay between the local and quasi-local quantities as follows: `Appreciating the fundamental nonlocality of the gravitational interaction yet believing in the basically local nature of physical interactions led to the idea of {\it quasi-local quantities}: quantities that take on values associated with a compact orientable spatial 2-surface'.

The plan of the paper is as follows. In the rest of this Introduction, I summarise the discussion of energy and momentum between Hilbert, Klein, Noether and Einstein (Section \ref{history}), illustrate Noether's theorems for Maxwell theory as my running example (Section \ref{Mxth}), and summarise Einstein's basic proposal of pseudo-tensors as candidates for energy and momentum and its problems (Section \ref{Einstein}). Section \ref{Einstein} lays out five problems that organise the subsequent discussion. Section \ref{history} has more of a historical character; the rest of the paper has a more philosophical and foundational character. Section \ref{NoeT} then gives Noether's two theorems and discusses two consequences of it, one of which I have not seen discussed in the literature (Section \ref{2con}). Section \ref{LEMGR} applies Noether's theorems to general relativity. Section \ref{BYQL} discusses the quasi-local Brown-York stress-energy tensor. Sections \ref{BYql} and \ref{hst} are slightly more technical than the rest of the paper, and can be skipped by readers only interested in the philosophical discussion. Section \ref{diffeos} discusses whether diffeomorphisms are redundancies of the description. Section \ref{Revisited} reviews proposals by Chang et al.~(1999) and Chen et al.~(2017, 2018, 2018a) and Nester (2004) that link the gravitational pseudo-tensor with a quasi-local expression that can be reproduced from the covariant Hamiltonian of general relativity, under assumptions that I will point out. Section \ref{scep} then compares the two proposals for quasi-local expressions and discusses the status of energy-momentum in general, and the pseudo-tensors in particular, in the philosophical literature. Sections \ref{BYQL} to \ref{scep} also discuss two themes that run through the philosophical literature on energy and momentum in general relativity: namely, covariance and background-independence.

\subsection{Noether's theorem and general relativity}\label{history}

Emmy Noether's (1918) two theorems, on the correspondence between symmetries of the action under infinitesimal transformations of a continuous group, and local conservation laws, are better known for their applications in classical mechanics and gauge theories than in general relativity. Yet Noether's motivation to study this problem came chiefly out of discussions between David Hilbert, Felix Klein, Albert Einstein, and herself about the interpretation of {\it general relativity,} between 1915 and 1918: specifically, about the nature of the conservation equations in GR.\footnote{For some of the scientific context of Noether's theorems, including her own biography, see Kosmann-Schwarzbach (2020:~pp.~1-9). A clear exposition of Noether's theorems in their original form is in Rowe (1999:~pp.~223-226). Norton (1993) is a useful reference about the history of some issues in general relativity that are complementary to the ones discussed in this Section: in particular, about general covariance.} Noether's theorem still gives the most general---and, as I will argue in Section \ref{NoeT}---best approach to the gravitational pseudo-tensor of general relativity and its associated superpotential: which was at the heart of the debate.

Since at least 1912, Einstein had been searching for an expression of the laws of gravitation, invariant under general coordinate transformations.\footnote{See e.g.~Sauer (2005:~pp.~806-808) and Kosmann-Schwarzbach (2011:~p.~37).} Already in his (1913:~p.~318) {\it Entwurf} paper with Marcel Grossmann, and in his (1914:~p.~99) `Formal Foundation' paper, which amongst other things gave a detailed development of tensor calculus and worked out the geodesic equation, he discussed at length (especially in his 1914) a covariant balance equation for the energy-momentum and stress tensor density, subject to external forces. When the external forces were set to zero, he got an analogue of the local conservation law of energy and momentum (pp.~100, 120):
\bea\label{Econs}
\pa_\m(\mathfrak{T}^\m{}_\n+t^\m{}_\n)=0~.
\eea
Here, $\mathfrak{T}$ is the energy-momentum tensor density of matter,\footnote{Compare this equation with Eq.~\eq{pseudot}.} and $t$ is the `energy tensor of the gravitational field' (later called the gravitational {\it pseudo-tensor}), in analogy with the Christoffel symbols, which are the components of the `field strength of the gravitational field'. He gave an explicit expression for $t^\m{}_\n$ in terms of the Christoffel symbols (p.~120) and noted that it is not a tensor, but it is only invariant under linear (i.e.~affine) transformations. At this point, Einstein wrote his tentative field equations explicitly in terms of the pseudo-tensor. Thus from very early on, the gravitational energy-momentum pseudo-tensor was, in Einstein's mind, an important ingredient of the theory that he was seeking.

In the spring of 1915, Hilbert and Klein invited Noether to G\"ottingen to work with them on the implications of the general theory of relativity (Kosmann-Schwarzbach, 2011: p.~45; Corry, 1999:~p.~177).\footnote{For the prehistory of this invitation, see Sauer (1999:~p.~539-540). The prehistory of Hilbert's acquaintance with Einstein is on pp.~527-539.} Although Noether was probably not an expert on variational methods (however, Hilbert {\it was} an expert on that),\footnote{See Sauer (1999:~pp.~535, 558, 569).} she was well-acquainted with the related methods for using formal differential operators to generate algebraic and differential invariants (Rowe 2019:~p.~6). Also, shortly after her arrival in 1915, Einstein gave a series of lectures on general relativity---he was staying at Hilbert's house during his visit. Thus by mid-1915, Hilbert himself was working intensely to understand Einstein's papers. Early on, Hilbert saw a connection between two topics: (i) the theory of invariants and geometry, and (ii) the problem of extending the special theory of relativity to general coordinate systems. This was of course a seed for Noether's theorems (and Hilbert himself would soon go on to give a special case of Noether's second theorem). Likewise, Klein would regard general relativity from the point of view of his Erlangen programme, which he had laid out with Sophus Lie, of geometry as the search for invariants of a group.\footnote{See Rowe (1999:~p.~215; 2019a:~p.~72) and Kosmann-Schwarzbach (2011:~pp.~26, 40, 64).}

In two of his papers from November 1915, Einstein gave a very general expression (1915:~p.~784) for the gravitational pseudo-tensor in terms of the Lagrangian density\footnote{See Eq.~\eq{tmn}, which is only a minor generalisation of this equation through the addition of a total-derivative term. However, at this point Einstein's Lagrangian was incomplete, since its variation gave only (part of) the the Ricci tensor, rather than the Einstein tensor. A problem in this paper, to which he returned in (1915a:~pp.~799-800), was the incompatibility between his restriction to coordinate systems where $\sqrt{-g}=1$ and the requirement that the energy-momentum tensor has non-zero trace. The full, covariant equations without any restrictions on the trace of the energy-momentum tensor appreared in Einstein (1915b:~p.~845).} and found the correct gravitational field equations (1915b:~p.~845). 

As is well-known, on 20 November 1915, i.e.~five days before Einstein submitted the paper that contained the final form of his field equations (1915b),\footnote{For the now accepted resolution of the priority issue, see Corry et al.~(1997:~pp.~1271-1273), Sauer (1999:~pp.~539-557), and Renn and Stachel (2007:~pp.~872-933).} Hilbert submitted his `Die Grundlagen der Physik', which is an attempt to axiomatise general relativity (and make a connection with Gustav Mie's unification programme).\footnote{Hilbert's reference to Mie in the introduction is as follows: `I would like to formulate---in the sense of the axiomatic method---a new system of basic equations of physics, out of two simple axioms, which are of ideal beauty, and which, I believe, contain the solution of the problems by both Einstein and Mie' (1915:~p.~395; see also pp.~396, 407). For more on Mie's influence on this paper by Hilbert, see Sauer (1999:~pp.~538, 552-555), Rowe (2019:~pp.~5-7), Renn and Stachel (2007:~pp.~857-867, 871-874, 899-900), Corry (1999:~p.~176).} Hilbert reproduced Einstein's field equations from his action and, among other results, he derived the four contracted Bianchi identities from the invariance of the action under arbitrary transformations of coordinates. Thus in this paper Hilbert derived a special case of Noether's second theorem. At this point, however, the nature of these relations was unclear. 
Even less clear was the meaning of the `energy equation', based on an `energy vector', that he derived.\footnote{Hilbert would later rewrite his two `Grundlagen der Physik' papers, (1916) and (1917), publishing them in sequence in Hilbert (1924).}

Noether also began to work on this problem. In 1916, she wrote an unpublished manuscript in which she studied Hilbert's energy vector, and used Hilbert's work to analyse Einstein's conservation law in general relativity. Although the manuscript is not extant, its content is partly known from two sources: namely, the exchange of letters between Klein and Hilbert published in the {\it Nachrichten von der Gesellschaft der Wissenschaften zu G\"ottingen, Mathematisch-Physikalische Klasse,} in 1917, and a partial transcription by R.~J.~Humm, who studied at Berlin and G\"ottingen at the time (Rowe 2019, 2019a). In this manuscript, she got the contracted Bianchi identities, and in fact she anticipated the main points of Klein's influential lecture of 22 January 1918 for the G\"ottingen {\it Gesellschaft.}

Einstein (1916a:~p.~1111) endorsed Hilbert's action in another paper, published in November 1916, and solved the conservation equation, Eq.~\eq{Econs}, in terms of what is now called a `superpotential',\footnote{See Eq.~\eq{superU} below.} i.e.~a quantity whose divergence is identically zero.\footnote{Such an expression already appeared in Einstein (1914:~p.~1077), but this was only for his old theory, in which $\sqrt{-g}=1$.} 

On 22 January 1918, 
Klein delivered his famous lecture at the {\it Mathematische Gesell\-schaft zu G\"ottingen} (i.e.~the G\"ottingen Mathematical Society), which elicited a reaction from Hilbert one week later (Rowe 1999:~p.~212; 2019:~p.~8). Klein's lecture would be published in the {\it Nachrichten von der Gesellschaft der Wissenschaften zu G\"ottingen,} in the unusual form of an exchange of letters with Hilbert (also, Klein and Hilbert lived a short distance from one another in the Wilhelm Weber Strasse!).\footnote{Klein (1917). For a reason why they chose this unusual form of publication, see Rowe (1999:~p.~212).} In it, he explicitly acknowledged Noether's help and her earlier contribution (Kosmann-Schwarzbach, 2011: p.~66):
\begin{quote}\small
You know that Miss Noether advises me continually regarding my work, and that in fact it is only thanks to her that I have understood these questions. When I was speaking recently to Miss Noether about my result concerning your energy vector, she was able to inform me that she had derived the same result on the basis of developments of your note... more than a year ago, and that she had then put all of that in a manuscript (which I was subsequently able to read). She simply did not set it out as forcefully as I recently did at the Mathematical Society.
\end{quote}
Like Noether had done before, Klein compares Hilbert's energy equation and Einstein's conservation law, Eq.~\eq{Econs}. Klein's main point was the disanalogy between Hilbert's energy equation (and Einstein's conservation law) and the conservation laws in classical mechanics. While, in classical mechanics, energy conservation follows as a consequence of the equations of motion, in general relativity Eq.~\eq{Econs} is an {\it identity,} i.e.~valid without requiring the equations of motion to be satisfied. He explained it thus:
\begin{quote}\small
Thus the above statement... cannot be regarded analogous to the conservation law of energy in ordinary mechanics. For when we write the latter as $d(T+U)/dt=0$, this differential equation is not satisfied identically [as the Bianchi identity is], but only as a consequence of the differential equations of mechanics.
\end{quote}
He concluded the note to Hilbert with a sceptical remark: `After all this, I can hardly believe that it is appropriate to characterise the very arbitrarily built quantities $t^\m{}_\n$ as `energy components of the gravitational field'{}'. This agreed with Noether's own view, which she had explained to Hilbert more than a year before. 

In his reply,\footnote{Klein (1917:~p.~477).} Hilbert completely agreed with Klein's comments about the lack of an analogy between the energy equation and conservation of energy in classical mechanics. He took this one step further, by expressing his belief that no true law of energy conservation exists in general relativity, and claiming that it is possible to {\it prove a theorem} to this effect---to which Klein responds that he `would be very interested in seeing the mathematical proof to which you alluded to in your reply' (p.~482). Such a proof would not be provided by Hilbert, but by Noether's second theorem, in the following year.


Meanwhile, in 1917-1918, Einstein and Klein corresponded about Klein's (1917) note, which Einstein (1918:~p.~673) praised for its clarity.\footnote{For an account of the correspondence, see Rowe (1999:~pp.~209-213).} Einstein disagreed that his conservation law for general relativity was an identity, since the equations of motion {\it are} used. To this, Klein (1918a:~p.~504) replied by saying that the divergence in Eq.~\eq{Econs} consists of two terms:\footnote{This can be seen by expressing the pseudo-tensor $t$ in terms of two contributions: the Einstein tensor and the superpotential: see Eq.~\eq{tGU} below.}

(1) the divergence of the Einstein field equations; 

(2) the divergence of the superpotential. 

Klein's point was that the vanishing of (2) is a mathematical identity, while (1) contains the {\it divergence} of the equations of motion, rather than the equations of motion themselves---and so, that this term is also zero, even before one takes its divergence in the pseudo-tensor. (This differs from classical mechanics, where the equations of motion, rather than their divergence, appear in the conservation equation). Einstein (1918a:~p.~513) disagreed: for the equations of motion were after all used, and the conservation equation used at least {\it part} of the field equations (i.e.~their contraction with the divergence).\footnote{The above summary of the argument in Klein (1918a:~p.~504) and Einstein (1918, 1918a:~p.~513) follows Brading (2005:~p.~126), which indeed seems to me a correct interpretation both of Klein's letter and of Einstein's (1918a:~p.~513) reply.\label{KElet}}

In the winter-spring of 1918, Noether understood the reasons for the difficulties in the interpretation of the conservation laws of general relativity. What she set out to investigate was indeed Klein and Hilbert's central problem of the analogy between the energy equations of classical mechanics and general relativity. On 23 July 1918, she presented her work at the G\"ottingen Mathematical Society. In short, her second theorem could account for the four identities that Hilbert had pointed out, so that the conservation laws were `improper', and thus disanalogous to classical mechanics. 

Noether had a deeper reason for the difference between general relativity and other theories: her theorems allowed her to `deduce... the proof of an assertion of Hilbert concerning the relationship between the lack of a proper law of energy and ``general relativity''' (Kosmann-Schwarzbach, 2011:~p.~19). Namely, given an action invariant under the group of translations, the energy relations are improper iff the action is invariant under an infinite group which contains the group of translations as a subgroup.\footnote{Noether's use of the word `subgroup' here appears to be stronger than the modern use. Her statement assumes not just the group of translations being a subgroup of an infinite group, but also its being {\it embedded} in a specific way: see Section \ref{2con}. I thank Bartlomiej Czech for pushing me on this point.}

Thus Noether interpreted her theorems along Hilbertian lines: `As Hilbert expresses his assertion, the lack of a proper energy law constitutes a characteristic of the general theory of relativity'. 

The paper by Klein (1918) clarifies the relationship between Einstein's and Hilbert's theories, and in particular concerning their conservation laws. In this paper, he derives what Brading and Brown (2003) call the `boundary theorem', from which it follows that the current is proportional to the field equations (and to the superpotential), so that its divergence is identically zero---as I already mentioned above. In doing so, he made essential use of Noether's work (Kosmann-Schwarzbach, 2011, p.~68):\footnote{Klein (1918:~p.~189).}
\begin{quote}\small
I would be remiss if I did not thank Miss Noether once again for her active participation in my new work. She has by herself completely set out in proper form the mathematical ideas that I use in connection with the physics problems related to [the action], which will be presented in a note to appear shortly in these {\it Nachrichten}.
\end{quote}

Thus by 1918, Noether's theorem was, in the eyes of Hilbert, Klein, and Noether, the proof that there is no proper conservation law in general relativity: which Klein interpreted---with Hilbert's and Noether's approval, one should add---as the energy equations' lacking physical content. 

I will return to Einstein's proposal for a conservation law in general relativity, and its problems, in more detail in Section \ref{Einstein}. In the next Section, I first illustrate Noether's two theorems in a well-known example.

\subsection{Illustrating Noether's theorems in Maxwell theory}\label{Mxth}

In this paper, I will use the Maxwell theory of electromagnetism as a running example to illustrate various aspects of Noether's theorems, since this theory already exemplifies some of the issues that appear in general relativity.\footnote{For an introduction to Noether's theorem for a physics audience, see e.g.~Ba\~nados et al.~(2016). For a mathematical treatment, see Olver (1986:~pp.~272-283, 334-336). Bessel-Hagen (1921:~pp.~44-51) was the first to apply Noether’s first theorem to electrodynamics, using the generalised theorem involving quasi-symmetries. He credits Klein (p.~33) with suggesting the problem: `Klein expressed a desire that Miss Emmy Noether's theorems on invariant variational problems, proved approximately two years ago, should be applied to the Maxwell equations'. I thank Harvey Brown for recommending the work by Bessel-Hagen.}

Thus let me recall how the theorems work for elementary electromagnetism without sources: take the following action, defined on Minkowski spacetime:
\bea\label{Mxact}
S[A]=\int\dd^4x\,{\cal L}(A;x)&=&-{1\over4}\int\dd^4x\,F_{\m\n}F^{\m\n}\nn
F_{\m\n}&:=&\pa_\m\,A_\n-\pa_\n\,A_\m~.
\eea
The field equations, derived by setting the variation of the action to zero, are of course:
\bea\label{Mxeom}
\pa_\n F^{\m\n}=0~.
\eea
I will give Noether's theorems in general form in Section \ref{NoeT}. I here give the specific cases that apply to Maxwell theory.

\ \\{\bf Noether's first theorem} asserts that the conservation of the energy-momentum tensor density $T_{\m\n}$ follows from the {\it global} translational symmetry of the action in Minkowski spacetime (and vice versa), i.e.~under a {\it constant} translation $\d x^\m=\e^\m$.\footnote{Under such a translation, we have a new field $A'_\m(x')$, where $x'=x+\e$, and $A_\m'(x')$ is related to $A_\m(x)$ by the usual tensor transformation rule for the components of one-forms, linearised in $\e$. In the variation, we evaluate $A_\m'$ at $x$.} Defining the variation 
\bea\label{varA}
\d A_\m:=A_\m'(x)-A_\m(x)=-\e^\n\,\pa_\n A_\m~,
\eea
Noether's first theorem then gives:
\bea\label{EMT}
T^\m{}_\n&:=&\d^\m_\n\,{\cal L}-{\pa{\cal L}\over\pa(\pa_\m A_\l)}\,\pa_\n A_\l=F^{\m\l}\,\pa_\n A_\l-{1\over4}\,\d^\m_\n\,F^{\l\s}F_{\l\s}~.
\eea
One readily finds the following conservation law:
\bea\label{cons}
\pa_\m T^\m{}_\n=\left(\pa_\m\,F^{\m\l}\right)\pa_\n\,A_\l\overset{{\sm{Eq.}}\,\eq{Mxeom}}{=} 0~.
\eea
The term between brackets vanishes only if the equations of motion, Eq.~\eq{Mxeom}, are satisfied: and thus the energy-momentum is conserved if and only if (a linear combination of) the equations of motion are satisfied, i.e.~it is a weak conservation law. 

Notice that the first term in the energy-momentum tensor Eq.~\eq{EMT} is not gauge invariant (it is also not symmetric in its two indices upon lowering one of them).\footnote{Goldberg (1958:~pp.~315, 319) discusses the desirability of symmetry for the discussion of angular momentum. See also Barbashov and Nesterenko (1983:~p.~559-560).} 
This lack of gauge invariance is unsurprising, given that we used a variation of the Lagrangian, Eq.~\eq{varA}, that is not gauge invariant. One can get a gauge-invariant energy-momentum tensor by constructing a gauge-invariant variation of the gauge potential, $A$: namely, by allowing this variation to contain a gauge transformation, i.e.~$\d A_\m=-\e^\n\,\pa_\n A_\m+\pa_\m\l$, where $\l$ is the gauge parameter. There is a unique gauge-invariant variation of this form, viz.~$\d A_\m=F_{\m\n}\,\e^\n$, where $\l=\e^\n A_\n$. In effect, this adds an `improvement term' to the energy-momentum tensor, i.e.~a term that is a total derivative in an antisymmetric tensor $U$, without modifying the equations of motion, such that the energy-momentum tensor is still conserved:\footnote{Improvement terms go back to Belinfante (1938:~p.~888). See also Anderson (1967:~p.~273).}
\bea\label{td}
T'{}^\m{}_\n:=T^\m{}_\n+\pa_\l\,U^{[\m\l]}{}_\n~.
\eea
The above variation gives $U^{[\m\l]}{}_\n=-F^{\m\l}\,A_\n$, which brings the energy-momentum tensor to the standard form, and is automatically symmetric in its indices:
\bea\label{Tp}
T'{}^\m{}_\n=F^{\m\l}\,F_{\n\l}-{1\over4}\,\d^\m_\n\,F^{\l\s}\,F_{\l\s}~.
\eea
This is of course the energy-momentum tensor that also appears on the right-hand side of the Einstein field equations. (Thus general relativity uniquely detects the energy-momentum tensor of its sources: it fixes the improvement term.)

\ \\{\bf Noether's second theorem} is also exemplified by the Maxwell theory,\footnote{For a discussion of Noether's theorems for Maxwell theory coupled to a complex scalar field, see Barbashov and Nesterenko (1983:~pp.~550-552) and Sus (2017:~pp.~273-274).} which has a {\it local} gauge symmetry:
\bea\label{gauge}
A_\m\mapsto A_\m+\pa_\m\l\,,
\eea
for an arbitrary {\it smooth function} $\l:\mathbb{R}^4\rightarrow\mathbb{R}$. Indeed, since $F_{\m\n}$ is itself invariant under this symmetry, the Maxwell action is trivially invariant.

It is instructive to study the gauge invariance when we couple Maxwell's action to a gauge-invariant source $J_\m$:
\bea\label{MxJ}
S[A,J]=\int\dd^4x\left(-{1\over4}\,F_{\m\n}\,F^{\m\n}+J^\m A_\m\right),
\eea
and require that the action is invariant up to a boundary term that does not modify the equations of motion, so that the transformed action gives the same equations of motion as the action Eq.~\eq{MxJ}.

The variation of this action under the gauge symmetry, Eq.~\eq{gauge}, gives:
\bea\label{gaugevar}
\d_\l S[A,J]=\int\dd^4x~J^\m\pa_\m\l=\int\dd^4x~\pa_\m(J^\m \l) -\int\dd^4x~(\pa_\m J^\m)\,\l~,
\eea
where in the last step I used partial integration. The first integral on the right contains a total derivative, which integrates to a boundary term. Although physics texts often require that $\l$ has compact support so that it vanishes at infinity and this boundary term is zero,\footnote{See e.g.~Maggiore (2005:~p.~69).} this is not necessary. For, as we will see in Section \ref{NoeT}, Noether's theorem can be generalised to admit a variation of the action with a non-zero boundary term, because such a term does not modify the equations of motion. Thus this term does not need to be zero and we simply keep it. The second term {\it must} be zero if the action is to be invariant up to a boundary term, which requires that the source is conserved:
\bea\label{consJ}
\pa_\m J^\m=0~.
\eea
Writing this out in components gives the equation for the {\it local conservation of charge}:
\bea\label{localC}
{\pa\rho\over\pa t}+\nabla\cdot{\bf J}=0~.
\eea
Integrating over the volume $V$, bounded by the two-surface $S=\pa V$ with outward pointing unit vector $\hat{\bf n}$, gives the time derivative of the total charge $Q_V$ enclosed by the volume:
\bea\label{Qcons}
{\dd Q_V\over\dd t}&=&{\dd\over\dd t}\int_V\dd^3x~\r\left(t,{\bf x}\right)\nn
&\overset{{\sm{Eq.}}\,\eq{localC}}{=}&-\int_V\dd^3x~\nabla\cdot{\bf J}=-\int_S\dd^2x~\hat{\bf n}\cdot{\bf J}=-I_S~,
\eea
where in the second line I used Gauss' theorem. $I_S$ is the current flowing {\it out} of the boundary $S$, so that this equation expresses the conservation of charge in the volume: if the charge in $V$ decreases, it is carried away through the boundary by a current $I_S$, and vice versa.

Thus, although gauge invariance is trivially satisfied by the purely electromagnetic part of the action, it implies a physical constraint on the gauge-invariant source $J$ to which it is coupled; the Maxwell theory can only be consistently coupled to a conserved current (I will return to this at the end of Section \ref{2N}).

I will now make a point that will recur in general relativity, and which was important in the discussions between Noether, Hilbert, Klein and Einstein, reviewed in Section \ref{history}. It is sometimes said that local gauge symmetry is a ``mere redundancy'', a consequence of our reformulation of the electomagnetic fields ${\bf E}$ and ${\bf B}$ in terms of the gauge potential $A_\m$ with no physical significance. However, it is not true that gauge symmetry has no physical significance. If it did not, it would indeed be very surprising that it could entail the local charge conservation law Eq.~\eq{consJ}. 

The point is that the local gauge symmetry reduces the number of physical degrees of freedom of the gauge potential $A_\m$ from three to two, thus leaving a field that only has transverse polarisations. Physically, the freedom to do a gauge transformation secures that the longitudinal polarisation of the photon does not contribute to any of the quantities that are usually deemed physical. This is of course crucial to get the correct phenomenology of the photon. Thus, despite the fact that a choice of gauge does not fix anything physical, gauge invariance does have physical content: it secures the correct number of degrees of freedom for the photon.

Recent discussions have also concluded that the alleged lack of physical significance of the conservation laws derived from the second theorem, as claimed by Klein (see Section \ref{history}), is misleading. For example, Brading and Brown (2003:~p.~99) have emphasised that symmetry transformations, such as local gauge symmetries, do have an empirical significance, because they restrict the kinds of terms that can appear in the action and in the equations of motion.\footnote{More precisely, they admit that local gauge symmetries can have {\it indirect,} but not {\it direct,} empirical significance, i.e.~as in Galileo's ship-type scenarios. While, in the philosophical literature, it is widely recognised that {\it global} gauge (i.e.~internal) symmetries applied to subsystems can have direct empirical significance thus understood, there is a debate about whether {\it local} gauge symmetries can have such significance. Early work by Kosso (2000:~pp.~95, 97), Brading and Brown (2004:~pp.~656-657), and Healey (2009) stated that local gauge (internal) symmetries have no direct empirical significance, but only indirect empirical significance (e.g.~through Noether's theorems). Their reasons for this are diverse---e.g.~for Brading and Brown, it is only `relative system transformations' (i.e.~subsystem transformations), but not local gauge symmetries, that have direct empirical significance. More recently, Greaves and Wallace (2014), and Teh (2016) have argued that local gauge symmetries, applied to subsystems, {\it can} have a direct empirical significance, through Galileo's ship-type scenarios. I will return to the discussion of the empirical significance of local symmetries, in the context of general relativity, in Section \ref{diffeos}. For a characterisation of gauge symmetries as transformations that only alter the description, but not the physical system itself, see e.g.~Horsk\'y and Novotn\'y (1969:~p.~420). For a more general discussion of symmetries in terms of `redundant structure', see e.g.~Ismael and van Fraassen (2003:~pp.~378-380), Caulton (2015:~p.~156-158), and Belot (2013:~p.~2).\label{des}} Indeed, as the above already shows, imposing local gauge symmetry implies, through Noether's theorem, that Maxwell theory can only be coupled to a {\it conserved} current.




\ \\Before we proceed, let us recall how one normally gets an expression analogous to the charge conservation law, Eq.~\eq{Qcons}, for energy and momentum conservation in general relativity. Such an expression can be derived {\it in the presence of spacetime symmetries,} in particular in spacetimes with timelike Killing vectors $K$:\footnote{I will discuss the general case in Section \ref{LEMGR}. Also, this equation can be relaxed to a conformal Killing equation, as in Eq.~\eq{confg} below.}
\bea\label{Killxi}
\na_\m K_\n+\na_\n K_\m=0~.
\eea
The idea here is to derive an analogue of the charge conservation equation Eq.~\eq{Qcons} for energy and momentum, starting from the covariant analogue of the energy and momentum conservation equation:
\bea\label{GRcon}
\nabla_\m  T^\m{}_\n=0~.
\eea
If $K$ is a timelike Killing vector satisfying Eq.~\eq{Killxi}, then we can define a vector current $P^\m:=T^{\m\n}\,K_\n$, such that:
\bea\label{naP}
\na_\m P^\m=\na_\m(T^{\m\n}\,K_\n)=(\na_\m T^{\m\n})\,K_\n+T^{\m\n}\,(\na_\m K_\n)\overset{{\sm{Eqs.}}\,\eq{Killxi},\eq{GRcon}}{=}0~,
\eea
where in the last equality I used the fact that $T^{\m\n}$ is symmetric, and then used Eq.~\eq{Killxi} to set this term to zero. We can now readily apply to $P$ the divergence theorem for a $d$-dimensional region of space(time) $\Si$ with boundary $\pa \Si$:
\bea\label{divth}
0=\int_\Si\dd^dx~\sqrt{-g}~\na_\m P^\m=\int_\Si\dd^dx~\pa_\m(\sqrt{-g}\,T^{\m\n}K_\m)=\int_{\pa \Si}\dd^{d-1}x~\sqrt{-h}~n_\m\,P^\m~,
\eea
where $g$ is the metric on $\Si$, $h$ is the induced metric on $\pa \Si$, and $n$ is the unit vector normal to $\pa\Si$.\footnote{After the second equality sign, I used: $\sqrt{g}\,\nabla_\m P^\m=\pa_\m(\sqrt{g}\,P^\m)$. For an overview of integration on manifolds, see Wald (1984:~pp.~430-434) and Landsman (2020:~pp.~133-135).}

Here, the Killing vector is important because it allows us to construct the {\it vector} quantity $P^\m$ from the tensor $T^\m{}_\n$. For, while Eq.~\eq{GRcon} does not give rise to a partial derivative that can be integrated (because it has one free index), the vanishing covariant divergence of $P$ in Eq.~\eq{naP}, which has no free indices, {\it does} allow for this, in Eq.~\eq{divth}. The result is a conservation law that is completely analogous to charge conservation in the Maxwell theory, as I will now show.

Take $\Si$ to be a three-dimensional spacelike volume $V$ with two-boundary $S$, and write out the index $\m=(0,i)$ in terms of timelike and spatial components, with coordinates $x^i$ that cover the whole boundary surface $S$, but are otherwise arbitrary:
\bea\label{Killingc}
\int_V\pa_\m(\sqrt{-g}\,T^{\m\n}K_\m)={1\over c}{\dd\over\dd t}\int_V\dd^3x~\sqrt{-g}\,\,T^{0\n}K_\n+\int_S\dd^2x~\sqrt{h}\,Nn_i\,T^{i\n}K_\n=0~.
\eea
Here, $N$ is the lapse function, and $h$ is the induced metric on $S$. In the last term, I used the divergence theorem Eq.~\eq{divth} applied to $V$ with boundary $S$, to give the integral over the surface $S$ of the momentum density, ${\cal P}_S=\sqrt{h}\,Nn_i\,T^{i\n}K_\n$, through the boundary $S$. The first term in the middle expression is the time derivative of the energy associated with the region $V$, with energy density: ${\cal E}_V=\sqrt{-g}~T^{0\n}K_\n$.  Thus the expression can be rewritten as:
\bea\label{E-P}
{1\over c}\,{\dd E_V\over\dd t}=-P_S~,
\eea
where $E_V$ is the total energy in $V$ and $P_S$ is the integrated momentum flux through $S$. This is analogous to the charge conservation law, Eq.~\eq{Qcons}. Notice that this interpretation is possible thanks to the presence of a timelike Killing vector $K$ in the spacetime, which defines the energy. As we will see in Section \ref{Einstein}, in the absence of Killing vectors, defining an analogous expression is problematic.

\subsection{Einstein's proposal, and five problems for energy and momentum in GR}\label{Einstein}


Having illustrated Noether's theorems in the example of Maxwell theory, I return, in this Section, to Einstein's idea of energy and momentum conservation laws in general relativity.\footnote{Janssen and Renn (2007:~pp.~842, 844) note that Einstein (1916a:~p.~1115) is an instantiation of Noether's second theorem, predating the theorem itself.} I will begin by listing five problems that are prima facie associated with this notion.\footnote{See for example the discussion in Hork\'y and Novotn\'y (1969:~pp.~426-430).}

The first problem is the one that motivated Noether's research into invariance and symmetries: namely, Klein's, Hilbert's and Noether's view that:

(i) {\it Energy and momentum conservation laws in GR are improper, i.e.~identities with no physical content.} This is a special property of the notion of {\it energy} in general relativity, and prompts the question whether this is the correct interpretation of Noether's theorem. It also bears on the conceptual relation between energy in general relativity and in classical mechanics. (I will return to the resolution of this problem at the end of Section \ref{NoeT}, after I discuss Noether's theorem).

In the literature, two further problems about the physical consequences of Noether's second theorem for general relativity have been discussed. Both concern the interpretation of the energy-momentum tensor:

(ii) When general relativity is coupled to matter, the Bianchi identity implies the analogue of the conservation law, Eq.~\eq{GRcon}. But $T^\m{}_\n$ is the energy-momentum tensor of {\it matter, rather than of the gravitational field.} So, the question is what is the definition of the energy and momentum {\it of the gravitational field itself}?\footnote{Even though the energy-momentum tensor will have this property for any field theory in a curved spacetime, there is a specific problem to GR here: namely, how to define the energy and momentum of the gravitational field from its dynamical principles.}

(iii) The conservation law, Eq.~\eq{GRcon}, is {\it not a conservation law in the usual sense.} For the energy-momentum tensor {\it cannot be straightforwardly integrated over space to give a conserved vector current} (though this {\it can} be done if there is a Killing vector, as we saw in Eqs.~\eq{naP}-\eq{divth}).\footnote{Note that this is not an exclusive property of general relativity, since it holds for the covariant divergence of the energy-momentum tensor in any field theory on a curved spacetime. However, in general relativity, because the gravitational field is dynamical, Eq.~\eq{GRcon} can be reinterpreted in terms of the gravitational energy-momentum pseudo-tensor $t^\m{}_\n$, as we will discuss.} Namely, this equation contains two additional terms that depend on the connection, i.e.~it is not a pure derivative like the expression Eq.~\eq{consJ} for charge conservation in Minkowski space.\footnote{In particular, in Eq.~\eq{divth}, it is possible to replace the covariant derivative in the first expression by the partial derivative in the middle expression because we have the covariant divergence of a {\it vector} quantity, $P^\m$. If we had simply integrated the covariant divergence of the tensor $T^\m{}_\n$, this would pick up additional terms involving the connection and it would not give a boundary term. This is also the origin of the extra term $t^\m{}_\n$, in Eq.~\eq{pseudot} below.} This suggests the question of the correct interpretation of the putative conservation law Eq.~\eq{GRcon}.

Einstein's (1913, 1914, 1915, 1915a, 1916a) idea to solve these problems was to use the particular form of the Christoffel symbols that appear in the covariant analogue of the conservation law for matter, Eq.~\eq{GRcon}, and to reinterpret the extra terms as a gravitational contribution to the energy and momentum. So, one can try to construct an energy-momentum tensor for {\it gravitation} from the Christoffel symbols---viz.~introduce a {\it pseudotensor} $t^\m{}_\n$ such that the conservation law contains an {\it ordinary derivative}. This can be done by defining $t^\m{}_\n$ so that Eq.~\eq{GRcon} takes the form of a total derivative:
\bea\label{pseudot}
\pa_\m(\sqrt{g}\,T^\m{}_\n+t^\m{}_\n)=0~.
\eea
As I already discussed in Section \ref{history}, Einstein interpreted $t^\m{}_\n$ as a `gravitational energy-momentum tensor' density. One can associate with this tensor a {\it superpotential}, ${\cal U}^{\m\l}{}_\n$, such that:
\bea\label{superU}
\sqrt{g}\,T^\m{}_\n+t^\m{}_\n=\pa_\l\,{\cal U}^{\m\l}{}_\n\,,~~~~\pa_\m\,\pa_\l\,{\cal U}^{\m\l}{}_\n=0~.
\eea
Einstein (1916a:~p.~244) obtained the superpotential from an action that differs from the Einstein-Hilbert action by a boundary term. 

Freud's (1939) expression for the superpotential is different from Einstein's, namely:
\bea\label{Freud}
{\cal U}^{\m\l}{}_\n=-\sqrt{g}\,g^{\b\s}\,\G^\a_{\b\g}\,\d^{\m\l\g}_{\a\s\n}~,
\eea
which automatically satisfies Eq.~\eq{superU}, because the symbol $\d^{\m\l\g}_{\a\s\n}$ is totally antisymmetric in its upper (and lower) indices. Another well-known expression is e.g.~Landau and Lifschitz's (1971:~p.~306)\footnote{For a critique of the Landau-Lifshitz pseudo-tensor, see Goldberg (1958:~pp.~316, 320) and Ohanian (2010:~p.~10).} pseudo-tensor (see also Bergmann, 1958:~p.~289).

However, even if Einstein's strategy appears to solve the two previous problems, (ii)-(iii), of finding an expression for the energy and momentum of the gravitational field, and of recasting Eq.~\eq{GRcon} with an ordinary derivative, so that we can integrate to get a conserved charge, it does so at the expense of introducing two new challenges:\footnote{These concerns can be found e.g.~in Hoefer (2000:~pp.~191-195). My (iv) corresponds to Duerr's (2019a:~Section 3.2) label (H2), and my (v) to his (H1). I will return to Duerr's objections in Section \ref{suphol}.}

(iv)~~{\it Non-uniqueness}: the ability to add an arbitrary superpotential ${\cal U}$ to the pseudotensor means that the pseudotensor is not unique  (I have so far mentioned three expressions: Einstein's, Freud's, and Landau and Lifschitz's).\footnote{Hork\'y and Novotn\'y (1969:~pp.~426-430) is an enlightening review of various approaches. See also Komar's (1958:~pp.~935-936) expressions.} In fact, there are an infinite number of pseudo-tensors compatible with Eq.~\eq{pseudot}. For define a gravitational (i.e.~matter-independent) pseudo-tensor with an arbitrary superpotential:
\bea\label{tGU}
t^\m{}_\n&:=&-{1\over\k}\,\sqrt{g}\,G^\m{}_\n+\pa_\l\,{\cal U}^{\m\l}{}_\n~,
\eea
where $\k=8\pi G_{\tn N}/c^4$, and $G^\m{}_\n$ is the Einstein tensor. Provided ${\cal U}$ is antisymmetric in its upper indices, 
\bea\nm
{\cal U}^{\m\l}{}_\n={\cal U}^{[\m\l]}{}_\n
\eea
(see e.g.~the Freud expression, Eq.~\eq{Freud}), the gravitational pseudo-tensor thus defined automatically satisfies the conservation law Eq.~\eq{pseudot} when the Einstein field equations are satisfied:\footnote{The expression that Einstein obtained was not antisymmetric: for a discussion, see Chen et al.~(2017:~p.~12).}

Thus the definition of the pseudo-tensor, Eq.~\eq{tGU}, is ambiguous. Nester (2004:~p.~264) summarises the problem thus: `there is no physical meaning here, only formal math---unless one has a good criterion for choosing the superpotential'. In my opinion, this is a serious problem that defenders of pseudo-tensors need to solve. 

(v)~~{\it Lack of covariance, and lack of appropriate transformation properties}: the pseudotensor is not a tensor, as can be seen from its definition in Eq.~\eq{pseudot}. For example, it can be made to vanish at any given point in appropriately chosen coordinates.\footnote{This argument was put forward to Einstein by Schr\"odinger and Bauer: for details, see Hork\'y and Novotn\'y (1969:~p.~430). See the responses by Pitts (2010:~p.~613) and Duerr (2019:~p.~8).} It is only invariant under affine transformations.

Thus Klein (1917) concluded: `After all this, I can barely believe that it is useful to designate [Einstein's] very arbitrarily built quantities $t^\m{}_\n$ as the components of the energy of the gravitational field.' Schr\"odinger (1950:~p.~104) called the gravitational pseudo-tensor a `sham tensor'. Twenty-three years later, Misner, Thorne, Wheeler (1973, p.~467) wrote: 

\begin{quote}\small
Anyone who looks for a magic formula for ``local gravitational energy-momentum'' is looking for the right answer to the wrong question. Unhappily, enormous time and effort were devoted in the past to trying to ``answer this question'' before investigators realized the futility of the enterprise.
\end{quote}

So, we have a strong warning from these authors that we should avoid looking for a general local expression for the energy and momentum of the gravitational field.

As I will mention in Section \ref{phildis}, beyond the lack of covariance, the problem is also that the pseudo-tensor does not have well-defined transformation properties: for example, it is not a geometric object, in the sense of Anderson (1967:~pp.~14-16, 424), Kucharzewski and Kuczma (1964:~pp.~15-18), and Trautman (1965:~p.~84-87):\footnote{See also Nijenhuis (1952:~pp.~23-30), Schouten (1954:~p.~67-68), and Bergmann (1958:~p.~287).} i.e.~roughly, such that its components in a given coordinate system, together with suitable transformation properties, are well-defined.\footnote{As Trautman (1965:~p.~86) mentions, spinors are not geometric objects either. In Section \ref{whatelse} I will explain what appears to be the real worry.} See the discussion in Pitts (2010:~pp.~603-604, 606) and Duerr (2019a:~Section 3.3).

In Section \ref{LEMGR}, I will propose a solution of the non-uniqueness problem (iv) based on an extension of Noether's theorem. Although the facts that I will use are known to the experts, I have not found this solution discussed in any detail in the literature on Noether's theorems that I studied---not as a solution of the alleged problem of non-uniqueness. In Section \ref{totd}, I will discuss a recent related proposal. In Section \ref{phildis}, I will assess how this approach deals with problem (iii) (in particular, the problem of the need for Killing vectors or similar structures) and with problem (v), i.e.~covariance.

Penrose (1982) made a new proposal to solve (ii) that does not require a local expression for the energy and momentum, thereby circumventing problems (iii)-(v), because he did not use the gravitational energy-momentum pseudo-tensor or other local expressions. He argued that one can still define {\it quasi-local} energy-momentum and angular momentum, i.e.~energy and momentum that are associated with a closed spacelike surface. It refers only to the geometry of the two-surface and the extrinsic curvature quantities for its embedding in the spacetime (Penrose, 1988:~p.~1). The idea has been widely accepted, and developed in various ways, in the literature on general relativity.\footnote{For some of the properties that the quasi-local energy should (presumably) have, and its relation to other definitions of energy such as the asymptotic ADM mass (for asymptotically flat spacetime regions) and Bondi mass (for asymptotically null regions), see for example Jaramillo and Gourgoulhon (2011:~pp.~97-102, 105-107) and Christodoulou and Yau (1988:~pp.~9-10).} I will return to it in Section \ref{BYQL}, when I discuss the Brown-York quasi-local stress-energy tensor.

\section{Noether's Theorems and the Energy-Momentum Tensor}\label{NoeT}

In this Section, I give Noether's first theorem (Section \ref{Noe1}), her second theorem (Section \ref{2N}), and I discuss two consequences (Section \ref{2con}) that, in my opinion, have not been sufficiently clarified in the literature.

\subsection{Noether's first theorem}\label{Noe1}

Noether's first theorem is a statement about the invariance of an action of the form:\footnote{For the connection between symmetries and conservation laws {\it without} the appeal of Lagrangian methods, see Brown (2020:~p.~3).}
\bea\label{Sf}
S[\f;x]=\int\dd^4x~{\cal L}(\f;x)~,
\eea
where $\f$ is the set of fields present in the problem (labelled by $a$, so one writes $\f_a$ if one wishes to indicate this) under an infinitesimal transformation of the dependent and-or independent variables, up to a boundary term.\footnote{This boundary term was not included in Noether's paper, and so this is a slight generalisation of her two theorems, first proven by Bessel-Hagen (1921:~p.~36), who attributes it directly to Noether: `I owe this generalization to an oral communication from Emmy Noether'. See also Kosmann-Schwarzbach (2020:~p.~11) and Brown (2020:~pp.~6, 8). This generalisation is standardly used in e.g.~Brading and Brown (2003:~p.~93) and Ba\~nados and Reyes (2016:~p.~17).} The physical significance of this invariance of the action up to boundary term is that the equations of motion derived from the new action, expressed in the new variables, will take the same form as in the old ones. Following Brading and Brown (2003:~pp.~91-97), on which the exposition in this Section is largely based,\footnote{For other treatments of Noether's theorems, closely related to the one given here, see Trautman (1962:~pp.~176-179), Horsk\'y and Novotn\'y (1969:~pp. 420-424), Barbashov and Nesterenko (1983:~pp.~541-544), Kastrup (1987:~pp.~116-117), and Brading and Brown (2000:~pp.~4-8).} I take the invariance to mean that, when the equations of motion are satisfied, the action is form-invariant up to a boundary term that depends on an arbitrary vector quantity $\L^\m$:
\bea\label{Sbar}
\d S=S[\f';x']-S[\f;x]=\int\dd^4x'\,{\cal L}(\f',x')-\int\dd^4x\,{\cal L}(\f,x)=\int\dd^4x\,\pa_\m\L^\m~.
\eea
Such a transformation, which leaves the action invariant up to a boundary term, is called a `quasi-symmetry'.\footnote{Echoing Bessel-Hagen (1921:~p.~36), Olver (1986:~p.~278) calls them `divergence symmetries'. See also Kosmann-Schwarzbach (2020:~p.~11).}

\ \\{\bf Noether's first theorem} states that, if the action Eq.~\eq{Sf} is invariant with respect to a continuous group of transformations $G_n$ that is generated by $n$ constant parameters $\xi^k$ (where $k=1,\ldots,n$), in the sense just indicated, then there are $n$ linearly independent combinations of the field equations that are equal to the total divergence of a current, as follows:
\bea\label{N1}
\sum_a\left({\pa{\cal L}\over\pa\f_a} -\pa_\m\,{\pa{\cal L}\over\pa(\pa_\m\f_a)}\right)\eta_{ak}&=&\pa_\m J^\m_k~.
\eea
Since the index $k$ labels the $n$ generators of $G_n$, there is one such equation for each generator. The index $a$ labels the various fields present in the problem, and $\eta_{ak}$ is a field-dependent expression, defined by the variation of the fields:
\bea\label{eta}
\d_\xi\,\f_a=\eta_{ak}\,\xi^k~.
\eea
The current is given by the following expression:
\bea\label{Ncurr}
J^\m_k=-\sum_a{\pa{\cal L}\over\pa(\pa_\m\,\f_a)}\,{\pa\d\f_a\over\pa\xi^k}-{\cal L}\,{\pa\d x^\m\over\pa\xi^k}+{\pa\L^\m\over\pa\xi^k}~.
\eea
The converse is also true: if the field equations satisfy Eq.~\eq{N1}, then this implies the invariance of the action under the corresponding group $G_n$. The theorem remains valid in the limiting case of an infinite number of parameters, $n\rightarrow\infty$.

The above current is weakly conserved, since the left-hand side of Eq.~\eq{N1} is a linear combination of the Euler-Lagrange field equations for $\f$. 

\ \\{\bf Example: Maxwell theory.} Let us illustrate the above for Maxwell theory, familiar from Section \ref{Mxth}.\footnote{Noether's theorems for Yang-Mills theory are discussed in Barbashov and Nesterenko (1983:~pp.~552-556).} Thus we take the group generated by $\xi^k$ to be the group of constant spacetime translations. Since $A$ is a vector field, the index $a$ labelling the different fields can be taken to be the spacetime index $\m$ of the vector field components $A_\m$. Since $\xi^k$ is the generator of translations, $\xi$ is also a vector field in Minkowski space, and the index $k=1,\ldots,n$, labelling the generators, is also a spacetime index, $\n=1,\ldots,4$. Thus the group $G_4$ here under consideration is the group of constant spacetime translations generated by $\d x^\m=\e^\m$. 

To see how $\eta_{ak}$ depends on the vector field $A$, we temporarily write it as $\eta_{\m\n}$ (not to be confused with the Minkowski metric!), and compare the general variation Eq.~\eq{eta} with the variation of the gauge field Eq.~\eq{varA} (which is obtained by expanding the field components $A_\m$ in terms of the constant coefficients, $\d x^\m=\e^\m$). The comparison gives $\eta_{\m\n}=-\pa_\n A_\m$. 

Let us check that the Noether current Eq.~\eq{Ncurr} reproduces the energy-momentum tensor of electromagnetism, Eq.~\eq{EMT}. Since $\d x^\m=\e^\m$, we have ${\pa\d x^\m\over\pa\e^\n}=\d^\m_\n$. The last term in Eq.~\eq{Ncurr} is zero, and the first term is $-{\pa{\cal L}\over\pa(\pa_\m A_\l)}\,\eta_{\l\n}={\pa{\cal L}\over\pa(\pa_\m A_\l)}\,\pa_\n A_\l=-F^{\m\l}\,\pa_\n A_\l$. Thus we get $J^\m_\n=-F^{\m\l}\,\pa_\n A_\l-{\cal L}\,\d^\m_\n=-T^\m{}_\n$, i.e.~the energy-momentum tensor Eq.~\eq{EMT}. The Noether current conservation equation Eq.~\eq{N1} of course reproduces Eq.~\eq{cons}.

We can repeat this for the gauge-invariant variation $\d A_\m=F_{\m\n}\,\e^\n$ leading to the energy-momentum tensor with the improvement term, i.e.~Eq.~\eq{td}. Comparing this variation with Eq.~\eq{eta}, we find that $\eta_{\m\n}=F_{\m\n}$. Evaluating the current Eq.~\eq{Ncurr}, we indeed get $J^\m_\n=-T'{}^\m{}_\n$, i.e.~we reproduce the standard energy-momentum tensor, Eq.~\eq{Tp}.

\subsection{Noether's second theorem and the boundary theorem}\label{2N}

In this Section, I will give Noether's second theorem and the identities that it implies. I will follow the terminology in Brading and Brown (2003:~pp.~100-104), who distinguish a set of formulas that they call the `boundary theorem', due to Klein (1918), from Noether's second theorem. (I will occasionally refer to both sets of formulas, collectively, as `Noether's second theorem', since both are entailed by Noether's general expression for the invariance of the action under a continuous group). 

The boundary theorem and Noether's second theorem, in generalised form, assume that the action Eq.~\eq{Sf} is invariant (up to a boundary term) with respect to a continuous group of transformations $G_{\infty n}$ that is generated by $n$ arbitrary {\it functions} $\xi^k(x)$ ($k=1,\ldots,n$). 

Before giving the theorems, it will be useful to introduce three important pieces of notation. I use the Einstein summation convention for repeated indices $a$ and, below, also for repeated $k$ indices:
\bea
E^a&:=&{\pa{\cal L}\over\pa\f_a}-\pa_\m\,{\pa{\cal L}\over\pa(\pa_\m\f_a)}\label{Ea}\\
J^\m_k&:=&-{\pa{\cal L}\over\pa(\pa_\m\f_a)}\,{\pa\d\f_a\over\pa\xi^k}-{\cal L}\,{\pa\d x^\m\over\pa \xi^k}+{\pa\L^\m\over\pa \xi^k}\label{Jmk}\\
U^{\m\n}_k&:=&{\pa{\cal L}\over\pa(\pa_\n\f_a)}\,\g^\m_{ak}+{\pa\L^\m\over\pa(\pa_\n \xi^k)}~.\label{defP}
\eea
The quantities here defined are, respectively, the Euler-Lagrange equations of motion, the Noether current, and the superpotential (more on this later). $\g^\n_{ak}$ is a (field-dependent) variation of the field, to be defined below. $\L^\m$ is a boundary term that may appear in the variation of the action, Eq.~\eq{Sbar}. Notice that we can also apply Noether's theorem to a single term in (i.e.~to a {\it part} of) the action, to require the invariance of that particular term. In that case, for example, $E^a$ is simply the functional variation of that term in the action with respect to the fields $\f_a$, and likewise for the other two quantities.

Using this notation, Noether's {\it first} theorem takes a particularly simple form:
\bea\label{N1notation}
E^a\,\eta_{ak}=\pa_\m J^\m_k~.
\eea

The expression $\g^\n_{ak}$ in Eq.~\eq{defP}, together with $\eta_{ak}$, are the (field-dependent) variations of the field with respect to $\xi$ and its first derivative (repeated $k$ indices are again summed over) relevant to Noether's second theorem and her boundary theorem:
\bea\label{varf}
\d\f_a=\eta_{ak}(\f;x)\,\xi^k(x)+\g^\m_{ak}(\f;x)\,\pa_\m\xi^k(x).
\eea
This variation is analogous to Eq.~\eq{eta}, where the second term is non-zero because in Noether's second theorem $\xi$ is a function of $x$.

\ \\{\bf The boundary theorem,} with this notation, takes the following simple form:
\bea
\pa_\m(E^a \g^\m_{ak})&=&\pa_\m J^\m_k\label{NB}\\
E^a\g^\m_{ak}&=&J^\m_k-\pa_\n U^{\m\n}_k\label{NB1}\\
U^{\m\n}_k&=&-U^{\n\m}_k~.\label{antis}
\eea
Again, the index $k=1,\ldots,n$ labels the functions $\xi^k(x)$ that generate $G_{\infty n}$, and so there is one such equation for each generator. Note that the first equation here follows from the second and third.\footnote{My notation, using the Einstein convention for the three different types of indices, also corrects errors in Brading and Brown (2003:~pp.~92, 95, 97, 101) here and in their equation corresponding to my Eq.~\eq{Ncurr}, where summation signs over $a$ (i.e.~their $i$) are incorrectly extended to terms that contain no index $a$.}

\ \\{\bf Noether's second theorem} reads:
\bea\label{N2}
E^a\,\eta_{ak}=\pa_\m(E^a\g^\m_{ak})
\eea
Again, the full statement is that, if the action Eq.~\eq{Sf} is invariant under the group $G_{\infty n}$ of variations as indicated, then the Euler-Lagrange equations satisfy the identities Eq.~\eq{N2}. The converse statement is here also valid.

Notice that both the boundary theorem and Noether's second theorem, Eqs.~\eq{NB}-\eq{N2}, take forms that differ from Noether's first theorem Eq.~\eq{N1notation}. However, it is possible to derive Noether's first theorem, which is the special case that the continuous group of transformations $G_n$ is a subgroup of $G_{\infty n}$, i.e.~$G_n\subset G_{\infty n}$. In this case, the generators $\xi^k$ of $G_n$ are constant, and the corresponding second term in Eq.~\eq{varf} vanishes. Combining the first of Eq.~\eq{NB} with Eq.~\eq{N2}, we recover Noether's first theorem, Eq.~\eq{N1notation}, with the $k$ generators thus identified.

\ \\{\bf Example: Maxwell theory.} It will be instructive to compare with our previous notation and reproduce the conservation law of Maxwell's action coupled to a gauge-invariant source, Eq.~\eq{MxJ}. The gauge symmetry modifies the field $A$ as: $\d A_\m=\pa_\m\l$. Since $\l$ is a function of $x$, Noether's second theorem is relevant here. Comparing with Eq.~\eq{varf}, since there is a single generator, namely $\l$, we see that $k=1$ and $\xi=\l$ (and thus, we can drop the index $k$). As before, the index $a$ is a spacetime index, $a=\n$. Further, $\d A_\m$ only depends on $\l$'s derivative and not on $\l$ itself, so $\eta_\m=0$ and $\g^\m_\n=\d^\m_\n$. In particular, this means that there is no non-trivial subgroup of transformations with {\it constant} parameters in this case, and so no Noether first theorem for the gauge transformations. Also, recall, from Eq.~\eq{gaugevar}, that the variation of the action was zero up to a boundary term, i.e.~a term with a total derivative $\int\dd^4x~\pa_\m\L^\m$. We find that $\L$ is given by:
\bea
\L^\m=J^\m\,\l~.
\eea
This confirms my claim in Section \ref{Mxth} that, with this generalised form of Noether's theorem, the gauge parameter $\l$ does not need to be zero at the boundary. It is straightforward to check the boundary theorem and Noether's second theorem, Eqs.~\eq{NB}-\eq{N2}. For the quantities defined in Eqs.~\eq{Ea} and \eq{defP}, we get the following expressions:
\bea
E^\m&=&J^\m-\pa_\n F^{\m\n}\nn
U^{\m\n}&=&F^{\m\n}~.
\eea
The first equation is of course the expression for the Euler-Lagrange equation derived from the action Eq.~\eq{MxJ}. 

With these expressions in hand, it is straightforward to verify that the boundary theorem, i.e.~Eqs.~\eq{NB}-\eq{antis}, is trivially satisfied, given the anti-symmetry of the field tensor. Interestingly, the left-hand side of the expression from Noether's second theorem, Eq.~\eq{N2}, is zero (because $\eta$ is zero), and so Noether's second theorem takes the form $\pa_\m E^\m=0$. Given the anti-symmetry of the field tensor, this is identically satisfied iff $J$ is conserved, {\it independent of the satisfaction of the equations of motion of Maxwell theory.} Thus imposing gauge symmetry implies, through Noether's theorem, that Maxwell theory can only be coupled to a conserved current $J$.



\subsection{Two consequences of Noether's second theorem}\label{2con}

I will first make a comment that I have not found in the literature on Noether's theorems, although it may well be known to the experts.\footnote{For example, Barbashov and Nesterenko (1993:~p.~562) mention, for the specific energy-momentum (pseudo-)tensors considered by Einstein and by Lorentz, that they are related by a total derivative, and so lead to the same equations of motion. But, a few paragraphs later, they categorise the non-uniqueness of the pseudo-tensor as a ``difficulty'', without mentioning the relevance of their previous remark, nor noting that the pseudo-tensor is unique after fixing the boundary terms of the action. Likewise, Ohanian (2010:~p.~10) has a formula for the pseudo-tensor in terms of the gravitational action, but he drops a total derivative term from it, and he does not comment on the effect of the dropped boundary term on the pseudo-tensor. Similar comments apply to Goldberg (1958:~p.~318), Bergmann (1958:~p.~288), Komar (1958:~pp.~934-935), Trautman (1962:~pp.~171, 174, 190-191), and Horsk\'y-Novotn\'y (1969:~pp.~425-427), all of whom explicitly survey and compare different expressions for the pseudo-tensor, mention the lack of uniqueness and sometimes propose a criterion for fixing a unique expression, but fail to mention the most obvious criterion: namely, that of fixing the boundary terms. That the superpotential is fixed, through Noether's theorem, by the choice boundary terms of the action, seems implicit in Wald (1993:~p.~3429), but he does not discuss or mention the gravitational pseudo-tensor or the superpotential at all. In Sections \ref{LEMGR} and \ref{hst}, I will further comment on the physical significance of fixing the boundary terms.\label{litbdytms}}
These remarks are made visible by the simplified notation above, and they will bear on the question of energy and momentum conservation discussed in the next Section. Then I will answer Klein's critique of the local energy-momentum conservation law in GR.

\ \\{\bf The superpotential is fixed, through Noether's theorem, by the boundary terms.} In studies of Noether's theorem in connection with general relativity, emphasis is often laid on the integration of Eq.~\eq{NB} (see Brading and Brown, 2003:~p.~102), as follows:
\bea\label{JU}
J^\m_k=E^a\g^\m_{ak}+\pa_\n {\cal U}^{\m\n}_k~,
\eea
where ${\cal U}^{\m\n}_k$ is an arbitrarily chosen superpotential, analogous to the one in Eq.~\eq{tGU}, whose divergence is automatically zero:
\bea\label{paU}
\pa_\m\,\pa_\n\,{\cal U}^{\m\n}_k=0~.
\eea
Indeed, this is how Einstein himself arrived at his result, Eq.~\eq{tGU}, for the gravitational pseudo-tensor.

However, note that Noether's theorem gives us, in Eq.~\eq{NB1}, an expression for the current that is not arbitrary but fixed:
\bea\label{JP}
J^\m_k=E^a\g^\m_{ak}+\pa_\n U^{\m\n}_k~,
\eea
where $U$ is {\it given} by Eq.~\eq{defP}. Since $U$ is antisymmetric in its upper indices, it indeed satisfies Eq.~\eq{paU}.\footnote{For diffeomorphism invariant theories, the superpotential corresponds, in Wald's (1993:~p.~3429) formulation, to the Noether charge ${\bf Q}$, which in a 4-dimensional spacetime is a 2-form.}

The compatibility of Eqs.~\eq{JU} and \eq{JP} requires that we must have:\footnote{This holds up to an antisymmetric, total derivative, term of the form $\pa^{[\m}w^{\n]}$, i.e.~the superpotential obtained by integrating Eq.~\eq{NB} must be compatible with Eq.~\eq{NB1}. I thank Henrique Gomes for pointing this out. Note that this ambiguity only appears if one integrates Eq.~\eq{NB} regardless of the other equations. Most important, this ambiguity does not affect the superpotential $U$ of Noether's theorem, which is completely fixed by Noether's theorem, through Eq.~\eq{NB1}.} ${\cal U}^{\m\n}_k=U^{\m\n}_k$. Since $U$ was defined in Eq.~\eq{defP}, and the boundary theorem depends on its having this exact form, this means that the superpotential is {\it not arbitrary,} but is fixed through Eq.~\eq{defP}. 

How exactly is the superpotential fixed by Noether's theorem? The first term in Eq.~\eq{defP} is fixed by the choice of the Lagrangian density and of the variation of the field. The second term (depending on $\L^\m$) is fixed by the boundary term that appears in the variation of the action. This means that {\it Noether's theorem determines the superpotential, $U$, once the action and its first variation are fixed.} 

In the next Section, I will return to the physical interpretation of this result, and to how it solves the problem (iv) of non-uniqueness in general relativity from Section \ref{history}.

\ \\{\bf A response to the Hilbert-Klein-Noether critique.} The above also casts light on the Klein-Hilbert worry, problem (i) from Section \ref{Einstein}, that energy and momentum conservation laws in GR are identities that do not require using the Einstein field equations, and hence they lack physical content. We can see that this follows from Eq.~\eq{JP}. Since the Noether current $J$, rather than its derivative, is linear in the equations of motion, the satisfaction of the equations of motion does not play a role in securing current conservation, i.e.~in the vanishing of the current's {\it derivative}. In other words, by the time one wants to prove the conservation of the current, the equations of motion have ``disappeared'' from it. They play a role in the current, but not in the conservation. This is unlike Noether's first theorem, Eq.~\eq{N1}, where the {\it divergence} of the current, rather than the current itself, is linear in the equations of motion. 

Noether herself took her theorems to confirm Hilbert's view that there is no proper energy conservation law in general relativity. She wrote (see Kosmann-Schwarzbach, 2011:~p.~20):
\begin{quote}\small
Given [an action] $S$ invariant under the group of translations, then the energy relations [i.e.~the conservation laws corresponding to translations] are improper [i.e.~the divergences vanish identically] iff $S$ is invariant under an infinite group which contains the group of translations as a subgroup...\footnote{The group of translations being a {\it subgroup} of an infinite group is not sufficient to get improper energy relations. For example, in the Maxwell theory in Minkowski space, the group of translations, $G_4$, is a sugroup of the total group containing both translations and local gauge transformations, i.e.~$G_4\times G_{\infty1}$, where $G_{\infty1}$ is the group of local gauge transformations: and yet, the energy relations of the Maxwell theory are {\it proper.} Thus, for {\it improper} energy relations, it is also necessary (and this is obviously what Noether had in mind) that the four constant generators of translations, $\e^\m$, are obtained by setting four corresponding generators of the infinite group, i.e.~the four functions $\xi^\m(x)$, to be constant. The latter condition is not satisfied in the Maxwell theory, but it is satisfied in general relativity.} As Hilbert expresses his assertion, the lack of a proper law of energy constitutes a characteristic of the ``general theory of relativity''.
\end{quote}

It is of course true that the details of the conservation are different in the two cases, because they satisfy different theorems. Noether's second theorem imposes more constraints. 

However, as we also saw above, the boundary theorem, combined with Noether's second theorem, does imply an identity that looks exactly like Noether's first theorem, Eq.~\eq{N1}. 
The two theorems differ in that, in the case of the boundary theorem, the equations of motion appear in the form of the current itself, while they do not appear in Noether's first theorem. Thus the disappearance of the equations of motion from the divergence of the Noether current in the boundary theorem is not a fact that is additional to its disappearance from the current itself (the two are connected through Noether's second theorem, Eq.~\eq{N2}). 

Let me make this reply to Klein (in particular) more precise. Recall his points (1) and (2), from Section \ref{history}, that the divergence of the total energy-momentum tensor density, i.e.~$\tau^\m{}_\n:=\sqrt{g}\,T^\m{}_\n+t^\m{}_\n$, has no physical meaning because it is an improper conservation law, i.e.~it consists of two terms:\footnote{Here, and in the rest of the paper, the notation $\sqrt{g}$ means $\sqrt{|\det(g)|}$.} (1) the divergence of the Einstein field equations, and (2) the divergence of the superpotential, which is identically zero. This follows from the boundary theorem, Eq.~\eq{NB1}, which gives: $\pa_\m J^\m_k=\pa_\m(E^a\g^\m_{ak})+\pa_\m\pa_\n U^{\m\n}_k$, where: (1) $E^a$ appears here under the divergence, (2) the second term vanishes identically. By (1), the equations of motion play no direct role in the vanishing of the divergence (as they do in e.g.~classical mechanics).\footnote{Klein's (1918a:~p.~504) argument is not quite as explicit as this. This reconstruction of the argument follows Brading (2005:~p.~126): see footnote \ref{KElet}.} Thus the equations of motion play no direct role in the vanishing of the divergence (as they do in e.g.~classical mechanics).

But using Noether's second theorem, i.e.~Eq.~\eq{N2}, for non-zero $\g^\m_{ak}$, we rewrite this equation as: $\pa_\m J^\m_k=E^a\eta_{ak}+\pa_\m\pa_\n U^{\m\n}_k$.\footnote{This is different from the case of the local gauge symmetry of the Maxwell theory, because $\g^\m_{ak}$ there is indeed zero, while it is non-zero in general relativity.} Since the second term is identically zero, we see that the Noether current $J$ is conserved if and only if (a linear combination of) the equations of motion are satisfied, in complete analogy with Noether's first theorem.\footnote{The (indirect) physical significance of Noether's second theorem has long been recognised by philosophers. Recently, a number of authors have also emphasised the (direct) physical significance of improper conservation laws for local gauge symmetries: see Teh (2016:~p.~99) and Sus (2017:~pp.~274-275).}

Slightly anticipating the results of the next Section, we can show this explicitly in the case of general relativity. Rewrite: $\tau^\m{}_\n=2E^\m{}_\n+\pa_\l U^{\m\l}_\n$, where $E^\m{}_\n={1\over2\k}\sqrt{g}(G^\m{}_\n-\k T^\m{}_\n)$ is proportional to the Einstein field equations. Using the contracted Bianchi identities, we find: $\pa_\m\t^\m{}_\n=E^{\a\b}\pa_\n g_{\a\b}+\pa_\m\pa_\l U^{\m\l}_\n$. Since the second term vanishes identically, the total energy-momentum (pseudo-) tensor density is conserved if and only if a certain linear combination of Einstein's equations are satisfied. This seems to be what Einstein (1918:~p.~513) had in mind in his reply that {\it part} of the field equations are used,\footnote{Also in classical mechanics, it is only {\it part,} i.e.~a linear combination, of the equations of motion that appear on the right-hand side. Consider, as an elementary example, the Lagrangian $L=\half m\dot x_i^2+V(x)$, where $i=1,2,3$. The time derivative of the Hamiltonian gives: $\mbox{d}H/\mbox{d}t=E_i\dot x_i$, where $E_i=m\ddot x_i+\pa_iV$, i.e.~the conservation law is proportional to a {\it linear combination} of the equations of motion.} and he was indeed right.\footnote{Klein's argument {\it can} go through if $E^a\g^\m_{ak}=0$ without use of the equations of motion, e.g.~if $\g^\m_{ak}=0$. In that case, it is true that the first term vanishes before its derivative is taken. However, this is not the case in general relativity.}

\section{Noether's Theorems in General Relativity}\label{LEMGR}

In this Section, I will apply Noether's theorem to general relativity coupled to matter.\footnote{For a discussion of Noether's theorem and Kretschmann's objection to Einstein's treatment of general covariance, see Brown and Brading (2002:~pp.~22-25). This paper also contains a treatment of Noether's theorem for electrodynamics with an arbitrary background metric.} I will first focus on the case of pure gravity, and then discuss adding matter.

Take the action Eq.~\eq{Sf} with Lagrangian density given by ${\cal L}={\cal L}_{\tn{E}}+{\cal L}_{\tn{m}}$, where ${\cal L}_{\tn{E}}$ is the Einstein Lagrangian, and ${\cal L}_{\tn{m}}$ is the matter Lagrangian. The Einstein Lagrangian is given by:
\bea\label{LE}
2\k\,{\cal L}_{\tn{E}}(g_{\m\n},\pa_\l g_{\m\n})&:=&\sqrt{g}~g^{\m\n}(\G^\l_{\m\n}\G^\l_{\l\s}-\G^\s_{\m\l}\G^\l_{\n\s})\nn
&=&\sqrt{g}~R-\pa_\m(\sqrt{g}\,(g^{\a\b}\G^\m_{\a\b}-g^{\m\n}\G^\l_{\n\l}))~,
\eea
and so the Einstein action differs from the usual Einstein-Hilbert action by a total derivative term that does not affect the equations of motion.\footnote{There are various boundary terms that one can add, depending on the choice of boundary conditions, and these lead to different expressions for the pseudo-tensor. For a treatment of Noether's second theorem for the Einstein-Hilbert action, see for example Hork\'y and Novotn\'y (1969:~p.~424). In this Section, I follow some of the historical literature in using the Einstein action. I will return to the issue of boundary terms and boundary conditions in Sections \ref{BYQL} and \ref{Revisited}.}

Let us first write out the notation for Noether's theorems applied to the Einstein term in the action. For, although we have a coupled gravity-matter system, we may of course require the diffeomorphism invariance of a single term. 

The field $\f_a$ is then the metric tensor, and the index $a$ is a symmetric pair of spacetime indices, $a=\a\b$, ranging from 1 to $4$, i.e.~the dimension of spacetime. Under a local translation $\d x^\m=\xi^\m(x)$, the metric transforms as follows:
\bea\label{dgab}
\d g_{\a\b}=-\nabla_\a\xi_\b-\nabla_\b\xi_\a=-\xi^\m\,\pa_\m\,g_{\a\b}-2g_{\n(\a}\,\pa_{\b)}\xi^\n~.
\eea
The index $k$ is also a spacetime index $\m$. Comparing with Eq.~\eq{varf}, we see that $\eta_{ak}$ is equal to $-\pa_\m g_{\a\b}$, and $\g^\m_{ak}$ is equal to $-2\d^\m_{(\a}\,g^{}_{\b)\n}$. 



\ \\{\bf Noether's second theorem,} Eq.~\eq{N2}, applied to the gravitational part of the action, gives the contracted {\it Bianchi identity} for the Einstein tensor:\footnote{This formulation follows Noether's (1918:~p.~241) original methods, reviewed in Section \ref{NoeT}. However, there is an important and useful reformulation of Noether's theorem, for diffeomorphism-invariant Lagrangians, in terms of differential forms: see Wald (1993:~pp.~3428-3429) and Wald and Zoupas (2000:~pp.~2-3).}
\bea\label{Bianchi}
\pa_\m(\sqrt{g}\,G^\m{}_\n)=\half\sqrt{g}\,G^{\a\b}\,\pa_\n g_{\a\b}~~\Leftrightarrow~~\nabla_\m G^\m{}_\n=0~,
\eea
Noether's second theorem, applied to the {\it matter} part of the action, gives the covariant analogue of the energy and momentum conservation equation, i.e.~Eq.~\eq{GRcon}. That equation is, as expected, obtained as an identity, without the use of the equations of motion.

Let us write down the Noether current, Eq.~\eq{Jmk}, $J^\m_\n=:-t^\m{}_\n$, for the gravitational part of the action:
\bea\label{tmn}
t^\m{}_\n=-{\pa{\cal L}_{\tn{E}}\over\pa(\pa_\m g_{\a\b})}\,\pa_\n\,g_{\a\b}+{\cal L}_{\tn{E}}\,\d^\m_\n-{\pa\L^\m\over\pa\xi^\n}~.
\eea
This gives a gravitational analogue of the matter energy-momentum tensor. In fact, as I shall discuss in a moment, it is not a tensor: it is the gravitational energy-momentum {\it pseudo-tensor}. 

\ \\{\bf The boundary theorem} gives the two following results. The first equation of the boundary theorem, Eq.~\eq{NB}, applied to the Einstein action, gives an expression for the divergence of the gravitational pseudo-tensor:
\bea\label{ptmn}
\pa_\m t^\m{}_\n=-{1\over2\k}\,\sqrt{g}\,G^{\a\b}\,\pa_\n g_{\a\b}=-{1\over\k}\,\pa_\m(\sqrt{g}\,G^\m{}_\n)~,
\eea
where in the last step I used the Bianchi identity, Eq.~\eq{Bianchi}. (The same result can be obtained by taking the divergence of Eq.~\eq{tmn} directly). Using Einstein's equations, $G_{\m\n}=\k\,T_{\m\n}$, we find Einstein's form of the conservation law, Eq.~\eq{pseudot}.
The second equation of the boundary theorem, Eq.~\eq{NB1}, reproduces the expression Eq.~\eq{tGU} for the pseudo-tensor in terms of the superpotential.

I will now argue that this form of the boundary theorem shows that the alleged non-uniqueness of the pseudo-tensor, i.e.~problem (iv) in Section \ref{Einstein}, only arises if one treats Eq.~\eq{pseudot} independently of Noether's theorems. For if one takes Eq.~\eq{pseudot} as the {\it definition} of the pseudo-tensor and integrates it, then the pseudo-tensor indeed admits an {\it arbitrary} superpotential, as in Eq.~\eq{tGU}.\footnote{For example, Lam (2011:~pp.~1017-1018) {\it defines} the energy-momentum pseudo-tensor in terms of an expression that automatically satisfies the conservation equation Eq.~\eq{pseudot}, as in Eq.~\eq{tGU}, and thus contains an arbitrary superpotential.}

But if one, correctly, applies Noether's theorem to the gravitational part of the action, Eq.~\eq{LE}, then both the pseudo-tensor and the superpotential are {\it fixed} by their Noether expressions in Eqs.~\eq{Jmk} and \eq{defP}, in terms of: 

(1) the Lagrangian density, 

(2) the boundary term $\L^\m$ that appears in the Noether variation of the action, Eq.~\eq{Sbar}. 

Noether's expression Eq.~\eq{defP} for the superpotential indeed shows that the superpotential is not arbitrary, but is fixed when (1) and (2), including the boundary terms of the action, are fixed. One may add boundary terms to the action without changing the equations of motion, and in this sense there {\it is} freedom to change the superpotential by changing the boundary terms. However, boundary terms often {\it do} change the boundary conditions, and so the boundary terms must be such that they give the desired boundary conditions. (Typically, the boundary conditions are not sufficient to uniquely determine the boundary terms; the finiteness of the on-shell action is also required to fix them: see Section \ref{BYQL}). 

This means that the freedom to choose the superpotential has a physical meaning: it is the freedom to change the action, without affecting the equations of motion, by boundary terms. The latter have a physical meaning, because they determine: 

(a) the boundary conditions in the variation of the metric that gives the equation of motion;

(b) the on-shell value of the (Euclidean) action, which e.g.~in thermodynamic approaches to gravity, especially in the context of black holes, is interpreted as a gravitational potential;\footnote{See e.g.~Hawking (1979:~pp.~772, 778-779).  Wald (1993:~p.~3431) and Hawking and Horowitz (1996:~pp.~1494-1495) contain reformulations of the first and second laws of black hole mechanics in terms of the Lagrangian and its variation. For some examples, see Wald and Zoupas (2000:~p.~8-14), Jacobson et al.~(1994:~pp.~6592-6593).}

(c) the boundary terms of the Hamiltonian are also determined by the boundary terms of the Lagrangian.\footnote{For the relation between the Lagrangian and Hamiltonian approaches to GR, see for example Wald (1984:~pp.~464-465)} Thus the significance of the boundary terms does not hinge on interpreting the Lagrangian physically.

Chang et al.~(1999) notice that, in a Hamiltonian approach, the freedom to change the superpotential is related to the freedom to change the boundary conditions, i.e.~analogous to my points (a) and (c) above. However, so far as I can see, they do not notice that it is in general {\it Noether's theorem} that completely fixes the superpotential from the variation of the action, in particular through Eq.~\eq{defP}.

In the absence of Killing vectors that allow us to construct the energy by integrating a scalar expression like Eq.~\eq{Killingc}, some authors use the pseudo-tensor to convert the covariant derivative in Eq.~\eq{GRcon} into a partial derivative and integrate. Thus these authors define a total momentum vector $P_\m$ as an integral over a spacelike volume $V$:\footnote{See Weinberg (1972:~p.~167), Landau and Lifshitz (1971:~p.~306), Chen et al.~(2018:~p.~3, 2018a:~p.~4). For the philosophical literature on this point, see Read (2020:~p.~217) and Hoefer (2000:~p.~194).} 
\bea\label{pmu}
P_\m:=\int_V\dd\Si_\n\left(T^\n{}_\m+{1\over\sqrt{g}}~t^\n{}_\m\right).
\eea
By choosing special coordinates, where the normal vector to the boundary $S$ of the volume $V$ points along the time direction, these authors derive:
\bea\label{dpdt}
{1\over c}{\dd P_\m\over\dd t}=-\int_S\dd^2x\,Nn_i\,(\sqrt{h}~T^i{}_\m+t^i{}_\m)~,
\eea
where, as before, $n_i$ is the outward-pointing unit vector normal to $S$, $N$ is the lapse function, and $h$ is the metric on $S$ induced from $g$ in these coordinates. This expression can be shown to reproduce the well-known ADM (for Arnowitt, Deser, and Misner) formulas for asymptotically flat spacetimes: see Arnowitt et al.~(1962:~p.~16), Wang (2015:~p.~3), Chen et al.~(2017:~p.~200), Chen et al.~(2018:~p.~3), Chen et al.~(2018a:~p.~158).\footnote{For a discussion of the history of the ADM mass formula, and various versions and derivations of it, see Ohanian (2010:~p.~9).} This expression is regarded as an analogue of the charge conservation law Eq.~\eq{Qcons}.

However, as I will discuss in Section \ref{Revisited}, the above definition is unsatisfactory: for it transforms covariantly only under affine transformations, and the analogy with charge conservation depends on a very special embedding of the volume $V$ in the spacetime and on the associated choice of coordinates.

\section{Quasi-local Stress-Energy Tensors in GR}\label{BYQL}

It is time to discuss other proposals for the energy-momentum and angular momentum of the gravitational field. This Section introduces two such. Section \ref{BYql} introduces the Brown-York quasi-local stress-energy tensor, which in Section \ref{hst} is then modified and gets a holographic interpretation that, I will argue, casts light on the nature of energy and momentum in GR. In Section \ref{diffeos}, I will discuss the transformations of the Brown-York stress-energy tensor under diffeomorphisms.

\subsection{The Brown-York quasi-local stress-energy tensor}\label{BYql}

At the end of Section \ref{Einstein}, where I discussed five problems with the gravitational pseudo-tensor, I mentioned Penrose's idea of giving up local expressions for energy and momentum in GR, and to consider quasi-local notions instead. This Section presents one such expression in some detail.

Brown and York's (1993) proposal is to define the quasi-local stress-energy tensor, associated with a spatially bounded region $M$, as the conjugate of the classical action with respect to an appropriately defined metric on the boundary of that region. 

The action to consider contains six terms. There is the usual Einstein-Hilbert action, coupled to matter, and possibly including a cosmological constant term. Furthermore, since the spacetime region $M$ is bounded, appropriate boundary terms are required in order to make the variational problem well-defined. Since $\pa M$ consists of two spacelike hypersurfaces $\Sigma_{\pm}$ to the past and future of $M$, and the timelike three-surface $N$ that spatially bounds $M$, there are three such boundary terms (where the first two are abbreviated into a single term, see Eq.~\eq{action} below).\footnote{Brown and York's (1993:~p.~1410) coordinates are such that the hypersurfaces $\Sigma_\pm$ and $N$ are orthonormal to each other on the two-surface where they intersect. Furthermore, no additional term in the action for this two-surface is required.} To make the variational problem of the Einstein-Hilbert action well-defined, a choice of boundary conditions for the metric on the boundary is required: Brown and York consider Dirichlet boundary conditions, i.e.~keeping the metric on the boundary fixed.\footnote{These boundary conditions also render the gravitational part of the action well-defined in case a cosmological constant term is present.} The appropriate boundary term for Dirichlet boundary conditions is the extrinsic curvature, denoted $K$ on the hypersurfaces $\Sigma_\pm$, and $\Theta$ on the three-surface $N$. Finally, a reference term is required that renders the value of the action and the stress-energy tensor well-defined (I will have more to say about this term later). Thus the action is:
\bea\label{action}
S={1\over2\k}\int_M\dd^4x\,\sqrt{-g}~R\!\!\!\!
\underset{\tn{extrinsic curvature of hypersurfaces $\Si_{\pm}$}}{\underbrace{+{1\over\k}\int_{\Sigma_\pm}\dd^3x\,\sqrt{h}~K}} ~~
\underset{\tn{extrinsic curvature of $N$}}{\underbrace{-{1\over \k}\int_N\,\sqrt{-\g}~\Theta}}
+S^{\tn{ref}}+S^{\tn{matter}}
\eea
Here, $h$ is the metric on the hypersurfaces $\Sigma_\pm$ embedded in $M$, and $\g$ is the metric on the three-boundary $N$ embedded in $M$. $K$ and $\Theta$ are their corresponding extrinsic curvatures. 

Varying this action with respect to the metric, we get:
\bea\label{vary}
\d S=-\int_M\dd^4x~E^{\m\n}\d g_ {\m\n}+\half\int_{\Sigma_\pm}\dd^3x~\sqrt{h}~P^{ij}\,\d h_{ij}+\half\int_N\dd^3x~\sqrt{\g}~\t^{ij}_{\tn{BY}}\,\d\g_{ij}~.
\eea
The first term contains the Einstein field equations including matter, and thus vanishes if the field equations are satisfied. The second and third terms define the momenta conjugate to the three-metrics $h$ on $\Sigma_\pm$ and $\g$ on $N$, respectively (with Dirichlet boundary conditions, as announced). $\t^{ij}_{\tn{BY}}$ is a genuine (quasi-local) energy-momentum tensor, because it contains both spacelike and timelike components. It also contains contributions from both the gravitational and the matter part of the action. The subscript `BY' emphasises that this is the energy-momentum tensor, as defined in the proposal by Brown and York. 

Restricting for the moment to the gravitational contribution to this last term, we get the following result for the Brown-York stress-energy tensor density:
\bea\label{BY}
\t^{ij}_{\tn{BY}}=-{1\over\k\sqrt{\g}}\left(\Theta\,\g^{ij}-\Theta^{ij}\right)+\sm{(subtracted terms)}~.
\eea
The subtracted terms subtract a potentially divergent piece, in cases where one is evaluating this tensor at infinity. These subtracted terms are obtained by subtracting to the action Eq.~\eq{action} the same action, evaluated on an appropriate reference spacetime. For example, if the solutions of the Einstein field equations that one considers are asymptotically flat, one usually subtracts from Eq.~\eq{action} the same action, but evaluated on a Minkowski solution. In this way, the `on-shell' value of the action (i.e.~the action evaluated on a particular solution) is finite, and also the Brown-York stress-energy tensor is finite. I will return to this procedure in the next Section, where I will mention an alternative way to resolve this problem that does {\it not} appeal to the reference spacetime.

The boundary stress-energy tensor (density) satisfies one of Einstein's equations: 
\bea\label{bE}
\na_i\t^{ij}_{\tn{BY}}=-T^{n j}\,,
\eea
where $T^{nj}$ is the matter energy-momentum tensor with one index projected normally, and the other tangentially, to $N$. This is a covariant balance equation for the quasi-local stress-energy tensor, where the contribution on the right is from the matter that passes through the boundary. 

If there is no matter passing through the boundary, and if the boundary three-metric $\g_{ij}$ possesses a Killing vector field $K_i$,\footnote{In the context of spaces with a cosmological constant, this condition can be weakened for $K_i$ to be only a {\it conformal} Killing vector, i.e.~one that satisfies Eq.~\eq{confg}.\label{cKv}} then $\t^{ij}$ defines a conserved charge, i.e.~we have the conservation law $\nabla_i(\t^{ij}_{\tn{BY}}K_j)=0$, which is analogous to Eq.~\eq{naP}. Thus by the same manipulations as in Section \ref{Mxth} we derive an analogue of Eqs.~\eq{Killingc}-\eq{E-P} on the boundary.

Note that, in the Brown-York proposal, the spacetime region $M$ can be finite and embedded in an ambient spacetime: and so, the three-surface $N$ that spatially bounds $M$ need not be at infinity. In the next Section, I will discuss the case $\L<0$, and $N$ will be taken to be the asymptotic boundary.\footnote{For studies of the Brown-York stress-energy tensor at surfaces that are not asymptotic boundaries, see e.g.~De Haro et al.~(2001a:~pp.~3174). For gauge-gravity dualities (both AdS-CFT and dS-CFT) with finite boundaries, see Harlow and Stanford (2011:~pp.~4-5) and Heemskerk and Polchinski (2011:~pp.~2-6).}

As any quasi-local energy-momentum must, the Brown-York expression agrees with the ADM energy at infinity  (see Section \ref{LEMGR}) for the special case of asymptotically flat spacetimes (Brown and York, 1993:~p.~1408).

\subsection{The holographic stress-energy tensor}\label{hst}

Consider the case of pure gravity with a negative cosmological constant, $\L<0$, , i.e.~no matter. (The Brown-York stress-energy tensor can be equally well be defined in the case $\L>0$).\footnote{See Balasubramanian et al.~(2001:~p.~3) and Anninos et al.~(2011:~pp.~4-5). Quite independent of quantum gravity or holographic ideas, the Brown-York energy-momentum tensor has been used to define notions of mass and momentum in spaces with a positive cosmological constant: see Balasubramanian et al.~(2001:~p.~4). The asymptotic diffeomorphism group at future null infinity of asymptotically de Sitter spaces, and the conservation equations for the associated charges, are considered in Anninos et al.~(2011:~pp.~3-6) and Strominger (2001:~pp.~4-5).\label{L>0}} According to AdS-CFT, the gravitational theory in the spacetime region $M$ with boundary $\pa M$ is dual to a quantum field theory on the boundary, $\pa M$.\footnote{For a general introduction to gauge-gravity dualities, see Ammon and Erdmenger (2015). For a conceptual introduction, see De Haro et al.~(2016). For a sample of the recent philosophical literature on dualities, see for example: Rickles (2011), Dieks et al.~(2015), Read (2016), Huggett and W\"uthrich (2020), De Haro (2020), Read and M\o ller-Nielsen (2020), Butterfield (2020).} As we will see, one can use gauge-gravity duality to reinterpret the Brown-York stress-energy tensor, evaluated on this asymptotic boundary, in terms of the dual CFT. 

Since we are interested in the Brown-York stress-energy tensor that is claimed to represent the energy and stress of the spacetime $M$, it will be useful to briefly mention how it fits within gauge-gravity duality.\footnote{In what follows, I will restrict to the case of a spacetime with a Euclidean signature, so that $M$ has a single boundary, $N$, with the topology of a three-sphere. For the case with Lorentzian signature, see Skenderis and van Rees (2008:~p.~2; 2009:~p.~6).}

Following the recent literature,\footnote{I here follow De Haro and Butterfield (2018:~pp.~335-339), De Haro (2020:~pp.~264-266), and Butterfield (2020:~p.~12).} a duality is, roughly, an isomorphism between two theory formulations. A theory can be construed as a triple of states, quantities, and dynamics. The statement of duality says that the sets of states, quantities, and the dynamics of the two theories are isomorphic. 

On the gravity side of the duality, a distinguished quantity should be the stress-energy tensor of the spacetime region $M$. The isomorphism should relate this to an appropriate quantity in the dual CFT. The question is: which stress-energy tensor has a dual in the CFT? If the Brown-York stress-energy tensor is to be a candidate for the energy and stress in the spacetime, then it ought to be related by the duality to a quantity in the CFT that has those properties. In fact, gauge-gravity duality gives us the following correspondence:

{\it The Brown-York stress-energy tensor, defined by the variation of the action (evaluated on the equations of motion) with respect to a boundary three-metric on $N$, and with appropriate boundary terms that render the action finite, is dual to the expectation value of the stress-energy tensor of the CFT defined on the background of that metric.}\footnote{I here omit technical details, such as that the boundary metric is defined only up to a conformal transformation. This point will resurface in Section \ref{diffeos}.}

This correspondence follows from the identification of the gravitational action Eq.~\eq{action} with a given set of matter fields, evaluated on a given solution of the Einstein field equations, with the generating functional of (connected) correlation functions in the dual CFT, usually called $W_{\tn{CFT}}$:
\bea\label{duality}
S[\g,\cdots]|_{\tn{EFE}}\equiv W_{\tn{CFT}}[\g,\cdots]~.
\eea
The left-hand side of this equation is the action Eq.~\eq{action}, evaluated for a metric and set of matter fields that satisfy Einstein's field equations (hence the subscript `EFE'). (Physicists call this an `on-shell' action). 
$\g$ is the metric on $\pa M$, and the dots indicate that there is also a dependence on the boundary values of the matter fields. The right-hand side of this equation is the generating functional of connected correlation functions, which is related to the CFT's partition function by: $Z_{\tn{CFT}}=\exp(-W_{\tn{CFT}})$. Again, this generating functional depends on the metric $\g$, which is now viewed as a curved background on which the CFT is defined, and other sources that the theory is defined with (these are dual to the boundary values of the matter fields on the left). 

On the right-hand side, the expectation value of the stress-energy tensor of the CFT is defined by the functional derivative with respect to the metric:
\bea\label{Tij}
\bra T_{ij}(x)\ket_{\tn{CFT}}={2\over\sqrt{\g}}{\d W_{\tn{CFT}}\over\d\g^{ij}(x)}~.
\eea
But substituting Eq.~\eq{duality} into Eq.~\eq{Tij}, and comparing with Eq.~\eq{vary}, we see that we indeed have the following correspondence:
\bea\label{BYQFT}
\t_{{\tn{BY}}}'{}^{ij}\equiv \bra T^{ij}\ket_{\tn{CFT}}~.
\eea
The dash on the Brown-York stress-energy tensor indicates that this tensor is modified, by replacing the `subtracted terms' in Eq.~\eq{BY} by local counterterms added to the action, according to a standard method in holographic renormalisation.\footnote{See Henningson and Skenderis (1998:~p.~5), Balasubramanian and Kraus (1999:~p.~415), De Haro et al.~(2001:~p.~603). For a review of holographic renormalisation, see Skenderis (2002).} In what follows, I will call $\bra T^{ij}\ket_{\tn{CFT}}$ `holographic stress-energy tensor'. 

The holographic stress-energy tensor can be calculated from Eq.~\eq{BY} for any given solution of Einstein's equations.\footnote{In general this is tedious, but there is a systematic procedure to do this calculation. 
See De Haro et al.~(2001:~pp.~602-607, 609-612), Skenderis (2002:~pp.~5855-5861). For an alternative method, see Papadimitriou and Skenderis (2004:~pp.~14-18).} 

The gravitational equation Eq.~\eq{bE} can be reproduced from the CFT perspective. For example, consider a two-dimensional CFT with a single (scalar) operator ${\cal O}$ which has a non-zero expectation value $\bra{\cal O}\ket$, coupling to a source $J$, the invariance of the CFT partition function under infinitesimal changes of coordinates,
\bea\label{naxi}
\d\g_{ij}=\nabla_i\xi_j+\na_j\xi_i~,
\eea
yields the following identity:
\bea\label{naT}
\nabla^j\bra T_{ij}\ket=\bra{\cal O}\ket\,\pa_iJ~.
\eea
One can check that this result is reproduced, on the gravity side with a three-dimensional spacetime, by the identity of Brown and York, Eq.~\eq{bE}, since, given the form of the normal energy-momentum tensor $T^{nj}$ for the dual scalar field, the above indeed reproduces Eq.~\eq{bE}.\footnote{See De Haro et al.~(2001:~p.~615).}

Noether's theorem in connection with asymptotic symmetries is discussed in Papadimitriou and Skenderis (2005:~pp.~15-19), who show, for various types of fields, that the holographic charges are the Noether charges.\\
\\
{\bf Implications of gauge-gravity duality for the gravitational energy.} Let us briefly discuss how these results bear on the problem of energy conservation. Gauge-gravity dualities 
give independent support to the claim that the modified Brown-York stress-energy tensor represents the quasi-local energy. We will consider the two main features of the holographic stress-energy tensor of the dual quantum field theory that bear on the status of the quasi-local energy in GR: 

(1) The duality map, Eq.~\eq{BYQFT}, relates the Brown-York stress-energy tensor to the holographic stress-energy tensor. This gives further support to the interpretation of the Brown-York stress-energy tensor as representing quasi-local stress and energy. For recall that a duality is an isomorphism of models of a single common core theory.\footnote{This `common core' view is discussed and-or endorsed by e.g.~Rickles (2011:~p.~66), Dieks et al.~(2015:~p.~209), De Haro (2017:~p.~116), Huggett (2017:~p.~84), De Haro and Butterfield (2018:~p.~321), Le Bihan and Read (2018:~p.~3).} The interpretation of two dual models need not be the same, but if it {\it is} the same,\footnote{That it {\it is} the same depends on our developing an {\it internal interpretation} of the dual models: for a discussion, and conditions under which this is possible, see Dieks et al.~(2015:~p.~209), De Haro (2020:~pp.~267-271), Read and M\o ller-Nielsen (2020:~p.~284-288).} then the two dual quantities on either side of Eq.~\eq{BYQFT} receive the same interpretation. Thus the right-hand side of this equation, i.e.~the holographic stress-energy tensor, can give us guidance about the interpretation of the side under discussion, i.e.~the gravitational, left-hand, side. And the interpretation of the left-hand side is indeed as energy, momentum, and stress. This is a further argument to the effect that the Brown-York stress-energy tensor represents energy, momentum, and stress.\footnote{Note that the internal interpretation of the common core theory, as currently worked out, does not support the interpretation of this energy as {\it gravitational} energy---even though this {\it can} be done on an external interpretation (see footnote \ref{borrowC}). For understanding the nature of the common degrees of freedom depends on having a good development of this common core theory, which we do not currently have. For a more detailed discussion, see De Haro (2020:~pp.~275-279).}

Thus, in AdS-CFT, the modified Brown-York stress-energy tensor in spacetime with $\L<0$ is {\it precisely} dual to the quantity that, in the CFT, is interpreted as describing energy and momentum: namely, the expectation value of the stress-energy tensor. The match is a consequence of the general AdS-CFT correspondence: namely, the identification of the gravitational action, evaluated on solutions, with the generating functional of the CFT, Eq.~\eq{duality}, and of the conformal metric on the boundary, $\g_{ij}$, with the metric of the dual quantum field theory. This precise correspondence, which has been illustrated in many examples, gives additional support to the statement that the Brown-York stress-energy tensor {\it is} indeed to be thought of as energy and momentum, since it plays that role in the boundary theory.

(2) The boundary conservation equation, Eq.~\eq{bE}, has a similar dual interpretation, and is not analogus to the usual (disputed) conservation laws of GR.\footnote{There are three important disanalogies between the boundary balance equation, Eq.~\eq{bE}, and the (disputed) conservation law, Eqs.~\eq{GRcon}-\eq{pseudot}, which are relevant to points (i)-(iv) from Section \ref{Einstein}. First, Eq.~\eq{bE} distinguishes between the quasi-local gravitational energy, $\t_{\tn{BY}}$, and matter energy-momentum, $T^{nj}$. Second, the covariant derivative is calculated using the boundary, metric-compatible, connection, which is non-dynamical, and fixed by the boundary conditions. Third, the right-hand side is in general non-zero, so that this is a balance equation. I will address points (iii) and (v), from Section \ref{Einstein}, in Section \ref{suphol}} This boundary conservation equation is dual to Eq.~\eq{naT}: and so, it can be interpreted as a balance equation for the energy and stress of the common core theory, and this again strengthens that interpretation in GR.\footnote{An internal interpretation (as in `Eq.~\eq{bE} is the energy balance equation') is usually {\it compatible} with a judiciously chosen external interpretation (as in `Eq.~\eq{bE} is the balance equation for the {\it gravitational} energy'), through the process of abstraction outlined in De Haro (2021:~Section 4.4). This is why, if the internal interpretation is correct, it bears on the interpretation on GR.\label{borrowC}}

\subsection{Diffeomorphisms and redundancy of the description}\label{diffeos}

In our discussion, in Section \ref{Revisited}, of the two main proposals for defining energy and momentum in general relativity, diffeomorphism invariance will play a crucial role. Therefore, we should understand the transformation properties of the Brown-York stress-energy under various diffeomorphisms: namely, those that are descriptive redundancies and those that are not.\footnote{My discussion here essentially summarises De Haro, Teh, and Butterfield (2015, 2017) and  De Haro (2017). For physics references on this topic, see Imbimbo et al.~(2000:~pp.~1130-1132), Skenderis (2001:~pp.~742-747).}

The asymptotic symmetries of gravity have been a central foundational topic in general relativity since at least the work of Arnowitt et al.~(1959, 1962), Bondi et al.~(1962), Penrose (1963, 1964), Newman et al.~(1966), Geroch (1972), Ashtekar et al.~(1978), among others. For a philosophical discussion of some aspects of asymptotic symmetries, see Belot (2018).

A central question is which asymptotic diffeomorphisms are true symmetries of a spacetime, and which asymptotic diffeomorphisms act non-trivially on the physical degrees of freedom of general relativity: and how these different kinds of diffeomorphisms are to be characterised. 

In spaces with a negative cosmological constant, Brown and Henneaux (1986) discovered that the asymptotic diffeomorphism of anti-de Sitter space is the conformal group of the asymptotic boundary.\footnote{The analysis by Brown and Henneaux has been extended to the case $\L>0$, including in higher dimensions: see Strominger (2001:~pp.~4-5) and Balasubramanian et al.~(2001:~pp.~7-8). For a more detailed analysis of diffeomorphisms in the case $\L>0$, see Anninos et al.~(2011:~pp.~3-4). As we will see in this Section, these results can be generalised to any space with $\L<0$ that is a solution of Einstein's equations, with arbitrary metric at infinity.\label{deSrefs}} As we will see in this Section, this asymptotic symmetry group acts non-trivially on the quasi-local energy-momentum tensor.

Thus the question of which subgroup of the diffeomorphism group is a symmetry of a spacetime is important from the point of view of quasi-local energy and momentum: for the quasi-local quantities will be either invariant or else have a well-defined transformation law under the diffeomorphisms that are true asymptotic symmetries of a spacetime, and they transform non-trivially under diffeomorphisms that are not true symmetries.

What do we mean by the `true asymptotic symmetries' of a spacetime? What we are looking for here is a criterion that distinguishes the symmetries (diffeomorphisms) that are true redundancies of a description, from those that entail a physical change in the solution. We can define redundancies as follows (De Haro et al., 2017:~p.~69):\footnote{There is an ongoing debate, following Greaves and Wallace (2015)---which itself argues against earlier work: see footnote \ref{des}---whether local gauge symmetries express mere redundancies of the description, or whether they can be empirically significant, as in Galileo's ship scenarios: see, for example, Friederich (2015:~p.~537), Murgueitio (2020:~p.~1), Murgueitio and Teh (2020:~p.~3), Gomes (2020:~p.~11), Gomes and Riello (2020:~pp.~3-5, 21). The approach I adopt here is, in effect, to construct a class of diffeomorphisms (the `visible' diffeomorphisms below) that are GR-analogues of the class of gauge symmetries that can have a direct empirical equivalence in gauge theories. The latter are called `non-interior, boundary-preserving symmetries' by Greaves and Wallace (2014:~p.~71), and their definition is further refined by Teh (2016:~p.~116).}

\begin{quote}\small
(Redundant): If a physical theory's formulation is redundant (i.e.~roughly: it uses more variables than the number of degrees of freedom of the system being described), one can often think of this in terms of an equivalence relation, `physical equivalence', on its states; so that gauge-invariant quantities are constant on an equivalence class, and gauge-symmetries are maps leaving each class (called a `gauge-orbit') invariant. Leibniz's criticism of Newtonian mechanics provides a putative example: he believed that shifting the entire material contents of the universe by one meter must be regarded as changing only its description, and not is physical state.
\end{quote}

I will introduce the notion of the `invisibility' of diffeomorphisms to make precise the idea of the asymptotic symmetries that are (Redundant). I will distinguish three relevant classes of diffeomorphisms that qualify differently under redundancy: `invisible' diffeomorphisms are always (Redundant), `large' diffeomorphisms are never (Redundant), and the intermediate category of `visible' diffeomorphisms can sometimes be (Redundant) and sometimes not, depending on the interpretation that the theory is given.\footnote{Belot (2018:~pp.~964-967) also distinguishes local translations that are redundant, from those that are physical: and he also connects this to the question of boundary conditions---although he does not appear to consider my intermediate category of `visible diffeomorphisms'.} The three classes are distinguished by the different boundary geometric structures that they preserve.\footnote{Class (1) is the class of `QFT-invisible' diffeomorphisms in De Haro (2017a:~pp.~1468, 1484), also called `(simple) gravity-invisible'. Class (2) is the class of `QFT-visible' diffeomorphisms, also called `weakly gravity-invisible' diffeomorphisms'. For the reasons why these classes are cases of (Redundant) or non-(Redundant), see De Haro, Teh, and Butterfield (2015, 2017) and De Haro (2017).} They are as follows:

(1)~~{\it Invisible diffeomorphisms}: these are diffeomorphisms under which the bulk action, Eq.~\eq{duality}, evaluated on a given solution, is invariant. Also the other quantities derived from the action, like the holographic stress-energy tensor Eq.~\eq{BYQFT}, are invariant. These diffeomorphisms fix the boundary manifold and the conformal metric.\footnote{It is sometimes said that the diffeomorphisms that are redundancies of the theory are those that ``go to unity'' at the boundary (this being the class by which the allowed diffeomorphisms have to be quotiented in order to obtain the asymptotic symmetry group). But this is imprecise, because a diffeormorphism can go to unity at the boundary while still inducing a non-trivial transformation of the boundary metric $\g_{ij}$, through its dependence on the parametrisation of the `radial direction'' $r$, which results in a rescaling of the boundary metric, and hence a diffeomorphism of type (2) below. Thus, going to unity at the boundary is not sufficient for (Redundant). Instead, the definition of invisible diffeomorphisms makes precise the sense in which diffeomorphisms are (Redundant).}

These diffeomorphisms count as (Redundant) symmetries, because all of the physical quantities, for a given solution of the equations of motion, are invariant under them. In particular, these diffeomorphisms allow us to set up an analogue of the hole argument in a space with a negative cosmological constant, which in turn leads to the conclusion that the diffeomorphisms in question are (Redundant).\footnote{See De Haro et al.~(2017:~p.~79).}

(2)~~{\it Visible diffeomorphisms}: these are diffeomorphisms that generate conformal transformations of the boundary metric, i.e.~$\g_{ij}\mapsto\ti\g_{ij}:=e^{-2\xi(x)}\,\g_{ij}$. The stress-energy tensor transforms by a conformal factor under them, and in the case of even boundary dimensions the stress-energy tensor (and the action) pick up an additional, anomalous term under them:
\bea\label{anomaly}
\bra T_{ij}[\ti \g](x)\ket=e^{(d-2)\,\xi(x)}\left(\bra T_{ij}[\g](x)\ket+{1\over\sqrt{\g}}{\d{\cal A}[\g]\over\d \g^{ij}(x)}\right),
\eea
where $\g_{ij}$ is the metric induced on the boundary, $d$ is the dimension of the boundary, and ${\cal A}[\g]$ is the anomalous term, which can be calculated from the action.\footnote{To simplify the discussion, I do not distinguish between conformal transformations and Weyl rescalings of the metric. For details, see De Haro (2017a:~p.~1479).} It is a local term written in terms of curvature invariants (for the explicit expressions, see Deser and Schwimmer (1993), Henningson and Skenderis (1998:~Section 3), De Haro et al. (2001:~Section 4.2.1)). The anomaly vanishes when $d=3$, i.e.~for a four-dimensional spacetime.\footnote{The anomaly can be understood as coming from the divergence of the gravitational action, Eq.~\eq{action}, at large distances. To cancel this divergence, the action is regularised and boundary counterterms are introduced, but this renormalisation procedure breaks covariance, albeit in a controlled way, i.e.~through the additional term Eq.~\eq{anomaly}.} Since the anomaly term vanishes in the case of a four-dimensional spacetime, which is the case of most interest, I will not discuss it further.\footnote{For spacetimes of odd dimension, the anomaly is proportional to powers of the {\it boundary} curvature that depend on the dimension, and so it is also zero if the boundary metric is flat.}

Visible diffeomorphisms are potentially physical symmetries, rather than redundancies of the description: namely, they act as conformal transformations of the boundary quantities. Whether they are physical depends on one's interpretation of the spacetime within one's theory.\footnote{In De Haro et al.~(2017:~p.~79), it was argued that, on an `internal' interpretation (typically, in the context of a theory of the whole universe), two states related by visible diffeomorphisms are physically equivalent, so that the transformations count as (Redundant) gauge symmetries. On the other hand, on an `external interpretation' (as e.g.~in a Galileo's ship scenario, where the inside of the ship is not the whole world, but can be related to other parts of the world, e.g.~the shore) these diffeomorphisms are not (Redundant), but express `the motion of the ship relative to the quay'.} 


(3)~~{\it Large diffeomorphisms}: these are diffeomorphisms that do not preserve sufficient boundary structure, and so they map solutions of Einstein's equations to {\it physically inequivalent} solutions of the equations (e.g.~equations with a different total mass or angular momentum). Such diffeomorphisms are therefore not (Redundant), but physical transformations that relate different states. In particular, they will change the value of the holographic stress-energy tensor.\footnote{For example, Grumiller et al.~(2016:~Eq.~(3.6)) report a metric that is claimed to be physically inequivalent to the corresponding metric in normal form.}

An analysis of the diffeomorphisms as detailed as the preceding one has, so far as I am aware, not been done for $\L>0$, and so one should proceed with some caution. However, much work has been done on this case as well: see Anninos et al.~(2011:~pp.~3-6), Strominger (2001:~pp.~4-5), Balasubramanian et al.~(2001:~pp.~7-8).

There are also a few general things one can say. As noted in Skenderis (2002:~Appendix A) and De Haro (2017a:~pp.~1490-1491), the geometric techniques that enabled the analysis summarised in Section \ref{diffeos} and just above, apply in the case $\L>0$ as well, due to the fact that the differential form of Einstein's equations in the radial variable was turned into a set of coupled algebraic relations. The results depend only on the asymptotic solutions, and can be continued from one case to the other, where the radial coordinate $r$ goes into a timelike coordinate $t$ (the anti-de Sitter boundary is thereby mapped to the timelike future infinity in de Sitter space).  

\ \\To summarise, in case (1) we have a hole-type argument that leads us to conclude that invisible diffeomorphisms correspond to redundancies of the theory. In case (3), such an argument is not available, because the diffeomorphisms in question change the physical properties of the spacetime, and we should not characterise spacetimes related by a large diffeomorphism as physically equivalent. Case (2) is an intermediate case, where the diffeomorphisms act as conformal transformations on the boundary quantities, and their nature as (Redundant) or as non-(Redundant) depends on one's interpretation (including further use) of the solutions, say as being cosmological models of the {\it whole} universe, or as being merely parts of the universe.

\ \\Let me say a bit more about this last case, (2), of visible diffeomorphisms, and how they give rise to transformations of the conformal boundary: namely, conformal transformations. Choose an open neighbourhood of the boundary, such that the boundary of $M$ is the surface $r=0$. A theorem of Fefferman and Graham (1985, 2012) guarantees that in such an open neighbourhood of the boundary, we can write the metric in {\it Poincar\'e normal form}:
\bea\label{Poin}
\dd s^2={\ell^2\over r^2}\left(\dd r^2+\g_{ij}(r,x)\,\dd x^i\,\dd x^j\right)~,
\eea
where $x^i$ are coordinates tangent to the boundary directions, and $\g_{ij}(r,x)$ is a metric along this direction, that in general depends on both $r$ and $x$. We obtain the earlier-used $\g_{ij}$ by setting $r=0$, that is: $\g_{ij}(x):=\g_{ij}(0,x)~.$ The diffeomorphisms we are interested in are those that ``fix the location of the boundary $r=0$''. They take the following form:
\bea
\d x^i=\xi^i(x)~,~~\d r=-r\,\xi(x)~.
\eea
Requiring that these diffeomorphisms satisfy the definition of `being visible', i.e.~fix: (i) the Poincar\'e normal form of the metric, i.e.~Eq.~\eq{Poin}, and (ii) the boundary metric, implies the following condition on the diffeomorphism generated by the boundary vector field $\xi$:\footnote{See Brown and Henneaux (1986), De Haro (2017a:~p.~1477).}
\bea\label{confg}
\nabla_i\,\xi_j(x)+\nabla_j\,\xi_i(x)&=&{2\over d}\,\g_{ij}(x)\,\na^k\,\xi_k(x)~,
\eea
where the covariant derivatives are with respect to the boundary metric $\g_{ij}(x)$. This equation is the {\it conformal Killing equation}, which is precisely the condition for the vector field $\xi$ to be a conformal transformation, rather than an isometry, of the boundary metric $\g_{ij}(x)$.\footnote{The conformal Killing equation reproduces the ordinary Killing equation, Eq.~\eq{Killxi}, in the special case $\na^k\xi_k=0$.} This gives us a very general relation between the symmetries of the metric in $M$, in Poincar\'e normal form Eq.~\eq{Poin}, and the conformal group on $\pa M$:

{\it The asymptotic symmetry group of the Poincar\'e normal form of the metric, Eq.~\eq{Poin}, on the manifold $M$, is the conformal group on the asymptotic boundary, $\pa M$, with boundary metric $\g$.}

This generalises the result by Brown and Henneaux (1986), for any solution of Einstein's equations with $\L<0$. 


These diffeomorphisms give us a conformal Killing vector $\xi$ of the boundary manifold. As I noticed after Eq.~\eq{bE}, especially in footnote \ref{cKv}, using the properties of the holographic stress-energy tensor, Eq.~\eq{Tij}, this can be used to derive a boundary analogue of the energy conservation equation Eq.~\eq{E-P}. 

\section{Gravitational Energy and Momentum, Revisited}\label{Revisited}

In this Section, I first review two recent discussions of pseudo-tensors that I aim to justify: one is from physics (Section \ref{totd}) and the other is philosophical (Section \ref{phildis}). I will agree with much of these defences of the pseudo-tensor approach---but I will nevertheless, in Section \ref{scep}, argue for the superiority, in the present state of knowledge, of the Brown-York approach.

\subsection{Superpotentials are boundary terms in the Hamiltonian of GR}\label{totd}

In recent work, Chang, Chen, Nester, and collaborators (1999, 2017, 2018, 2018a) have defended a rehabilitation of pseudo-tensors that thereby casts light on the physical nature of pseudo-tensors. Their work can be seen as arguing for a resolution of problem (iv) in Section \ref{Einstein}. Namely, they argue that pseudo-tensors:

(a)~~{\it Do provide a description of energy and momentum conservation.} They do this, in particular, by establishing a relationship between pseudo-tensors and the Hamiltonian of general relativity. The superpotential, $U^{\m\n}_\l$, can be identified with a total divergence in the Hamiltonian of general relativity, and so different choices of the superpotential correspond to different boundary terms/conditions. Thus problem number (iv) is thereby resolved, because fixing the boundary terms determines the pseudotensor uniquely.

(b)~~\,{\it Give, for apropriate choices of boundary conditions, the expected values of the total energy and momentum as measured at infinity,} i.e.~the ADM and Bondi formulas,\footnote{See for example Jaramillo and Gourgoulhon (2011:~pp.~97-102, 105-107) and Christodoulou and Yau (1988:~pp.~9-10).} thus further strengthening the contention, in the resolution of problem (iv), that pseudo-tensors are physical.

(c)~~{\it Have well-defined values in each system of coordinates.} In particular, the pseudo-tensors give quasi-local quantities, which are naturally identified with the two-surfaces that bound regions of space, and give correct results at infinity, as in (b). 

Here I will focus on (a), which is the most distinctive part of the approach by Chang, Chen, Nester, and collaborators. Then I will more briefly comment on (b) and (c). Let me state, in words, the main statement that goes into arguing for (a):

\ \\{\bf Pseudotensor-Hamiltonian correspondence:} {\it The quasi-local energy-momentum over a hypersurface $\Si$ with constant vector $N$, associated with the pseudo-tensor $t^\m{}_\n$ and superpotential $U^{\n\l}_\m$, corresponds to a Hamiltonian $H(\Si,N)$ whose boundary term is given by the expression $B(N)$ (and vice versa). The choice of the superpotential $U^{\n\l}_\m$ in the pseudo-tensor determines a boundary term in the Hamiltonian (and vice versa).}

Working this out in the forward direction, i.e.~showing that the quasi-local energy-momentum gives rise to the Hamiltonian in the way stated, goes in three main steps:

(1) Follow Einstein's derivation of the pseudo-tensor $t^\m{}_\n$, in Section \ref{Einstein}, and express it in terms of the superpotential $U^{\m\l}_\n$, in Eq.~\eq{tGU}. 

(2) ``Integrate Einstein's derivation'' to get a pseudo-tensor, analogously to the steps in Eqs.~\eq{GRcon}-\eq{divth}. Here, Chen, Nester et al.~construct an energy-momentum quantity $P(N)$ by integrating the total local energy-momentum density (including its matter and gravitational contributions) over a three-dimensional spacelike hypersurface $\Si$. Upon the use of the Einstein field equations, the result is a boundary term, i.e.~an integral over $\pa\Si$. Thus in this step they associate to each pseudo-tensor, and solution of Einstein's field equations, a quasi-local quantity---and this is their preferred formulation of the problem. 

The details are as follows: they define energy and momentum by integrating Eq.~\eq{pseudot} with an arbitrary vector $N$:
\bea\label{PN}
P(\Si,N)&:=&\int_\Si\dd\Si_\n\,(T^\n{}_\m+t^\n{}_\m)N^\m \nn
&=&{1\over\k}\int_\Si\dd\Si_\n\,[(-G^\n{}_\m+\k\,T^\n{}_\m)N^\m+\half\pa_\l(N^\m\,U^{\n\l}_\m)]\nn
&=&\int_\Si {\cal H}_\m\,N^\m+\oint_{\pa\Si}B(N)~,
\eea
where I have introduced the notation ${\cal H}_\m=\dd\Si_\n({1\over\k}\,G_\m{}^\n-T_\m{}^\n)$ and  $B(N)=-{1\over4\k}\,\dd S_{\n\l}\,N^\m\,U^{\n\l}_\m$ for the first and second terms, respectively, in the last line. Going from the first to the second line uses Eq.~\eq{tGU}, and going from the second to the third line uses Gauss' law to rewrite the total derivative, integrated over the hypersurface $\Si$, as an integral over the boundary $\pa\Si$. 

In this derivation, $N$ is taken to be an arbitrary vector that is {\it constant} with respect to the given system of coordinates. Such a vector always exists, although it will be different for each system of coordinates. This will be an important point of discussion in Section \ref{scep}, in connection with point (c) above.\footnote{Nester (2020, personal communication) argues that, in Eq.~\eq{PN}, `the volume integrand is proportional to the field equations, and so vanishes on-shell. The value of the integral comes only from the 2-surface boundary term. This means that we can extend our formula, modifying the vector field $N$ within the region, without changing the value of the integral, as long as one keeps $N$ constant on the boundary. Thus it is only necessary for the vector field to match a translational Killing field of the Minkowski coordinate reference geometry on the boundary'. Although this is of course true, the independence of the value of the vector field $N$ in the region inside the volume $V$ is only valid in the last two lines of Eq.~\eq{PN}, i.e.~for the Hamiltonian expression. However, the step from the first to the second line of Eq.~\eq{PN} does {\it not} hold if $N$ is not constant, because the second term is no longer a total derivative. In other words, the above direct relation between the integrated total energy-momentum pseudo-tensor (i.e.~energy-momentum plus gravitational pseudo-tensor) and the Hamiltonian expression only holds if $N$ is everywhere constant in a system of coordinates adapted to the boundary. (Nester's main interest is of course the Hamiltonian formalism, and less so the pseudo-tensor. But, in so far as these papers aim to {\it justify} the use of pseudo-tensors through the Hamiltonian approach, my point is that this only works for {\it constant} vector fields $N$ and special systems of coordinates, as just argued).\label{NesterP}} For the moment, I will set this aside.

As announced above, and as we see from the middle expression in Eq.~\eq{PN}, the volume term in the Hamiltonian vanishes when Einstein's field equations are satisfied, and we are left only with the boundary term $B(N)$, integrated over a 2-surface. Thus we have succeeded in constructing, from the local energy and momentum, a quasi-local quantity.\footnote{This is true at least formally. See Section \ref{scep} for a discussion.}

(3) Recognise that the expression Eq.~\eq{PN}, which is already written in suggestive notation, takes exactly the form of the Hamiltonian of general relativity, associated with a finite region $\Si$ and constant vector $N$:
\bea
P(\Si,N)=-H(\Si,N)~.
\eea
The Hamiltonian is the generator of spacetime displacements of the hypersurface $\Si$ along a vector field $N^\m$, and ${\cal H}_\m$ indeed has the form in Eq.~\eq{PN}, and vanishes if the field equations are satisfied. 

$B(N)$ is the boundary term of the Hamiltonian, which gives the {\it quasi-local energy and momentum,} i.e.~it is an expression for the energy and momentum on the 2-surface $\pa\Si$. This boundary term, as it appears in the above calculation starting from the pseudo-tensor, contains the {\it superpotential.} Given this correspondence, between superpotentials and boundary terms, the Hamiltonian approach casts light on the physical significance of the various choices of superpotentials. In particular, the variation of the boundary term, $\d B(N)$, must be set to zero in order to get the Einstein field equations in their Hamiltonian form. Filling in various the choices of superpotential $U^{\n\l}_\m$ in this variation, one finds that each choice of superpotential leads to a specific class of boundary conditions. For example, the Freud expression, Eq.~\eq{Freud}, corresponds to keeping $\sqrt{g}\,g^{\m\n}$ fixed at the boundary,\footnote{The condition is $\d(\sqrt{g}\,g^{\m\n})=0$, which gives $\d g^{\m\n}=\half g^{\m\n}g_{\a\b}\,\d g^{\a\b}$. This {\it allows,} but does not require, the Dirichlet boundary condition $\d g^{\m\n}=0$, and thus it is slightly more general than a Dirichlet boundary condition.} while keeping the vector field $N$ constant. Thus the Freud superpotential allows Dirichet boundary conditions. 

Other choices of superpotential lead to other boundary conditions, e.g.~Neumann boundary conditions where the momentum conjugate to the metric is held fixed. For each choice of superpotential there is a corresponding boundary condition, and a quasi-local energy expression. 

To show that the pseudotensor-Hamiltonian correspondence also works in the opposite direction, one reverses the above steps: begin with the Hamiltonian, and show how its boundary terms correspond to various superpotentials in the energy-momentum pseudo-tensor.

The expression Eq.~\eq{PN} for the Hamiltonian, and in particular the various choices of its boundary term, can be motivated further by using the Hamiltonian formalism. To do this, Chen, Nester et al.~construct the Hamiltonian of general relativity in form notation. In this formalism, the various boundary terms $B(N)$, that naturally enforce Dirichlet or Neumann boundary conditions in the variation, take simple forms in terms of three geometrical quantities and their corresponding conjugate momenta: namely, the metric, and the orthonormal frame and coframe fields. These relatively simple forms of the boundary terms reproduce, when worked out in tensor components, the various expressions for the superpotential found in the literature. Thus the form notation for the Hamiltonian explains the relatively long tensor formulas for the superpotential through simple boundary terms. 

\subsection{Recent philosophical discussions of energy conservation in GR and pseudo-tensors}\label{phildis}

The philosophical discussion of energy conservation in general relativity revolves around two main problems, mainly to do with problems (v) (lack of covariance, and lack of appropriate transformation properties) and (iii) (non-integrability) in Section \ref{Einstein}. (Surprisingly, Lam (2011:~p.~1017) and Read (2020:~p.~215) mention the problem (iv) of non-uniqueness, which in my view is pressing, only in passing; other papers do not appear to mention it.)\footnote{Duerr (2019:~pp.~2, 6, 9-10) does stress the importance of this ambiguity, which he calls `baneful' (p.~9).} The first is about coordinate-dependence; the second is about background-dependence. I give them new labels, (A) and (B), to cover other aspects of recent discussions:\\

(A)~~{\it Lack of covariance, and lack of appropriate transformation properties.} The known expressions for the local energy and momentum in general relativity are not tensorial. In particular, the pseudo-tensors Eq.~\eq{tGU} are coordinate-dependent, and so they do not have good local (or global) definitions. (As I mentioned in Section \ref{Einstein}, pseudo-tensors are only invariant under affine transformations).

For example, Hoefer (2000:~pp.~194-195) has stressed that the expressions for energy and momentum, even in their integrated form Eq.~\eq{dpdt}, are coordinate-dependent, and so are ill-defined---as I already mentioned at the end of Section \ref{LEMGR}. 

One immediate comment to make about (A) is that it is diffeomorphism invariance, rather than covariance, that is most important here: 
and that, as we saw in Section \ref{diffeos}, the various kinds of diffeomorphisms require a different treatments. Second, although the lack of covariance can be technically and physically inconvenient, the main worry is the pseudo-tensor's lack of appropriate transformation properties.\\

(B)~~{\it ``Background-dependence''.} The conservation of energy and momentum in general relativity requires the introduction of ``background structures'': such as a flat metric (usually, at infinity) or timelike Killing vectors in the spacetime, see Eq.~\eq{Killxi} and the discussion following.\footnote{Absolute objects are defined e.g.~in Anderson (1967:~pp.~83-84; 1971:~p.~166). However, as we will discuss below, these background structures are not absolute objects in this sense.} Without such additional structures, the existing expressions for the energy-momentum tensors do not express genuine conservation laws, and do not give unique answers for the physical energy and momentum. While both Hoefer (2000) and Lam (2011) admit that expressions like Eq.~\eq{dpdt} do yield unambiguous results when such structures are added, they ultimately reject them, because they amount to introducing undesired ``background structures''.\footnote{Background-independence is obviously a large topic that I cannot enter in detail into here, nor will that be required. For, as I mentioned before, the ``background structures'' discussed in this context are not the same as the ones discussed in the literature on background-independence. Thus my comments will be restricted to the specific discussion of the role of background structures in Noether's second theorem. For more general treatments of background-independence: see, for example, Pooley (2017:~pp.~24-29) and Read (2016a). For my own views about background-independence, see De Haro (2017).\label{bi}}

For example, Hoefer (2000:~p.~194) criticises the calculations of the energy-momentum conservation equation Eq.~\eq{dpdt} that evaluate integrals at infinity to get the total energy and mass of the spacetime, because they assume that the spacetime is asymptotically flat and that the coordinate system at infinity is Minkowskian. This does not give a genuine conservation principle for general relativity, the most important reason being that fixing the asymptotics in this way `goes against the most important and philosophically progressive approach to spacetime physics: that of downplaying coordinate-dependent notions and effects, and stressing the intrinsic, covariant and coordinate-independent as what is important'. Thus points (A) and (B) are, for Hoefer, really part of a {\it single} worry: namely, coordinate-dependence.\footnote{Furthermore, Curiel (2019:~p.~91) proves a theorem that there exists no local, divergence-free, covariant two-tensor other than a multiple of the Einstein tensor, and thus no such local candidates for a gravitational energy-momentum tensor.}

Lam (2011:~p.~1010) is more tolerant about coordinate-dependence: he concedes that `there does not seem to be any contradiction' in relying on `background-dependent (coordinate-dependent) entities'. However, he worries that there is no unique definition of pseudo-tensors, because the possibility to add a superpotential to them means that `one cannot say how much energy there is in a given spatial region', i.e.~problem (iv), non-uniqueness.\footnote{Pitts (2010:~pp.~605-606) proposes to restore the covariance of the gravitational energy-momentum expressions by embracing their multiplicity: namely, by constructing a single geometric object whose infinitely many components are the components of a given pseudotensor with respect to all possible coordinate systems, so that they transform into each other under coordinate transformations. In the text quoted, Lam is commenting on Pitts' proposal, but similar remarks could apply to the pseudo-tensor itself.}

But his main worry is about the introduction of ``background structures''. For example, although the standard integral formulation of the conservation equation with a Killing vector, Eq.~\eq{Killingc},
is a genuine energy-momentum conservation law for the matter fields, it requires the introduction of the Killing vector $K$, which Lam considers is an undesired ``non-dynamical background''. And likewise for the introduction of asymptotically flat metrics at infinity in order to reproduce the ADM and Bondi formulas: these `depend crucially on the background structures inserted in the form of asymptotic symmetries' (p.~1022). 

\ \\Read (2020) has recently responded (at least partially) to the critiques (A) and (B), by arguing that {\it a genuine stress-energy tensor does exist in general relativity}, in both: (i) a weak sense applicable in a certain class of models of the theory (namely, those that admit appropriate Killing vectors), (ii) a strong sense, applicable in all models. 

The above distinction between cases (i) and (ii) is underpinned by the existence of a timelike Killing vector in the spacetime. Namely, in response to (B), Read (2020:~p.~222) argues that the Killing vector $K$ that gives the well-defined energy conservation law Eq.~\eq{E-P}, should not be seen as ``background structure''. For the existence of a Killing vector is merely a property of the metric: if the metric has an isometry, then a Killing vector exists. This simply means that the models with a Killing vector are a subset of all the dynamically possible models of the theory: no additional structures are introduced into the theory.

Read's advocacy of energy-momentum conservation appeals to functionalism (i.e. briefly, the idea that there exists in GR a quantity that fulfils the functional role of gravitational stress-energy: cf.~p.~224) and to the possibility of being a realist about pseudo-tensors. The possibility of realism is motivated by {Einstein's (1918b:~pp.~449-451) own position that quantities can be coordinate-dependent, but still well-defined.} Just as the connection coefficients are frame-dependent and are intimately tied to the gravitational field\footnote{See Einstein (1916:~p.~802), and also the discussion in Norton (1985:~p.~233).} (recall that the Newtonian gravitational potential arises from the Christoffel symbols), it is {\it prima facie} plausible to interpret the gravitational pseudo-tensor as representing gravitational energy, momentum, and stress in a given frame of reference. Thus Read proposes that gravitational stress-energy is always defined with respect to a given frame of reference, analogously to how the relative velocity of a body is evaluated in the rest frame of another body (but see the objections in the next Section).

He argues that in case (i), i.e.~in those models admitting timelike Killing vectors, in which the conservation law Eq.~\eq{Killingc} exists, the quantity $t^\m{}_\n$ does fulfil the role of a gravitational energy-momentum tensor, i.e.~it fulfils a role analogous to that of gravitational energy as traditionally conceived, and bears an appropriate relation to the gravitational degrees of freedom. He proposes, based on (a) the pseudo-tensor's fulfilling this functional role, on (b) the larger explanatory apparatus that the advocates of gravitational energy have available to them, and on (c) the simple and practical character of functionalism involved: that, for this subset of models of the theory, the gravitational stress-energy does exist, and should be labelled as genuine energy, momentum, and stress. Thus a frame-independent gravitational stress-energy tensor does, in this sense, exist in general relativity.

In case (ii), although a frame-independent stress-energy tensor is not available, one can still be a realist about the pseudo-tensor. Namely, realism about the pseudo-tensor is compatible with functionalism, since the pseudo-tensor may be defined, in each frame, such that it is conserved. In particular, he argues from the broader idea that it is acceptable to give explanations of phenomena by appealing to coordinate-dependent effects. If one accepts this realist line, then the gravitational stress-enegy exists in the stronger sense, which is applicable to all the models of the theory.

While I will {\it in part} agree with Read's two arguments in response to (A) and (B), I will discuss, in the next Section, whether these arguments conclusively establish the existence of local energy and momentum in general relativity---since indeed they do not offer a satisfactory response to (A). 

\section{A Comparison of Two Proposals}\label{scep}

In this Section, I will compare the two proposals for energy and momentum from the previous two Sections. In Section \ref{whatelse}, I discuss three worries about the gravitational pseudo-tensor that, in my opinion, should be addressed before pseudo-tensors are taken to be part of the ontology of general relativity. In Section \ref{suphol}, I will defend the holographic stress-energy tensor from two general worries, articulated in the philosophical literature on energy and momentum in general relativity, and already mentioned in the previous Section. In Section \ref{worries3}, I will discuss further questions that one might have about the physical interpretation of quasi-local quantities.

\subsection{What else is required for us to accept pseudo-tensors?}\label{whatelse}

In this Section I make three qualifications about Read's (2020) rebuttal of (A) and (B).\footnote{A fourth point is that pseudo-tensors are non-unique because one can add an arbitrary superpotential to them, and so it is not a priori clear how to interpret them. However, as I discussed in Section \ref{LEMGR}, this is clarified by Noether's second theorem, which establishes a clear relation between the superpotential and the boundary term in the first variation of the action, see Eq.~\eq{defP}.} My second qualification is in the neighbourhood of a more technical objection by Duerr (2019a:~Section 3.3) (I will return to the resolution of a number of Duerr's other objections at the end of Section \ref{suphol}). My third qualification leads in to Section \ref{suphol}.

The first qualification is that, while Read (2020) frames his defence of pseudo-tensors in terms of scientific realism, this is not necessary. For also the empiricist, e.g.~van Fraassen (1980), could argue along Read's functionalist lines, and conclude that pseudo-tensors are {\it physical quantities,} which describe properties of some entities in the world---in this case, an unobservable entity, viz.~gravitational energy and momentum. Agreed: this depends upon the empiricism under consideration. But the point is that 
the significance of pseudo-tensors goes beyond the question of whether there is a justified {\it realism} about them: rather, it is a matter of whether the semantics of GR must invoke such an entity.

Thus while I agree that there is no a priori reason why coordinate-dependent quantities could not be physical, and also that, in general, {\it some} coordinate-dependent quantities {\it are} physical, I think that more work is required to admit these specific coordinate-dependent quantities---pseudo-tensors---into one's theory. This leads in to my next two points, about further conditions that are required to ascribe to pseudo-tensors a semantics that is physically acceptable. 

The second qualification is that, to accept pseudo-tensors as good physical quantities, one {\it requires that they have well-defined transformation properties under (different kinds of) diffeomorphisms.} (Of course, pseudo-tensors transform covariantly only under {\it affine} transformations). Even if one is in principle sympathetic to the idea of coordinate-dependent quantities, surely some relationship between the quantities in the different coordinates, at least for special classes of coordinates, must be required before such quantities can be accepted in the theory.\footnote{For example, when the number of spacetime dimensions is odd, under visible diffeomorphisms the quasi-local stress-energy tensor picks up an anomalous term that is in general non-zero, i.e.~Eq.~\eq{anomaly}. Although this breaks covariance, it is a well-defined transformation law, and it is admissible.} 
As the analogy below should make clear, the real physical worry is that different coordinate systems would be, quite literally, incommensurable. This worry is, of course, in the neighbourhood of a closely related one: namely, that pseudo-tensors are not geometric objects\footnote{However, as Trautman (1965:~p.~86) points out, spinors are not geometric objects either.} (see Anderson (1967:~p.~424) and Duerr (2019a:~Section 3.3)). Therefore, it is not clear that the gravitational pseudo-tensor gives well-defined values for the energy in the various coordinate systems.\footnote{In this context, Nester argues that the Hamiltonian is a fully covariant object, and that its significance is independent of the coordinates. But this is not entirely satisfying, for two reasons: (i) As I already mentioned in footnote \ref{NesterP}, the direct relation Eq.~\eq{PN} between the Hamiltonian and the pseudo-tensor only holds for everywhere {\it constant} vector fields $N$: and so, a putative covariance of the Hamiltonian does not automatically apply to the pseudo-tensor (which is expressly known to {\it not} be  non-covariant). (ii) As I mentioned in my reply to (A) in Section \ref{phildis}, covariance is not what matters the most, because it can be easily achieved by introducing additional quantities. Specifically, the Hamiltonian depends on a choice of vector field $N$, and the Hamiltonian approach is only covariant relative to a given such choice. But since spacetimes do not come naturally equipped with vector fields, the Hamiltonian approach is dependent on such a choice, which makes it liable to either: (a) introducing background structures, or (b) treating different models differently.\label{Hamco}}
This is something that Nester's work, in my opinion, does not clarify.

Thus Hoefer's (2000) argument that the quantities are not well-defined still seems to have {\it some} force, because a satisfactory definition of energy and stress would involve either giving their transformation rules or else giving a physical interpretation of the non-covariance (e.g.~as in the case with boundaries, see Section \ref{worries3}). Lacking these, the realist story about pseudo-tensors seems less convincing.

Consider the analogy with special relativity, where the 4-momentum vector $p$ is a physical quantity that transforms covariantly under Lorentz transformations, but its components are not invariant. Choose a coordinate system where we write $(p^\m)=(E/c,{\bf p})$, where $E$ is the energy and ${\bf p}$ the spatial momentum as measured in those coordinates. Then we know that $E^2/c^4$ and ${\bf p}^2$ are not invariant quantities, whereas their difference, $-E^2/c^4+{\bf p}^2$, is invariant (and likewise for covariant quantities). The energy and the momentum transform into each other under Lorentz transformations, and so their definition depends on the coordinates chosen. Despite this coordinate-dependence, $E$ and ${\bf p}$ {\it are} physical quantities, and we can: 

(1)~~define energy and momentum for each choice of coordinates (these definitions being incompatible in the sense that what is defined as energy in one system is not the energy in another system), and furthermore these definitions can be made operational with respect to the corresponding frame, e.g.~a preferred measuring apparatus; 

(2)~~give well-defined transformation laws for $E$ and ${\bf p}$ from one coordinate system to another: namely, the Lorentz transformation law.

Property (2) seems to fail in the case of pseudo-tensors, and thus there is a disanalogy with other coordinate-dependent quantities that we are familiar with. The point is not that pseudo-tensors fail to be covariant, but rather that there is no well-defined transformation law between them.\footnote{As I showed in Section \ref{diffeos}, in spaces with boundaries, different kinds of diffeomorphisms need to be distinguished, and for some diffeomorphism one may obtain a modified transformation law, involving the anomalous term in Eq.~\eq{anomaly}.} Even if it were still possible to define energy and momentum in every coordinate system, as in (1), these different notions would be completely unrelated to each other---different coordinate systems would be, quite literally, incommensurable. 

The point is that it is not quite clear that one can simply {\it drop} the requirement of well-defined transformation properties under diffeomorphisms and put nothing in its place: a coherent alternative for it seems required. At any rate, this is an objection that the programme needs to address.\footnote{Pitts (2010) is one such proposal. Duerr (2019a:~Section 3.3) notices, that, while `nothing is {\it inherently} baneful about coordinate-dependent objects... [f]or pseudotensors... the coordinate-dependence is ``vicious'' (Pitts): in general, the preferred coordinate transformations do {\it not} pick out the characteristic invariants of the spacetime. The spacetime symmetries do not align with the pseudotensor's symmetry group'.}

A third qualification is that, even if Read's arguments indeed allow taking pseudo-tensors as physical quantities, they do not require this. Pseudo-tensors {\it might} be the correct definition of energy and momentum in general relativity---or they might not: for there are alternative proposals---in particular, the quasi-local proposals. Thus the different proposals need to be compared. I do this in the following Section.

\subsection{Quasi-local expressions: superpotentials vs.~the Brown-York stress-energy tensor}\label{suphol}

In this Section, I will defend the `modified' version of the Brown-York quasi-local stress-energy tensor from Section \ref{BYQL} as a preferred quasi-local definition of energy, momentum, and stress in general relativity. I will partly do this by comparison with the other main proposal, from Section \ref{Revisited}: namely, the pseudo-tensor, and the quasi-local Hamiltonian expression obtained from it. I will argue that the former is so far better developed, and thus better able to answer various queries. Hence my favouring, on balance, the Brown-York quasi-local stress-energy tensor. But since 
 some of the details about the pseudo-tensor approach have not been fully worked out, the latter approach, with all its open questions, should still be considered.

By the `modified' version of the Brown-York quasi-local stress-energy tensor', or simply the `Brown-York stress-energy tensor', I will mean the quasi-local Brown-York stress-energy tensor from Section \ref{BYQL}, modified, where this technique is available (i.e.~in the case $\L\not=0$, regardless of its sign),\footnote{The modified Brown-York stress-energy tensor, in the case $\L>0$, has been worked out in the references in footnote \ref{L>0}.} by the inclusion of counterterms: as opposed to the older reference field configuration subtraction (see (B)-(b) below). This is all, in principle, regardless of its possible holographic interpretation.


The quasi-local perspective has been almost entirely missed in recent philosophical discussions of energy conservation. Lam (2011:~p.~1022) mentions the existence of quasi-local stress-energy tensors, but then dismisses them, on two grounds: 

(1)~~A `divide and conquer' argument to the effect that there are different expressions in the literature for such tensors, and that `the links betwen [them] are not always clear... there is for the time being no consensus on how to build a `quasi-local' notion of energy'.

(2)~~The quasi-local expressions allegedly also depend on particular background structures, `such as the dependence on some particular embedding or on some particular boundary conditions'. 

I will respond to both arguments below. Let me here say, about (1), that it is not a strong argument: for the mere {\it existence} of more than one expression, without further analysis or elaboration, is not a good reason to reject them collectively, but rather to further investigate which one is correct, or whether they are related. Furthermore, at least in cases with a non-zero cosmological constant, and {\it pace} Szabados (2009) which Lam cites, it is not really true that there is no consensus in---a large portion of---the physics community about which is the preferred expression\footnote{The literature here is too large to survey. Some examples are e.g.~the textbook Ammon and Erdmenger (2015:~p.~210), Anninos et al.~(2011:~p.~5), Balasubramanian and Kraus (1999:~p.~413), and Skenderis (2001, 2002).} (by itself, this need of course not be a sufficient reason to join the consensus, but neither should the reasons for the consensus be ignored). 


My comparison here will be along two main lines, which I will label (A) and (B), that parallel the philosophical discussion in Section \ref{phildis} (and go back to points (iii) and (v) in Section \ref{Einstein}). The first is a modification of point (A) in Section \ref{phildis}, namely: diffeomorphism-invariance; the second is, like (B) in Section \ref{phildis}, about ``background-dependence'', or the requirement that the theory should not contain any ``background structures''. Then, at the end of the Section, I will ask how the two approaches, the pseudo-tensor/Hamiltonian and the Brown-York approaches, are related.

\ \\(A)~~{\it The issue of diffeomorphism invariance.} I begin with the Brown-York approach. As I discussed in some detail in Section \ref{diffeos} for the case of spacetimes with a negative cosmological constant, the properties of the modified Brown-York stress-energy tensor under diffeomorphisms of different kinds are well-understood, and can be summarised in three points: 

(1) The modified Brown-York stress-energy tensor is {\it invariant} under diffeomorphisms that express true redundancies (in the sense of (Redundant)), i.e.~invisible diffeomorphisms.

(2) It is {\it covariant} under visible diffeomorphisms, i.e.~those that generate a conformal transformation on the boundary (see Eq.~\eq{confg}).\footnote{As I said before, I am ignoring the anomaly term in Eq.~\eq{anomaly}, which vanishes identically in four-dimensional spacetimes.} 

(3) It is {\it not invariant} under those diffeomorphisms that express physical transformations and are not (Redundant), i.e.~those that give {\it physically inequivalent} solutions of Einstein's field equations. This is as one would expect (see also the discussion in the next Section).

Thus the modified Brown-York stress-energy tensor has correct transformation properties under these different kinds of diffeomorphisms. 

Also, up to a possible anomaly of the diffeomorphism group of the type discussed in Section \ref{diffeos}, it is clear from the Brown-York construction that their stress-energy tensor transforms {\it covariantly} with respect to arbitrary reparametrisations, $\xi_i(x)$, of the boundary manifold (i.e.~a visible diffeomorphism) for any value of $\L$. This shows that the modified Brown-York stress-energy tensor is not liable to the allegation (A) from Section \ref{phildis}. 

\ \\By contrast, in the case of pseudo-tensors and the quasi-local expressions derived from them in Section \ref{Revisited}, an analysis along these lines has, so far as I know, not been carried out. There are, however, some remarks in the literature. For example, Chang et al. (1999:~p.~3) write:
\begin{quote}\small
[T]he energy-momentum defined by such a pseudotensor does not really depend on the local value of the reference frame, it is actually {\it quasi-local}---it depends (through the superpotenial) on the values of the reference frame (and the fields) only on the boundary of a region.
\end{quote}
The quote is perhaps somewhat optimistic. First, as I noticed in Section \ref{totd}, there are assumptions that go into deriving the quasi-local expression Eq.~\eq{PN} from the pseudo-tensor, which surely do not involve just the boundary itself, but at least an open neighbourhood of it. Furthermore, the `dependence on the values of the reference frame' of the quasi-local energy that Chang et al.~mention seems incompatible with its being covariant, since the reference frame dependence comes in through the superpotential, which is not a tensor, also not when evaluated at the boundary. In other words, we appear to be in the situation that I discussed in Section \ref{phildis}, where there is a strong dependence on the frame of reference chosen at the boundary, and apparently no rule to compare one system to another.\footnote{One might think that the Hamiltonian formalism comes to the rescue here. However, two problems need to be addressed: first, the Hamiltonian depends on a choice of vector field $N$, and there is no natural way to choose this vector field in a general manifold; second, relating the pseudo-tensor to the Hamiltonian requires a {\it constant} vector field $N$, as already discussed in footnote \ref{NesterP}.} At any rate, these authors do not clarify how the coordinate-dependence disappears. This contrasts with the Brown-York stress-energy tensor, which as I discussed in (2) does have satisfactory properties under changes of frame on the boundary. Also, Chang et al.~(1999) do not make a distinction between the diffeomorphisms (1)-(3), whereas this is key to understand the transformation properties of the quasi-local energy.

In Chen et al.~(1999:~Section 2) and Nester (2004:~pp.~272-275), a `covariant Hamiltonian formalism' is worked out, in which as I already mentioned the various superpotentials are reproduced from relatively simple geometric expressions. Apart from the physical significance that one may ascribe to such covariance (see footnote \ref{Hamco}), it is not clear how the quasi-local energy and momentum transform under (non-affine) changes of frame on the boundary, i.e.~whether there is any way to compare quantities between different boundary frames.

Either way, although the Hamiltonian approach to pseudo-tensors seems promising, it needs more work before a full judgment about diffeomorphism invariance can be made.

\ \\(B)~~{\it ``Background-dependence''.} Here, I begin with the pseudo-tensor/Hamiltonian approach. I already endorsed, in Section \ref{phildis} (B), Read's response to claims that pseudo-tensors require the introduction of background structures. Here I will defend the Brown-York approach, as regards background-independence. This will be three points, (a), (b), (c), of which (a) is a brief response to certain criticisms of AdS-CFT:\footnote{For a fuller treatment and references, see De Haro (2017:~pp.~113-116); cf.~footnote \ref{bi}.}

(a)~~It has occasionally, and informally, been claimed that the methods in AdS-CFT are ``background-dependent'', because they allegedly require the use of anti-de Sitter boundary conditions. This is part of a more general criticism, in Section \ref{phildis}-(B), that defining asymptotic expressions for energy and momentum requires fixing the structure at infinity (typically, a flat metric). Here, there are two points to be made.

First, the AdS-CFT correspondence, and in particular the calculation of the modified Brown-York stress-energy tensor from Section \ref{hst}, works for an {\it arbitrary,} smooth boundary metric $\g$: and so, the claim that AdS-CFT requires `AdS boundary conditions' is incorrect. This means that the space is not even {\it asymptotically} AdS: it is asymptotically {\it locally} AdS, i.e.~there exists, in an open neighbourhood of the boundary, always a choice of coordinates where the space looks like AdS---locally! But the space is in general {\it not} AdS: it is only required that the boundary metric have conformal Killing vectors, as in Eq.~\eq{confg}, if the CFT is to have a non-trivial conformal symmetry group. Thus the boundary metric is taken to have conformal symmetries, but it does not need to have any pre-determimed form, like e.g.~being flat or being a round sphere. Examples in AdS-CFT have been worked out for boundary manifolds of many different kinds. 

Second, the above objection perhaps originates in a confusion about the role of boundary conditions in AdS-CFT. As I discussed in Section \ref{BYql}, the original Brown-York definition of the stress-energy tensor did have a dependence on a given, non-dynamical, background spacetime, through the introduction of the reference term in the action, Eq.~\eq{action}. But as I also noted after Eq.~\eq{BY}, this is {\it not} how the tensor is rendered finite in AdS-CFT. Rather, one adds covariant counterterms that are {\it intrinsic} to the boundary. The renormalisation method, from Section \ref{hst}, of adding boundary counterterms in the action, has the advantage of removing another kind of background: namely, the reference spacetime. Thus there is no reference spacetime that one needs to compare with. 

(b)~~The role of boundary conditions in the approach that I am advocating here thus differs from the idea of a reference field configuration or background subtraction. For the boundary conditions required are merely those that are dictated by the fact that Einstein's equation is a second-order differential equation. For example, in globally hyperbolic spacetimes, one standardly specifies data on an initial Cauchy surface in order to solve Einstein's equations (think of the analogy with the hole argument!).\footnote{There are of course also disanalogies: see Stachel (2014:~pp.~14-15, 50).} But this is not in any sense a dependence on a background: rather, the specification of the initial data is the specification of a solution among the set of all possible solutions. If the spacetime had a boundary somewhere at a finite distance, in order to specify a solution one would, in addition, need to specify boundary conditions. This is what one does when one solves differential equations in spaces with boundaries: it has nothing to do with imposing backgrounds, but rather with choosing one's solution of interest out of the total set.

Thus, the background-independence of the modified Brown-York stress-energy tensor is ``as good as it gets'' in general relativity.\footnote{Janssen (2014:~pp.~198-202) discusses Einstein's struggle, in his debate with de Sitter, with the need to impose boundary conditions at infinity, and in connection his (failed) attempts to incorporate Mach's principle into general relativity. See also Norton (1993:~pp.~806-809). In De Haro (2017:~p.~115), I argue that there are two distinct senses of `background-independence': only one of which is implemented in general relativity---hence the `as good as it gets'.}

The situation is quite different for the pseudo-tensor and its quasi-local energy and momentum. First of all, these methods have only been worked out using reference spacetimes, both in the cases $\L=0$ and $\L\not=0$. Furthermore, to my knowledge, the asymptotics considered is only the simplest one: namely, Minkowski space and (A)dS space. Thus only cases with a flat boundary metric $\g$ have been considered. This may be only a matter of technical development: but it is of course yet to be seen whether the pseudo-tensor gives sensible (e.g.~finite!) results in other cases.\footnote{It also seems a sign of lack of communication between, at least some parts of, the communities that the intrinsic renormalisation methods have not been applied, or even noticed, by those working on pseudo-tensors. For example, the review by Szabados (2009), written 12 years after gauge-gravity dualities were discovered by Maldacena, makes a single passing mention of it, and takes no notice of how holographic renormalisation improves on the existing methods, by not appealing to a reference spacetime (see the discussion after Eq.~\eq{BY}).} 

Nester (2004:~p.~178) describes the problem as follows: `The pseudotensor coordinate reference frame ambiguity now reappears in the guise of the reference field configration, which must be specified on the boundary'. 

Despite their differences, both approaches emphasise the physical importance of choosing boundary terms to define energy and momentum. In the superpotential approach, boundary conditions/terms resolve the ambiguity in the specification of the superpotential. Also in the Brown-York approach are boundary terms central, since the stress-energy tensor is defined directly in terms of boundary quantities. 

This dependence of the energy and momentum on boundary conditions is natural: for example, in classical mechanics, the momentum of a bullet of course depends on the initial velocity with which it is shot. 

(c)~~Nor should the requirement that the {\it boundary} spacetime has conformal Killing vectors be seen as a dependence on a background, in the usual way. As Read (2020:~p.~222) mentions, choosing a Killing vector field is making a choice of a model among all the possible dynamical models. 

\ \\There is also a question of whether the pseudo-tensor and the Brown-York approach are directly related. It is unclear that they are, and a direct calculation of the Freud superpotential for the case $\L<0$, which includes Dirichlet boundary conditions, does not exhibit a clear or direct relationship to the Brown-York stress-energy tensor. Nester (2004:~p.~178) claims that the Hamiltonian formalism establishes a relation between the two: `Quasilocally, with certain choices for the reference fields and displacement, [our quasi-local expressions] reduce to the famous expressions of Brown and York'. However, the exact relationship remains obscure.

As already discussed in Section \ref{hst}, AdS-CFT gives, through its holographic correspondence with the expectation value of the stress-energy tensor of the dual quantum field theory, Eq.~\eq{BYQFT}, independent support to the modified Brown-York stress-energy tensor as being the gravitational energy and momentum. 

The advantage of the quasi-local approach is perhaps best illustrated by mentioning how it solves several problems that Duerr (2019a:~Section 3.3) discusses in connection with the pseudo-tensor (and, as I said earlier, I agree with much of Duerr's critique). 

The first has already been discussed: it is the lack of covariance, and the pseudo-tensor's lacking appropriate transformation properties, i.e.~my (v) from Section \ref{Einstein}, which is Duerr's label (H1). The quasi-local stress-energy is both a geometric object and a (quasi-local) tensor under the relevant diffeomorphisms (see Section \ref{diffeos}), and so this problem does not appear. 

Second, asymptotic flatness (Duerr 2019a:~Section 3.3) is not a requirement of this approach: since, as I have stressed, the quasi-local stress-energy tensor is well-defined for both positive and negative values of $\L$, and {\it for arbitrary boundary conditions on the metric,} i.e.~not just the purely de Sitter or purely anti-de Sitter boundary conditions. In particular, this can be done for de Sitter-type solutions, which, in cosmological settings, are more realistic than the asymptotically flat ones. Furthermore, the divergences can be correctly renormalised by the method of holographic renormalisation. This is one of Duerr's main points of critique, and so this is a clear advantage of the quasi-local approach. 

Third, as a general point (not specific to quasi-local energy): one of Duerr's (2019a:~Section 3.2) main criticisms to Read (2020) is that the latter {\it assumes,} rather than argues, that the ambiguity of the pseudo-tensor can be resolved. But we have seen that Read's ``assumption'' is in fact vindicated by the details of Noether's second theorem. For the ambiguity is only apparent: it is fixed by a choice of boundary conditions, and of corresponding boundary terms (and this important point is common to all three approaches I have discussed). 

Gauge-gravity dualities also cast further light on the interpretation of these boundary conditions. Each choice of boundary conditions corresponds to a different CFT (and also a different common core theory): namely, a CFT where a different kind of source is held fixed, with a different Lagrangian, and with a different path integral. The choices between such CFT's are different physical choices: which explains why their energies are different.

\subsection{Interpreting quasi-local quantitites}\label{worries3}


Although, in previous Sections, I have addressed various objections to the quasi-local energy-momentum expressions, one can still ask about the physical meaning of such quasi-local expressions. After all, the traditional view in general relativity is that physical quantities are closely associated with geodesics and, in particular, with point coincidences, that are used to delimit these quantities. Indeed, the view goes back to Einstein himself.\footnote{See Einstein (1916:~p.~776) and Kretschmann (1917:~pp.~576-577, 579). However, elsewhere Einstein (1915c:~p.~228) considers physically real events as consisting of more general {\it spatio-temporal coincidences.} For a discussion of Einstein's (1916) paper and its prehistory, see Sauer (2005). The point-coincidence argument, and how Einstein sees in it an implicit objection to the hole argument, are discussed on p.~816.} In more substantivalist terms,\footnote{For a review of substantivalist and anti-substantivalist arguments, see Pooley (2012:~pp.~47-63).} it is {\it points} that are fundamental in general relativity, and other meaningful geometric quantities ultimately reduce to quantities defined on points. But the pseudo-quantities, especially the ones that---like the Brown-York stress-energy tensor---are directly defined quasi-locally, without a local definition, would seem to give some physical significance to {\it spacelike two-surfaces,} possibly located at infinity: and it is not immediately clear what this means, both in terms of the theory's ontology (should we include surfaces in the theory's ontology? And, if so, which ones?) and in empirical terms (what is the empirical content of such terms?).

I will here offer a number of arguments to the effect that surfaces and boundaries {\it should} be counted as part of general relativity's ontology, and that they do have an empirical meaning. 

First, notice that the fact that {\it points} are regarded, by some substantivalists at least, as fundamental, does not exclude the reality of {\it finite surfaces,} which, just like the manifold itself, are made of points. Also {\it boundaries} can be understood in terms of neighbourhoods of the manifold,\footnote{The boundary of a manifold $M$ is the set of points $p$ such that $p$ has no neighbourhood homeomorphic with an open set of $\mathbb{R}^n$, where $n$ is the dimension of the manifold. 
} and so as such what we call `the existence of a boundary' is a topological property of the manifold, which just is part of the ontology of the theory. Thus, once we have a semantics for points and for neighbourhoods in general relativity (and the latter are required for both topological and geometric reasons), this very semantics gives a meaning to surfaces and boundaries. This point can also be argued from two philosophical directions:

(a) An ontology that accepts points as objects should surely accept sets, classes, or mereological fusions, of points. For example, this is standard in ``free lunch'' supervenience metaphysics: see Lewis (1986:~pp.~211-212), Armstrong (1997:~pp.~12-13), Sider (2001:~pp.~xvi, 7, 59). 

(b) As Butterfield (2011) argues, in his campaign against {\it pointillisme,} not all classical physical quantities are defined at a point (where `defined at a point' is here taken in the sense of Lewis' (1986a:~pp.~ix-xvi; 1994) Humean supervenience). Classical continuum mechanics gives us many examples of `spatial extrinsicality: the way that an ascription ``reaches across space'''. In particular, boundaries and quasi-local energy and momentum are as admissible in general relativity as they are in classical mechanics: `continuum mechanics embraces spatial extrinsicality due to surface forces, acting at a point, {\it with respect to a surface through that point} (see Butterfield, 2011:~pp.~354, 356; my emphasis). 

Second, there is no contradiction with any fact about general relativity if it turns out that energy and momentum can only be defined quasi-locally but not everywhere locally. 
There is no a priori reason why energy and momentum should exist everywhere locally on the manifold (see e.g.~Lam, 2011:~p.~1023). So, if some quasi-local notion of energy and momentum is well-defined, rather than being regarded with suspicion, it should be a welcome possibility. 

Third, there are several physical reasons to consider boundaries, and stress-energy tensors as physical quantities defined on them. One reason is that one is interested in considering open systems, i.e.~finite regions of the universe. In such cases, the non-vanishing of the right-hand side of Eq.~\eq{bE} is interpreted as matter entering or leaving the region through the boundary. These open systems are very important for several reasons: as modelling a real system (e.g.~a black hole, and the gravitational waves that it emits), or as a way to generate new solutions by gluing a given solution to other solutions through the boundary (for example, solutions with different values of the cosmological constant, with so-called `domain walls' in between: see e.g.~Freedman et al.~(1999:~pp.~376-377), Ammon and Erdmenger (2015, p.~302-304); for gluing of spacetimes in a cosmological context, see Carrera and Giulini (2010:~pp.~170, 181)). One then needs to match the conditions of the two solutions at the boundary, and the boundary metric and its conjugate variable, viz.~the stress-energy tensor, are the two canonical variables to control at the boundary.\footnote{
If the Brown-York stress-energy tensor is evaluated not at conformal infinity, but at some finite distance, then this can have several other applications in string physics. For example, the universe is modelled in some models as the four-dimensional boundary of a five-dimensional space, placed at finite distance from the centre of the space (so that we `live on a D-brane': see e.g.~Randall and Sundrum (1999:~p.~4690)). In that case, the Brown-York stress-energy tensor of the five-dimensional space acts as an extra source for the matter energy-momentum appearing in the four-dimensional Einstein field equations on the D-brane (De Haro et al.~2001a: p.~3171).}

More generally, in semi-classical descriptions of quantum gravity theories, where the solutions of general relativity coupled to matter contain objects such as strings and D-branes, higher-dimensional surfaces such as the ones considered here ``stand ready'' for an interpretation in terms of extended material membranes, i.e.~as a natural generalisation of the notion of a geodesic. In other words, such surfaces must be considered, if general relativity is to be the semi-classical limit of these theories of quantum gravity.





\section{Summary and Conclusion}


In Section \ref{Einstein}, I identified five putative, inter-dependent, problems for local definitions of energy and momentum in general relativity: 

\hspace{1.5mm}(i)~~the Hilbert-Klein-Noether worry that the conservation laws of GR are {\it identities,} with no physical content; 

\hspace{1mm}(ii)~~the lack of a {\it gravitational energy-momentum tensor,} analogous to the matter energy-momentum tensor; 

(iii)~~the lack of integral conservation laws, because of the lack of a {\it timelike Killing vector field} in a general spacetime; 

(iv)~~the {\it non-uniqueness} of the gravitational energy-momentum pseudo-tensor; and 

\hspace{1mm}(v)~~its lacking {\it covariance,} and its lacking appropriate geometric properties. 

The Hilbert-Klein-Noether worry, (i), was addressed in Sections \ref{2N} and \ref{2con}, where I argued 
that the conservation of energy and momentum in GR {\it does} have physical content. 

The lack of a gravitational energy-momentum tensor, (ii), can be resolved in two ways: either by introducing the gravitational pseudo-tensor, or by defining quasi-local quantities. I then analysed two quasi-local proposals, and specialised questions (iii) to (v) to these proposals. 

About the lack of a timelike Killing vector field, (iii), in a general spacetime: while I accept that this is {\it prima facie} a problem for local expressions, I noted that the modified Brown-York stress-energy tensor only requires the existence of a {\it conformal} Killing vector field {\it at the quasi-local surface} (usually: the asymptotic boundary). 

I also argued that the putative non-uniqueness, (iv), of the gravitational energy-momentum pseudotensor, i.e.~the ability to add an ``arbitrary'' superpotential, is fixed by Noether's theorem directly in terms of the variation of the action, as one sees from Eqs.~\eq{defP}-\eq{Sbar}. Specifically, it is fixed by a choice of boundary terms. So far as I know, this possibility has not been used in the literature before, to solve the problem of non-uniqueness.\footnote{See footnote \ref{litbdytms} for a brief discussion of the literature on this point.} In more detail: one can understand the ability to add an abitrary term to the superpotential as the ability to add an arbitrary boundary term to the action without changing the equations of motion; (this then agrees with results by Chang et al.~(1999:~p.~1898) in the Hamiltonian formalism).

An important point is that standard arguments in gauge-gravity dualities, which interpret the duality in terms of a common core shared between two isomorphic models, further support the statement that the Brown-York stress-energy tensor correctly represents the quasi-local energy and stress of general relativity, in the corresponding spacetimes.

Finally, point (v) about covariance, and the lack of appropriate transformation properties, was re-phrased in terms of two different notions: (A) diffeomorphism invariance, and (B) background-independence. This was done in Section \ref{phildis} onwards.

About (A): I summarised, in Section \ref{diffeos}, the properties of the modified Brown-York stress-energy tensor under three relevant kinds of diffeomorphisms. (1) The modified Brown-York stress-energy tensor is correctly invariant under `invisible', i.e.~(Redundant), diffeomorphisms. (2) It is correctly {\it non-invariant} under `large', i.e.~always non-redundant, diffeomorphisms, because they relate different models. (3) `Visible' diffeomorphisms correctly give conformal transformations at infinity. These diffeomorphisms thus correspond to the conformal transformations of the conformal field theory. (The results were obtained for $\L<0$, but I argued that they can be extended to $\L>0$).

About (B): I agreed with Read's (2020) defence of the background-independence of expressions involving Killing vectors: although, as mentioned under (iii), the Killing vectors required for the modified Brown-York stress-energy tensor to give an integral conservation law satisfy a weaker condition. Also, I noted that the Brown-York stress-energy tensor can be defined for any smooth boundary metric $\g$, and I clarified the role of the boundary conditions.

In sum, the modified Brown-York stress-energy tensor gives the clearest analysis and resolution, so far, of the above problems. As for admitting the gravitational pseudo-tensor itself into the ontology of general relativity: I argued that there are valid points in Read's (2020) defence of pseudo-tensors, although important questions remain open (especially, about its transformation properties, and about which proposal is correct). The Chang et al.~(1999) and Chen et al.~(2017, 2018) quasi-local proposal, and its relation to the Hamiltonian of GR, also seems promising, and it clarifies the differences between the different pseudo-tensors used in the literature: but it also faces additional questions (especially, about the status of the vector field $N$).

\section*{Acknowledgements}
\addcontentsline{toc}{section}{Acknowledgements}

I thank Bryan Roberts and Nic Teh for organising the conference `The Philosophy and Physics of Noether's Theorems' at the University of Notre Dame, London Global Gateway, and James Read, Bryan Roberts, and Nic Teh, for their invitation to contribute to this volume. I thank James Nester for detailed correspondence and for his patience answering my questions, and James Read, Henrique Gomes, Harvey Brown, Jeremy Butterfield, and Jeroen van Dongen for comments. I also thank audiences at the University of Notre Dame, London Global Gateway, the Hong Kong University of Science and Technology, and the Tsinghua Sanya International Mathematics Forum in Hainan, China. 


\section*{References}
\addcontentsline{toc}{section}{References}

\small

Ammon, M., Erdmenger, J.~(2015). {\it Gauge/Gravity Duality. Foundations and Applications}. Cambridge: Cambridge University Press.

\ \\Anderson, J.~L.~(1967). {\it Principles of Relativity Physics.} New York, London: Academic Press.

\ \\Anderson, J.~L.~(1971). `Covariance, Invariance, and Equivariance: A Viewpoint'. {\it General Relativity and Gravitation,} vol. 2, 2, pp.~161-172.

\ \\Anninos, D., Ng, G.S., Strominger, A. (2011). `Asymptotic Symmetries and Charges in De Sitter Space', {\it Classical and Quantum Gravity} 28, 175019. [arXiv:1009.4730 [gr-qc]].

\ \\Armstrong, D.~M.~(1997). {\it A World of States of Affairs.} Cambridge: Cambridge University Press.

\ \\Arnowitt, R., Deser, S., and Misner, C.~W.~(1959). `Dynamical structure and definition of energy in general relativity'. {\it Physical Review}, 116 (5), p.~1322.

\ \\Arnowitt, R., Deser, S., Misner, C.~W.~(1962). `The Dynamics of General Relativity'. In: Witten, L.~(Ed.), {\it Gravitation: An Introduction to Current Research,} Chapter 7. New York: Wiley. Reprinted in: `Republication of: The dynamics of general relativity'. {\it General Relativity and Gravitation}, 2008, 40 (9), pp.~1997-2027. arXiv:gr-qc/0405109.

\ \\Ashtekar, A.~and Hansen, R.~O.~(1978). `A unified treatment of null and spatial infinity in general relativity. I. Universal structure, asymptotic symmetries, and conserved quantities at spatial infinity'. {\it Journal of Mathematical Physics,} 19, pp.~1542-1566.

\ \\Balasubramanian, V., de Boer, J.~and Minic, D.~(2001). `Mass, Entropy, and Holography in Asymptotically de Sitter Spaces'. {\it Physical Review,} D, 65, 123508, pp.~1-15.

\ \\Balasubramanian, V.~and Kraus, P.~(1999). `A Stress Tensor for Anti-de Sitter Gravity'. {\it Communications in Mathematical Physics,} 208, pp.~413-428. arXiv: hep-th/9902121.

\ \\Ba\~nados, M.~and Reyes, I.~(2016). `A Short Review of Noether's Theorems, Gauge Symmetries and Boundary Terms'. {\it International Journal of Modern Physics,} D, 25 (10), 1630021. 

\ \\Barbashov, B.~M.~and Nesterenko, V.~V.~(1983). `Continuous Symmetries in Field Theory'. {\it Fortschritte der Physik,} 31 (10), pp.~535-367.

\ \\Belinfante, F.~J.~(1939). `On the Spin Angular Momentum of Mesons'. {\it Physica} IV, 9, pp.~887-898.

\ \\Belot, G.~(2013). `Symmetry and Equivalence'. In: {\it The Oxford Handbook of Philosophy of Physics,} Batterman, R.~(Ed.), chapter 9. Oxford: Oxford University Press.

\ \\Belot, G.~(2018). `Fifty Million Elvis Fans Can't Be Wrong'. {\it No$\hat u$s,} 52 (4), pp.~946-981.

\ \\Bergmann, P.~G.~(1958). `Conservation Laws in General Relativity as the Generators of Coordinate Transformations'. {\it Physical Review,} 112 (1), pp.~287-289.

\ \\Bessel-Hagen, E.~(1921). `\"Uber die Erhaltungss\"atze der Elektrodynamik'. {\it Mathematische Annalen,} 84, pp.~258-276. English translation by M.~Albinus and N.~H.~Ibragimov: `On Conservation Laws of Electrodynamics', in Archives of ALGA, volume 3 (2006), pp.~33-51. ALGA Publications. https://www.researchgate.net/publication/311456950$\_$Archives$\_$of$\_$ALGA$\_$vol$\_$3. The page numbers in the text refer to the English translation.

\ \\Bondi, H., van der Burg, M.~G.~J., Metzner, A.~W.~K.~(1962). `Gravitational Waves in General Relativity. VII. Waves from Axi-Symmetric Isolated Systems', {\it Proceedings of the Royal Society of London. Series A, Mathematical and Physical Sciences}, 269 (1336), pp.~21-52.

\ \\Brading, K.~(2005). `A Note on General Relativity, Energy Conservation, and Noether's Theorems'. In: {\it The Universe of General Relativity,} Kox, A.~J.~and Eisenstaedt, J.~(Eds.), pp.~125-135. Boston and Berlin: Birkh\"auser. 

\ \\Brading, K.~and Brown, H.~R.~(2000). `Noether's Theorems and Gauge Symmetries'. 	arXiv:hep-th/0009058.

\ \\Brading, K.~and Brown, H.~R.~(2003). `Symmetries and Noether's Theorems'. In: Brading, K.~and Castellani, E.~(Eds.), {\it Symmetries in Physics. Philosophical Reflections,} pp.~89-109. Cambridge: Cambridge University Press.

\ \\Brading, K.~and Brown, H.~R.~(2004). `Are Gauge Symmetry Transformations Observable?' {\it The British Journal for the Philosophy of Science,} 55, pp.~645-665.


\ \\Brown, H.~R.~(2005). {\it Physical Relativity. Space-time Structure from a Dynamical Perspective.} Oxford: Oxford University Press.

\ \\Brown, H.~R.~(2020). `Do Symmetries ``Explain'' Conservation Laws? The Modern Converse Noether Theorem vs.~Pragmatism'. This volume. 

\ \\Brown, H.~R.~and Brading, K.~A.~(2002). `General Covariance from the Perspective of Noether's Theorems'. PhilSci: 821. http://philsci-archive.pitt.edu/821.

\ \\Brown, J.~D.~and Henneaux, M.~(1986). `Central Charges in the Canonical Realization of Asymptotic Symmetries: An Example from Three-Dimensional Gravity', {\it Communications in Mathematical Physics} 104, pp.~207-226.

\ \\Brown, J.~D.~and York, Jr., J.~W.~(1993). `Quasilocal Energy and Conserved Charges Derived from the Gravitational Action'. {\it Physical Review D,} 47, pp.~1407-1419.

\ \\Butterfield, J.~N.~(2011). `Against Pointillisme: A Call to Arms'. In: {\it Explanation, Prediction, and Confirmation,} Dieks, D., Gonzalez, W.~J., Hartmann, S., Uebel, T., Weber, M.~(Eds.), pp.~347-365. Dordrecht: Springer.

\ \\Butterfield, J.~N.~(2020). `On Dualities and Equivalences Between Physical Theories'. Forthcoming in {\it Space and Time after Quantum Gravity}, Huggett, N.~and W\"uthrich, C.~(Eds.).

\ \\Carrera, M.~and Giulini, D.~(2010). `Influence of Global Cosmological Expansion on Local Dynamics and Kinematics'. {\it Reviews of Modern Physics,} 82 (1), pp.~169-208. 

\ \\Caulton, A.~(2015). `The Role of Symmetry in the Interpretation of Physical Theories'. {\it Studies in History and Philosophy of Modern Physics,} 52, pp.~153-162.

\ \\Chang, C.~-C., Nester, J.~M., and Chen, C.-M~(1999). `Pseudotensors and Quasilocal Energy-Momentum'. {\it Physical Review Letters,} 83.10: 1897. arXiv:gr-qc/9809040.

\ \\Chen, C.-M., Nester, J.~M., Tung, R.-S.~(2017). `Gravitational Energy for GR and Poincar\'e Gauge Theories: a Covariant Hamiltonian Approach'. In: Ni, W.-T. (Ed.), {\it One Hundred Years of General Relativity. From Genesis and Empirical Foundations to Gravitational Waves, Cosmology and Quantum Gravity,} Volume 1, pp.~187-261. arXiv:1507.07300 [gr-qc].

\ \\Chen, C.-M., Liu, J.-L, Nester, J.~M.~(2018). `Gravitational Energy is Well-Defined'. {\it International Journal of Modern Physics D,} 27 (14), 1847017. arXiv: 1805.07692. (Page numbers refer to the arXiv version).

\ \\Chen, C.-M., Liu, J.-L., Nester, J.~M.~(2018a). `Quasi-Local Energy from a Minkowski Reference'. {\it General Relativity and Gravitation,} 50, 158, pp.~1-14. arXiv:1811.05640 [gr-qc].

\ \\Christodoulou, D.~and Yau, S.-T.~(1988). `Some Remarks on the Quasi-Local Mass'. In: Isenberg, J.~A.~(Ed.), {\it Mathematics and General Relativity,} pp.~9-14. Providence, Rhode Island: American Mathematical Society.

\ \\Corry, L.~(1999). `Hilbert and Physics (1900-1915)'. In: {\it The Symbolic Universe. Geometry and Physics 1890-1930,} J.~Gray (Ed.), pp.~145-188. Oxford: Oxford University Press.

\ \\Corry, L.~(2004). {\it David Hilbert and the Axiomatization of Physics (1898-1918). From Grundlagen der Geometrie to Grundlagen der Physik.} Dordrecht: Springer.

\ \\Corry, L., Renn, J.~and Stachel, J.~(1997). `Belated Decision in the Hilbert-Einstein Priority Dispute'. {\it Science,} 278 (5341), pp.~1270-1273.

\ \\Curiel, E.~(2019). `On Geometric Objects, the Non-Existence of a Gravitational Stress-Energy Tensor, and the Uniqueness of the Einstein Field Equation'. {\it Studies in History and Philosophy of Modern Physics,} 66, pp.~90-102.

\ \\De Haro, S. (2017). `Dualities and Emergent Gravity: Gauge/Gravity Duality', {\it Studies in History and Philosophy of Modern Physics}, 59, pp.~109-125. arXiv:1501.06162 [physics.hist-ph].

\ \\De Haro, S.~(2017a). `The invisibility of Diffeomorphisms'. {\it Foundations of Physics,} 47, pp.~1464-1497. arXiv:1709.01949 [gr-qc].

\ \\De Haro, S.~(2020). `Spacetime and Physical Equivalence'. In: {\it Beyond Spacetime. The Foundations of Quantum Gravity,} Huggett, N., Matsubara, K.~and Wüthrich, C.~(Eds.), pp.~257-283. Cambridge: Cambridge University Press. arXiv:1707.06581 [hep-th].

\ \\De Haro, S.~(2021). `The Empirical Under-Determination Argument Against Scientific Realism for Dual Theories'. {\it Erkenntnis,} https://doi.org/10.1007/s10670-020-00342-0.

\ \\De Haro, S., Skenderis, K., and Solodukhin, S. (2001). `Holographic reconstruction of spacetime and renormalization in the AdS/CFT correspondence", {\it Communications in Mathematical Physics}, 217 (3), pp.~595-622. arXiv:hep-th/0002230.

\ \\De Haro, S., Skenderis, K.~and Solodukhin, S.~N.~(2001a). `Gravity in Warped Compactifications and the Holographic Stress-Tensor'. {\it Classical and Quantum Gravity,} 18, pp.~3171-3180. arXiv:hep-th/0011230.

\ \\De Haro, S.~, Mayerson, D.~and Butterfield, J.~N.~(2016). `Conceptual Aspects of Gauge-Gravity Duality'. {\it Foundations of Physics,} 46, pp.~1381-1425.

\ \\De Haro, S., Teh, N. and Butterfield, J.~N.~(2015). `On the Relation between Dualities and Gauge Symmetries'. {\it Philosophy of Science,} 83 (5), pp.~1059-1069.

\ \\De Haro, S., Teh, N. and Butterfield, J.~N.~(2017). `Comparing dualities and gauge symmetries', {\em Studies in History and Philosophy of Modern Physics}, 59 (2017), pp.~68-80. 	arXiv:1603.08334. 

\ \\De Haro, S.~and Butterfield, J.N.~(2018). `A Schema for Duality, Illustrated by Bosonization'. In: {\it Foundations of Mathematics and Physics One Century After Hilbert,} Kouneiher, J.~(Ed.), pp.~305-376. Springer. arXiv:1707.06681 [physics.hist-ph].


\ \\Dieks, D., Dongen, J. van, Haro, S. de~(2015). `Emergence in Holographic Scenarios for Gravity'. {\it Studies in History and Philosophy of Modern Physics,} 52 (B), pp.~203-216. arXiv:1501.04278 [hep-th].

\ \\Duerr, P.~M.~(2019). `Fantastic Beasts and Where (Not) to Find them: Energy Conservation and Local Gravitational Energy in General Relativity'. {\it Studies in History and Philosophy of Modern Physics,} 65, pp.~1-14.

\ \\Duerr, P.~M.~(2019a). `Against `functional gravitational energy': a critical note on functionalism, selective realism, and geometric objects and gravitational energy'. {\it Synthese,} \\ https://doi.org/10.1007/s11229-019-02503-3.



\ \\Einstein, A.~(1914). `Die formale Grundlage der allgemeinen Relativit\"atstheorie'. {\it K\"oniglich Preu$\b$ische Akademie der Wissenschaften} (Berlin). {\it Sitzungsberichte,} pp.~1030-1085. In: {\it The Collected Papers of Albert Einstein,} Vol.~6, Doc.~9, pp.~72-130.  Princeton: Princeton University Press.

\ \\Einstein, A.~(1915). `Zur allgemeinen Relativit\"atstheorie'. {\it K\"oniglich Preu$\b$ische Akademie der Wissenschaften} (Berlin). {\it Sitzungsberichte,} pp.~778-786. In: {\it The Collected Papers of Albert Einstein,} Vol.~6, Doc.~21, pp.~214-224. Princeton: Princeton University Press.

\ \\Einstein, A.~(1915a). `Zur allgemeinen Relativit\"atstheorie (Nachtrag)'. {\it K\"oniglich Preu$\b$ische Akademie der Wissenschaften} (Berlin). {\it Sitzungsberichte,} pp.~799-801. In: {\it The Collected Papers of Albert Einstein,} Vol.~6, Doc.~22, pp.~225-229. Princeton: Princeton University Press.

\ \\Einstein, A.~(1915b). `Die Feldgleichungen der Gravitation'. {\it K\"oniglich Preu$\b$ische Akademie der Wissenschaften} (Berlin). {\it Sitzungsberichte,} pp.~844-847. In: {\it The Collected Papers of Albert Einstein,} Vol.~6, Doc.~25, pp.~225-229. Princeton: Princeton University Press.

\ \\Einstein, A.~(1915c). Letter to Paul Ehrenfest. In: {\it The Collected Papers of Albert Einstein,} Vol.~8A, Doc.~173, pp.~228-229.  Princeton: Princeton University Press.

\ \\Einstein, A.~(1916). `Die Grundlage der allgemeinen Relativit\"atstheorie'. {\it Annalen der Physik,} 49, pp.~769-822. In: {\it The Collected Papers of Albert Einstein,} Vol.~6, Doc.~30, pp.~284-339. Princeton: Princeton University Press.

\ \\Einstein, A.~(1916a). `Hamiltonsches Prinzip und allgemeine Relativit\"atstheorie'. {\it K\"oniglich Preu$\b$ische Akademie der Wissenschaften} (Berlin). {\it Sitzungsberichte,} pp.~1111-1116. In: {\it The Collected Papers of Albert Einstein,} Vol.~6, Doc.~41, pp.~410-416. Princeton: Princeton University Press. 

\ \\Einstein, A.~(1918). Letter to F.~Klein, 13 March 1918. In: {\it The Collected Papers of Albert Einstein,} Vol.~8B, Doc.~480, pp.~673-675. Princeton: Princeton University Press. 

\ \\Einstein, A.~(1918a). Letter to F.~Klein, 24 March 1918. In: {\it The Collected Papers of Albert Einstein,} Vol.~8B, Doc.~492, pp.~697-700. Princeton: Princeton University Press. 

\ \\Einstein, A.~(1918b). `Der Energiesatz in der allgemeinen Relativit\"atstheorie', {\it K\"oniglich Preu$\b$ische Akademie der Wissenschaften} (Berlin). {\it Sitzungsberichte,} pp.~448-459. In: {\it The Collected Papers of Albert Einstein,} Vol.~7, Doc.~9, pp.~63-77. Princeton: Princeton University Press.

\ \\Einstein, A.~and Grossman, M.~(1913). `Entwurf einer Verallgemeinerten Relativit\"tstheorie und einer Theorie der Gravitation'. Teubner, Leipzig. In: {\it The Collected Papers of Albert Einstein,} Vol.~4, Doc.~13, pp.~302-343. Princeton: Princeton University Press.

\ \\Fefferman, C.~and Graham, C.R. (1985). `Conformal invariants',  in {\it Elie Cartan et les Math\'ematiques d'aujourd'hui}, Ast\'erisque, 95.

\ \\Fefferman, C.~and Graham, C.R.~(2012). `The Ambient Metric', Annals of Mathematics Studies, number 178.  Princeton and Oxford: Princeton University Press. http://arxiv.org/abs/0710.0919.

\ \\Freedman, D.~Z., Gubser, S.~S., Pilch, K., Warner, N.~P.~(1999). `Renormalization Group Flows from Holography: Supersymmetry and a c-Theorem'. {\it Advances in Theoretical and Mathematical Physics,} 3 (2), pp.~363-417. arXiv:hep-th/9904017.

\ \\Freud, Ph.~(1939). `\"Uber die Ausdr\"ucke der Gesamtenergie und des Gesamtimpulses eines Materiellen Systems in der Allgemeinen Relativit\"atstheorie'.  {\it Annals of Mathematics,} 40 (2), pp.~417-419.

\ \\Friederich, S.~(2015). `Symmetry, Empirical Equivalence, and Identity'. {\it The British Journal for the Philosophy of Science,} 66, pp.~537-559.

\ \\Geroch, R.~(1972). `Structure of the Gravitational Field at Spatial Infinity', {\it Journal of Mathematical Physics}, 13, pp.~956-968.

\ \\Goldberg, J.~N.~(1958). `Conservation Laws in General Relativity'. {\it Physical Review,} 111 (1), pp.~315-320.

\ \\Gomes, H.~(2020). `Holism as the Empirical Significance of Symmetries'. arXiv:1910.05330 [physics.hist-ph], https://arxiv.org/abs/1910.05330.

\ \\Gomes, H.~and Riello, A.~(2020). `Large Gauge Transformations, Gauge Invariance, and the QCD $\theta$-Term'. 	arXiv:2007.04013 [physics.hist-ph], https://arxiv.org/abs/2007.04013.

\ \\Greaves, H.~and Wallace, D.~(2014). `Empirical Consequences of Symmetries'. {\it The British Journal for the Philosophy of Science,} 65, pp.~59-89. 

\ \\Grumiller, D.~and Riegler, M.~(2016). `Most general AdS$_{3}$ boundary conditions'. {\it The Journal of High-Energy Physics}, 1610, 023, pp.~1-22. arXiv:1608.01308 [hep-th].

\ \\Harlow, D.~and Stanford, D.~(2011). `Operator Dictionaries and Wave Functions in AdS/CFT and dS/CFT'. arXiv:1104.2621 [hep-th].

\ \\Hawking, S.~W.~(1979). `The Path-Integral Approach to Quantum Gravity'. In: {\it General Relativity. An Einstein Centenary Survey,} Hawking, S.~W.~and Israel, W.~(Eds.). Cambridge: Cambridge University Press.

\ \\Hawking, S.~W.~and Horowitz, G.~T.~(1996). `The Gravitational Hamiltonian, Action, Entropy, and Surface Terms'. {\it Classical and Quantum Gravity,} 13, pp.~1487-1498.


\ \\Healey, R.~(2009). `Perfect Symmetries'. {\it The British Journal for the Philosophy of Science,} 60, pp.~697-720.

\ \\Heemskerk, I.~and Polchinski, J.~(2011). `Holographic and Wilsonian Renormalization Groups'. {\it Journal of High-Energy Physics,} 6, 31, pp.~1-27. 	arXiv:1010.1264 [hep-th].

\ \\Henningson, M.~and Skenderis, K.~(1998). `The Holographic Weyl Anomaly'. {\it The Journal of High-Energy Physics,} 7, 23, pp.~1-11. arXiv: hep-th/9806087.

\ \\Hilbert, D.~(1916). `Die Grundlagen der Physik (Erste Mitteilung). {\it Nachrichten von der Gesellschaft der Wissenschaften zu Göttingen, Mathematisch-physikalische Klasse,} Heft 3, pp.~395-407.

\ \\Hilbert, D.~(1917). `Die Grundlagen der Physik (Zweite Mitteilung). {\it Nachrichten von der Gesellschaft der Wissenschaften zu Göttingen, Mathematisch-physikalische Klasse,} pp.~53-76.

\ \\Hilbert, D.~(1924). `Die Grundlagen der Physik'. {\it Nachrichten von der Gesellschaft der Wissenschaften zu Göttingen, Mathematisch-physikalische Klasse,} pp.~1-32.

\ \\Hoefer, C.~(2000). `Energy Conservation in GTR'. {\it Studies in History and Philosophy of Modern Physics,} 31 (2), pp.~187-199.

\ \\Horsk\'y, J.~and Novotn\'y, J.~(1969). `Conservation Laws in General Relativity'. {\it Czechoslovak Journal of Physics,} B 19, pp.~419-442.

\ \\Huggett, N.~(2017). `Target Space $\not=$ Space'. {\it Studies in History and Philosophy of Modern Physics,} 59, pp.~81-88.

\ \\Huggett, N.~and W\"uthrich, C.~(2020). {\it Out of Nowhere: Duality,} Chapter 7. Oxford: Oxford University Press (forthcoming). http://philsci-archive.pitt.edu/17217.

\ \\Imbimbo, C., Schwimmer, A., Theisen, S., Yankielowicz, S.~(2000). `Diffeomorphisms and Holographic Anomalies', {\it Classical and Quantum Gravity,} 17, pp.~1129-1138. [hep-th/9910267].

\ \\Ismael, J.~and van Fraassen, B.~(2003). `Symmetry as a Guide to Superfluous Theoretical Structure'. In: {\it Symmetries in Physics. Philosophical Reflections,} Brading, K.~and Castellani, E.~(Eds.), pp.~371-392. Cambridge: Cambridge University Press. 

\ \\Jacobson, T., Kang, G.~and Myers, R.~C.~(1994). `On Black Hole Entropy'. {\it Physical Review D,} 49, pp.~6587-6598. arXiv:gr-qc/9312023.

\ \\Janssen, M.~(2014). ``No Success Like Failure...': Einstein's Quest for General Relativity'. In: {\it The Cambridge Companion to Einstein,} Janssen, M.~and Lehner, C.~(Eds.), pp.~167-227. Cambridge: Cambridge University Press.

\ \\Janssen, M.~and Renn, J.~(2007). `Untying the Knot: How Einstein Found his Way Back to Field Equations Discarded in the Zurich Notebook'. In: {\it The Genesis of General Relativity,} Renn, J.~and Schemel, M.~(Eds.), Volume 2, {\it Einstein's Zurich Notebook: Commentary and Essays,} pp.~839-925. Dordrecht: Springer. 

\ \\Jaramillo, J.~L.~and Gourgoulhon, E.~(2011). `Mass and Angular Momentum in General Relativity'. In: Blanchet, L., Spallicci, A., Whiting, B.~(Eds.), {\it Mass and Motion in General Relativity,} pp.~87-124. Springer.

\ \\Kastrup, H.~(1987). `The Contributions of Emmy Noether, Felix Klein and Sophus Lie to the Modern Concept of Symmetries in Physical Systems'. In: Doncel, M.~G., Hermann, A., Michel, L., Pais, A.~(Eds.), {\it Symmetries in Physics (1600-1980),} proceedings of the 1st International Meeting on the History of Scientific Ideas held at Sant Feliu de Gu\'ixols, Catalonia, Spain. 

\ \\Klein, F.~(1917). `Zu Hilberts erster Note \"uber die Grundlagen der Physik'. {\it Nachrichten von der Gesellschaft der Wissenschaften zu Göttingen, Mathematisch-physikalische Klasse,} pp.~469-482.

\ \\Klein, F.~(1918). `\"Uber die Differentialgesetze f\"ur die Erhaltung von Impuls und Energie in der Einsteinschen Gravitationstheorie'. {\it Nachrichten von der Gesellschaft der Wissenschaften zu Göttingen, Mathematisch-physikalische Klasse,} pp.~171-189.

\ \\Klein, F.~(1918a). Letter to A.~Einstein, 20 March 1918. In: {\it The Collected Papers of Albert Einstein,} Vol.~8, Doc.~487, pp.~685-690. Princeton: Princeton University Press. 

\ \\Komar, A.~(1958). `Covariant Conservation Laws in General Relativity'. {\it Physical Review,} 113 (3), pp.~934-936.

\ \\Kosmann-Schwarzbach, Y.~(2011). {\it The Noether Theorems.} New York Dodrecht Heidelberg London: Springer.

\ \\Kosmann-Schwarzbach, Y.~(2020). `The Noether Theorems in Context'. This volume. \\arXiv:2004.09254 [math.HO].

\ \\Kosso, P.~(2000). `The Empirical Status of Symmetries in Physics'. {\it The British Journal for the Philosophy of Science,} 51 (1), pp.~81-98.

\ \\Kretschmann, E.~(1917). `\"Uber den physikalischen Sinn der Relativit\"atspostulate, A. ~Einsteins neue und seine urspr\"ungliche Relativit\"atstheorie'. {\it Annalen der Physik,} 53 (16), pp.~575-614.

\ \\Kucharzewski, M.~and Kuczma, M.~(1964). `Basic Concepts of the Theory of Geometric Objects'. Warsaw: Pa\'nstwowe Wydawnictwo Naukowe.

\ \\Lam, V.~(2011). `Gravitational and Nongravitational Energy: The Need for Background Structures'. {\it Philosophy of Science,} 78, pp.~1012-1023.

\ \\Landau, L.~D.~and Lifshitz, E.~M.~(1971). {\it The Classical Theory of Fields.} Oxford and New York: Pergamon Press, Third Revised English Edition.

\ \\Landsman, K.~(2020). {\it Foundations of General Relativity: From Geodesics to Black Holes.} Forthcoming at Nijmegen: Radboud University Press.

\ \\Lehmkuhl, D.~(2019). `The Equivalence Principle(s)'. Forthcoming in: {\it Routledge Companion to the Philosophy of Physics.} Routledge.

\ \\Lehmkuhl, D., Schiemann, G., and Scholz, E.~(2017). {\it Towards a Theory of Spacetime Theories'}. Springer. 

\ \\Lewis, D.~(1986). {\it On the Plurality of Worlds.} Oxford: Blackwell.

\ \\Lewis, D.~(1986a). {\it Philosophical Papers,} Volume II. New York, Oxford: Oxford University Press.

\ \\Lewis, D.~(1994). `Humean Supervenience Debugged'. {\it Mind,} 103 (412), pp.~473-490.

\ \\Maggiore, M.~(2005). {\it A Modern Introduction to Quantum Field Theory.} Oxford: Oxford University Press.

\ \\Misner, C.~W., Thorne, K.~S., Wheeler, J.~A.~(1973). {\it Gravitation.} San Francisco: W.~H.~Freeman and Company.

\ \\Murgueitio Ram\'irez, S.~(2020). `A puzzle concerning local symmetries and their empirical significance'. Forthcoming in {\it The British Journal for the Philosophy of Science.} http://philsci-archive.pitt.edu/17002.

\ \\Murgueitio Ram\'irez, S.~and Teh, N.~J.~(2020). `Abandoning Galileo's Ship: The quest for non-relational empirical significance'. Forthcoming in {\it The British Journal for the Philosophy of Science.} http://philsci-archive.pitt.edu/17429.

\ \\Nester, J.~M.~(2004). `General Pseudotensors and Quasilocal Quantities'. {\it Classical and Quantum Gravity,} 21, pp.~S261-280.

\ \\Newman, E.~T.~and Penrose, R.~(1966). `Note on the Bondi-Metzner-Sachs Group', {\it Journal of Mathematical Physics}, 7, pp.~863-870.

\ \\Nijenhuis, A.~(1952). {\it Theory of the Geometric Object}. PhD thesis, University of Amsterdam.

\ \\Noether, E.~(1918). `Invariante Variationsprobleme'. {\it Nachrichten von der Gesellschaft der Wissenschaften zu Göttingen, Mathematisch-physikalische Klasse,} pp.~235-257. 

\ \\Norton, J.~(1985). `What Was Einstein's Principle of Equivalence?' {\it Studies in History and Philosophy of Science, Part A,} 16 (3), pp.~203-246.

\ \\Norton, J.~(1993). `General Covariance and the Foundations of General Relativity: Eight Decades of Dispute'. {\it Reports on Progress in Physics,} 56, pp.~791-858.

\ \\Ohanian, H.~C.~(2010). `The Energy-Momentum Tensor in General Relativity and in Alternative Theories of Gravitation, and the Gravitational vs.~Inertial Mass'. arXiv preprint arXiv:1010.5557, https://arxiv.org/abs/1010.5557.

\ \\Olver, P.~J.~(1986). {\it Applications of Lie Groups to Differential Equations.} Berlin Heidelberg New York: Springer-Verlag. 

\ \\Papadimitriou, I.~and Skenderis, K.~(2004). `AdS/CFT Correspondence and Geometry'. {\it IRMA Lectures in Mathematics and Theoretical Physics,} 8, pp.~73-101. hep-th/0404176 [hep-th].

\ \\Papadimitriou, I.~and Skenderis, K.~(2005). `Thermodynamics of Asymptotically Locally AdS Spacetimes'. arXiv:hep-th/0505190.

\ \\Pauli, W.~(1958). {\it Theory of Relativity.} Oxford: Pergamon Press. New York: Dover Edition, 1981.

\ \\Penrose, R.~(1963). `Asymptotic Properties of Fields and Space-Times', {\it Physical Review Letters}, 10 (2), pp.~66-68.

\ \\Penrose, R.~(1964). `Conformal treatment of infinity', In: {\it Relativity, groups and topology}, pp.~565-584. DeWitt, B.~and DeWitt, C.~(Eds). New York and London: Gordon and Breach. Republished in: {\it General Relativity and Gravitation} (2011) 43, pp.~901-922.

\ \\Penrose, R.~(1982). `Quasi-Local Mass and Angular Momentum in General Relativity'. {\it Proceedings of the Royal Society of London,} A 381, pp.~53-63.

\ \\Penrose, R.~(1988). `Aspects of Quasi-Local Angular Momentum'. In: Isenberg, J.~A.~(Ed.), {\it Mathematics and General Relativity,} pp.~1-8. Providence, Rhode Island: American Mathematical Society.

\ \\Pitts, J.~B.~(2010). `Gauge-Invariant Localization of infinitely many Gravitational Energies from All Possible Auxiliary Structures'. {\it General Relativity and Gravitation,} 42, pp.~601-622.


\ \\Pooley, O.~(2012). `Substantivalist and Relationist Approaches to Spacetime'. In: Batterman, R.~(Ed.), {\it The Oxford Handbook of Philosophy of Physics,} Chapter 15, pp.~1-48. Oxford: Oxford University Press.

\ \\Pooley, O.~(2017). `Background Independence, Diffeomorphism Invariance, and the Meaning of Coordinates'. In: Lehmkuhl et al.~(2017), pp.~105-143.


\ \\Randall, L.~and Sundrum, R.~(1999). `An Alternative to Compactification'. {\it Physical Review Letters,} 83 (23), pp.~4690-4693. arXiv:hep-th/9906064.

\ \\Read, J.~(2016). `The Interpretation of String-Theoretic Dualities'. {\it Foundations of Physics,} 46, pp.~209-235.

\ \\Read, J.~(2016a). {\it Background Independence in Classical and Quantum Gravity} [Master's thesis]. University of Oxford. https://ora.ox.ac.uk/objects/uuid:b22844af-aac9-4adc-a6c5-1e2815c59655

\ \\Read, J.~(2020). `Functional Gravitational Energy'. {\it The British Journal for the Philosophy of Science,} 71, pp.~205-232.

\ \\Read, J.~and M\o ller-Nielsen, T.~(2020). `Motivating Dualities'. {\it Synthese,} 197, pp.~263-291.

\ \\Renn, J.~and Stachel, J.~(2007). `Hilbert's Foundation of Physics: From a Theory of Everything to a Constituent of General Relativity'. In: {\it The Genesis of General Relativity,} Renn, J.~and Schemel, M.~(Eds.), Volume 4, {\it Gravitation in the Twilight of Classical Physics. The Promise of Mathematics,} pp.~857-973. Dordrecht: Springer. 

\ \\Rickles, D.~(2011). `A Philosopher Looks at String Dualities'. {\it Studies in History and Philosophy of Modern Physics,} 42, pp.~54-67.

\ \\Rowe, D.~E.~(1999). `The G\"ottingen Response to General Relativity and Emmy Noether's Theorems'. In: {\it The Symbolic Universe. Geometry and Physics 1890-1930,} J.~Gray (Ed.), pp.~189-233. Oxford: Oxford University Press.

\ \\Rowe, D.~E.~(2019). `Emmy Noether on Energy Conservation in General Relativity'. \\arXiv:1912.03269 [physics.hist-ph].

\ \\Rowe, D.~E.~(2019a). `On Emmy Noether's Role in the Relativity Revolution'. {\it The Mathematical Intelligencer,} 41, pp.~65-72. 

\ \\Sauer, T.~(1999). `The Relativity of Discovery: Hilbert's First Note on the Foundations of Physics'. {\it Archive for History of Exact Sciences,} 53, pp.~529-575.

\ \\Sauer, T.~(2005). `Albert Einstein, Review Paper on General Relativity Theory (1916)'. In: {\it Landmark Writings in Western Mathematics, 1640-1940.} I.~Grattan-Guinness (Ed.), pp.~802-822. Amsterdam: Elsevier. 

\ \\Schouten, J.~A.~(1954). {\it Ricci-Calculus}. Berlin Heidelberg: Springer-Verlag.

\ \\Schr\"odinger, E.~(1950). {\it Space-time Structure}. Cambridge: Cambridge University Press. 

\ \\Sider, T.~(2001). {\it Four-Dimensionalism. An Ontology of Persistence and Time.} Oxford: Oxford University Press.

\ \\Skenderis, K.~(2001). `Asymptotically Anti-de Sitter Space-times and their Stress-Energy Tensor', {\it International Journal of Modern Physics} A, 16 (5), pp.~740-749. [hep-th/0010138].

\ \\Skenderis, K.~(2002). `Lecture Notes on Holographic Renormalization'. {\it Classical and Quantum Gravity,} 19, pp.~5849-5876. arXiv:hep-th/0209067.

\ \\Skenderis, K.~and van Rees, B.~C.~(2008). `Real-Time Gauge/Gravity Duality'. {\it Physical Review Letters,} 101, 081601, pp.~1-4. arXiv:0805.0150 [hep-th].

\ \\Skenderis, K.~and van Rees, B.~C.~(2009). `Real-Time Gauge/Gravity Duality: Prescription, Renormalization and Examples'. {\it The Journal of High-Energy Physics,} 5, 85, pp.~1-70. arXiv:0812.2909 [hep-th].

\ \\Stachel, J.~(2014). `The Hole Argument and Some Physical and Philosophical Implications'. {\it Living Reviews in Relativity,} 17 (1), pp.~1-66.

\ \\Strominger, A.~(2001). `The dS / CFT correspondence', {\it Journal of High-Energy Physics}, 0110, 034.  [hep-th/0106113].

\ \\Sus, A.~(2017). `The Physical Significance of Symmetries from the Perspective of Conservation Laws'. In: Lehmkuhl et al.~(2017), pp. 267-286.

\ \\Szabados, L.~B.~(2009). `Quasi-Local Energy-Momentum and Angular Momentum in General Relativity'. {\it Living Reviews in Relativity,} 12, 4.

\ \\Teh, N.~J.~(2016). `Galileo's Gauge: Understanding the Empirical Significance of Gauge Symmetry'. {\it Philosophy of Science,} 83, pp.~93-118.

\ \\Trautman, A.~(1962). `Conservation Laws in General Relativity'. In: {\it Gravitation: An Introduction to Current Research.} Witten, L.~(Ed.), pp.~169-198. New York and London: John Wiley and Sons.

\ \\Trautman, A.~(1965). `Foundations and Current Problems in General Relativity'. In: {\it Lectures on General Relativity,} Trautman, A., Pirani, F.~A.~E., Bondi, H.~(Eds.), pp.~1-248. Englewood Cliffs, New Jersey: Prentice-Hall. 

\ \\van Fraassen, B.~C.~(1980). {\it The Scientific Image.} Oxford: Oxford University Press.

\ \\Wald, R.~M.~(1984). {\it General Relativity.} Chicago: The University of Chicago Press.

\ \\Wald, R.~M.~(1993). `Black Hole Entropy is the Noether Charge'. {\it Physical Review D,} 48 (8), pp.~R3427-3431. arXiv:gr-qc/9307038.

\ \\Wald, R.~M.~and Zoupas, A.~(2000). `General Definition of ``Conserved Quantities'' in General Relativity and Other Theories of Gravity'. {\it Physical Review D,} 61, 084027, pp.~1-16. arXiv:gr-qc/9911095.

\ \\Wang, M.-T.~(2015). `Four Lectures on Quasi-Local Mass'. arXiv:1510.02931.

\ \\Weinberg, S.~(1972). {\it Gravitation and Cosmology: Principles and Applications of the General Theory of Relativity.} New York London Sydney Toronto: John Wiley and Sons.

\end{document}